\documentclass{article}

\usepackage{arxiv}
\usepackage{subfigure}
\usepackage[utf8]{inputenc} 
\usepackage[T1]{fontenc}    
\usepackage{hyperref}       
\usepackage{url}            
\usepackage{booktabs}       
\usepackage{subcaption}
\usepackage{amsfonts}       
\usepackage{nicefrac}       
\usepackage{microtype}      
\usepackage{gensymb}
\usepackage{amsmath}
\usepackage{cleveref}       
\usepackage{lipsum}         
\usepackage{graphicx}
\usepackage{natbib}
\usepackage{doi}
\usepackage{tikz} 
\usepackage{calc}

\usepackage{multirow}
\usepackage{array}

\title{GraphDOP: Towards skilful data-driven medium-range weather forecasts learnt and initialised directly from observations}

\newbox{\orcid}\sbox{\orcid}{\includegraphics[scale=0.06]{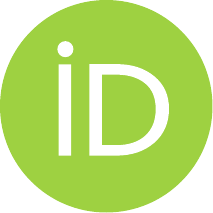}} 

\author{
    {\href{https://orcid.org/0009-0007-7798-6524}{\usebox{\orcid} Mihai Alexe}}
    \And {\href{https://orcid.org/0000-0002-6070-2544}{\usebox{\orcid} Eulalie Boucher}}
    \And {\href{https://orcid.org/0000-0002-3662-5382}{\usebox{\orcid} Peter Lean}}
    \And {\href{https://orcid.org/0000-0003-1869-3426}{\usebox{\orcid} Ewan Pinnington}}
    \And {\href{https://orcid.org/0000-0003-2808-0463}{\usebox{\orcid} Patrick Laloyaux}}
    \And {\href{https://orcid.org/0000-0002-5193-3041}{\usebox{\orcid} Anthony McNally}}
    \And {\href{https://orcid.org/0000-0003-3952-586X}{\usebox{\orcid} Simon Lang}}
    \And {\href{https://orcid.org/0000-0002-1132-0961}{\usebox{\orcid} Matthew Chantry}}
    \And {Chris Burrows}
    \And {\href{https://orcid.org/0000-0002-6673-5400}{\usebox{\orcid} Marcin Chrust}}
    \And {\href{https://orcid.org/0000-0002-3003-3888}{\usebox{\orcid} Florian Pinault}}
    \And {\href{https://orcid.org/0009-0003-6920-3068}{\usebox{\orcid} Ethel Villeneuve}}
    \And {\href{https://orcid.org/0000-0001-5302-6093}{\usebox{\orcid} Niels Bormann}}
    \And {\href{https://orcid.org/0000-0003-4810-9593}{\usebox{\orcid} Sean Healy}}
    \And
    European Centre for Medium-Range Weather Forecasts (ECMWF)
}

\renewcommand{\shorttitle}{GraphDOP: Towards skilful medium-range forecasts learnt directly from observations}

\begin{document}

\maketitle

\begin{abstract}
We introduce GraphDOP, a new data-driven, end-to-end forecast system developed at the European Centre for Medium-Range Weather Forecasts (ECMWF) that is trained and initialised exclusively from Earth System observations, with no physics-based (re)analysis inputs or feedbacks. GraphDOP learns the correlations between observed quantities - such as brightness temperatures from polar orbiters and geostationary satellites - and geophysical quantities of interest (that are measured by conventional observations), to form a coherent latent representation of Earth System state dynamics and physical processes, and is capable of producing skilful predictions of relevant weather parameters up to five days into the future.
\end{abstract}

\section{Introduction}

In recent years, data-driven approaches to numerical weather prediction (NWP) have taken the field by storm, with several global models demonstrating forecast skill scores comparable or superior to that of leading physics-based NWP systems across a wide range of weather variables and lead times \citep{pathak2022fourcastnet,lam2022graphcast,bi2023accurate,bodnar2024aurora,lang2024aifs}. Without exception, these data-driven models have been trained on reanalysis products such as ECMWF's ERA5 \citep{hersbach2020era5}. To produce a forecast, the models must be started from a weather (re)analysis valid at the initial time of the forecast. 

A (re)analysis is the product of data assimilation, a family of algorithms that aim to optimally combine the best available estimate of the current global atmospheric state - e.g., a previous short-range forecast from a physics-based weather model - with information obtained from Earth System observations. For example, the ECMWF runs four-dimensional variational data assimilation (4D-Var; see, e.g., \cite{rabier20004dvar}) to produce high-resolution gridded analyses (currently ca. 9 km along the horizontal and 137 vertical levels) that are then used to initialise the medium-range predictions of both the physics-based Integrated Forecast System (IFS; \cite{ecmwf4dvar48r1doc}) and the data-driven Artificial Intelligence Forecast System (AIFS; \cite{lang2024aifs}). The current operational version of 4D-Var assimilates over 20 million observations during each 12-hour cycle. While an undeniable success \citep{4dvarecmwfhistory}, 4D-Var is also a computationally expensive procedure that requires careful tuning and precise specifications of complex spatio-temporal background and observation error covariances, observation operators, tangent and adjoint model linearisations, instrument-specific variational bias corrections and forecast model error representations. The question then arises \citep{mcnally2024dop}: can machine learning offer an alternative?

Notably, \cite{vaughan2024aardvark} propose Aardvark, a hybrid end-to-end data-driven forecasting system operating at a spatial resolution of 1.41 degrees. Aardvark is pre-trained on the ERA5 reanalysis and then fine-tuned on gridded observations. It shows good forecasting performance at the medium-range when initialised with a limited set of observation data. Other recent works have attempted to use the tangent linear and adjoint of established data-driven models in a traditional data assimilation procedure \citep{tian2024exploringusemachinelearning,Xu2024GSIpangu}, or, more ambitiously, to emulate the entire data assimilation procedure using data-driven methods \citep{rozet2023scorebaseddataassimilation,huang2024diffdadiffusionmodelweatherscale,li2024fuxien4dvar,xu2024fuxida,xiao2024fengwu4dvarcouplingdatadrivenweather,sun2024fuxiweatherdatatoforecastmachine,xiang2024adafartificialintelligencedata}. However, as of yet, none of these methods are able to calculate a weather analysis at comparable resolution and quality to what is routinely produced by traditional data assimilation methods at operational weather centres such as the ECMWF. We also note that previous observation-driven machine learning approaches have focused primarily on nowcasting applications (e.g., \citep{Agrawal2019,sonderby2020metnet,ravuri2021nature,andrychowicz2023,Zhang2023}) that often cover a limited area domain and emphasise high-resolution precipitation and near-surface observations such as 2-metre temperature or 10-metre winds.

Over the past year, ECMWF has been exploring a radically different data-driven approach to learning a \textit{medium-range} weather forecast \textit{exclusively} from Earth System observations, called Artificial Intelligence Direct Observation Prediction, or AI-DOP \citep{mcnally2024dop,mcnally2024}. AI-DOP seeks to learn the dynamics of the atmosphere and the relationships between observed quantities (e.g., satellite brightness temperatures) and physical quantities such as temperature and winds, using only historical time series of satellite and conventional observations. In stark contrast to the medium-range data-driven forecast systems mentioned above, AI-DOP operates \textit{solely on inputs and outputs in observation space, with no gridded climatology and/or NWP (re)analysis inputs or feedbacks}.

Here, we report on results obtained with an end-to-end graph neural network (GNN) forecast model, henceforth referred to as GraphDOP, that learns a latent representation of the atmospheric state from the correlations between measured quantities (e.g., brightness temperatures, bending angles, backscatter coefficients, radar altimeter data, etc.) and relevant geophysical parameters.

We show that GraphDOP is capable of producing skilful forecasts of surface and upper-air weather parameters up to five days into the future. Two-metre temperature (t2m) forecasts from GraphDOP are competitive with those produced by the operational IFS system, with GraphDOP having smaller t2m forecast departures than IFS over the Tropics at lead times of 5 days. The forecasts produced by GraphDOP in two particular weather scenarios - a rapid freezing event in the Arctic and Hurricane Ian - offer compelling evidence that GraphDOP is able to exploit heterogeneous (and largely indirect) Earth System observations and learn a consistent representation of coupled Earth System dynamics and processes.

A concurrent and closely related research effort at ECMWF is exploring the use of end-to-end transformer neural networks for direct observation prediction, with a manuscript \citep{lessig2024dop} currently in preparation for publication.

\section{Datasets}
\label{ref:section-datasets}

\begin{figure}[!ht]
\centering
\subfigure{
    \includegraphics[height=.25\linewidth]{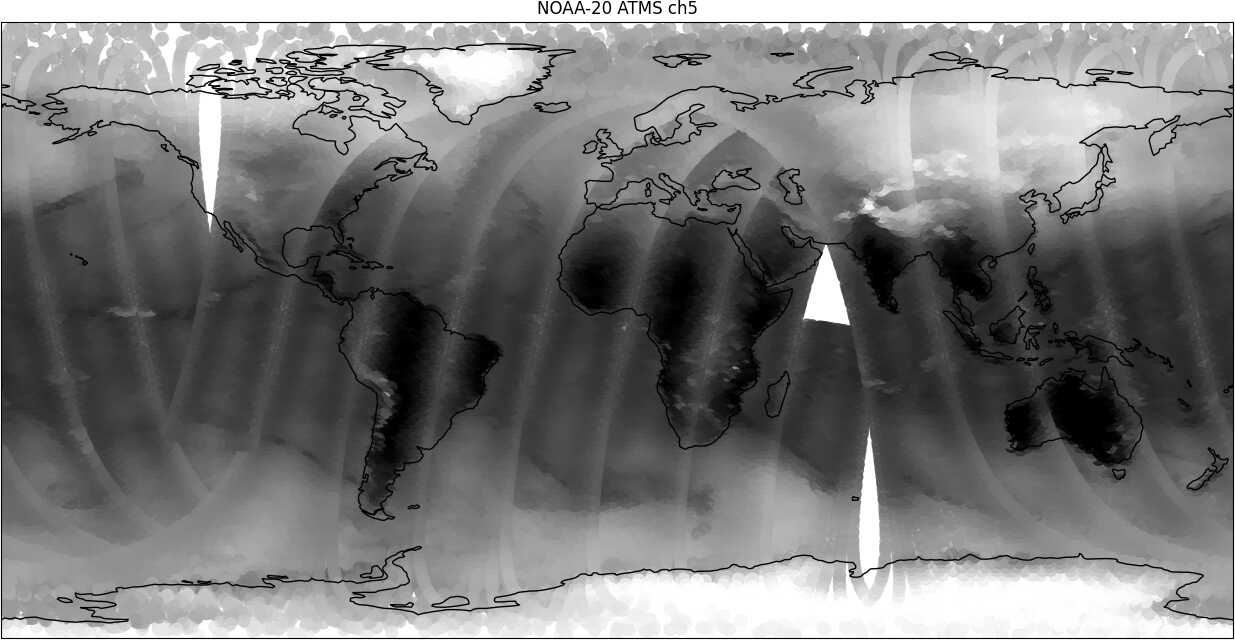}
    \hfill
    \includegraphics[height=.25\linewidth]{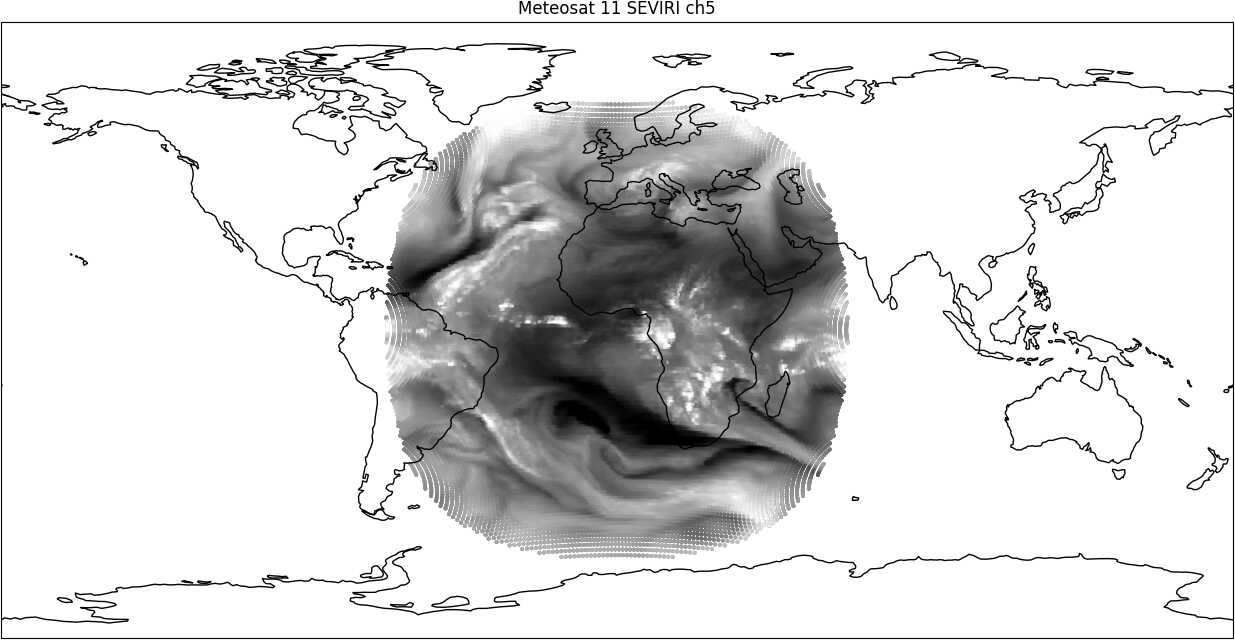}
}
\subfigure{
    \includegraphics[height=.25\linewidth]{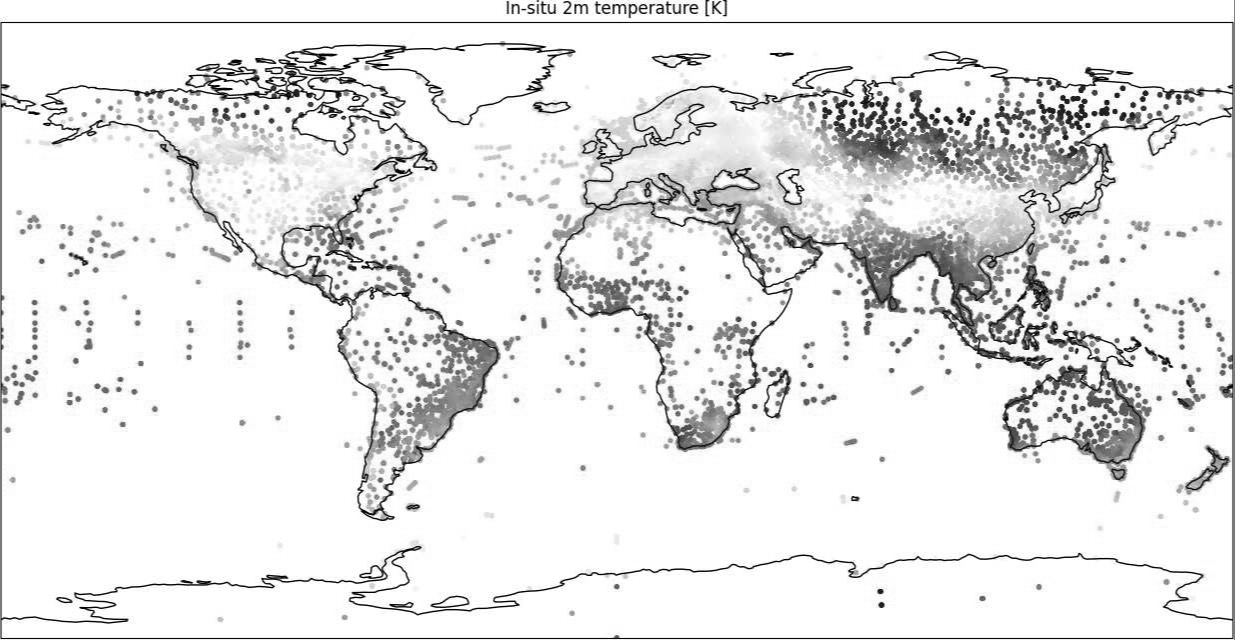}
    \hfill
    \includegraphics[height=.25\linewidth]{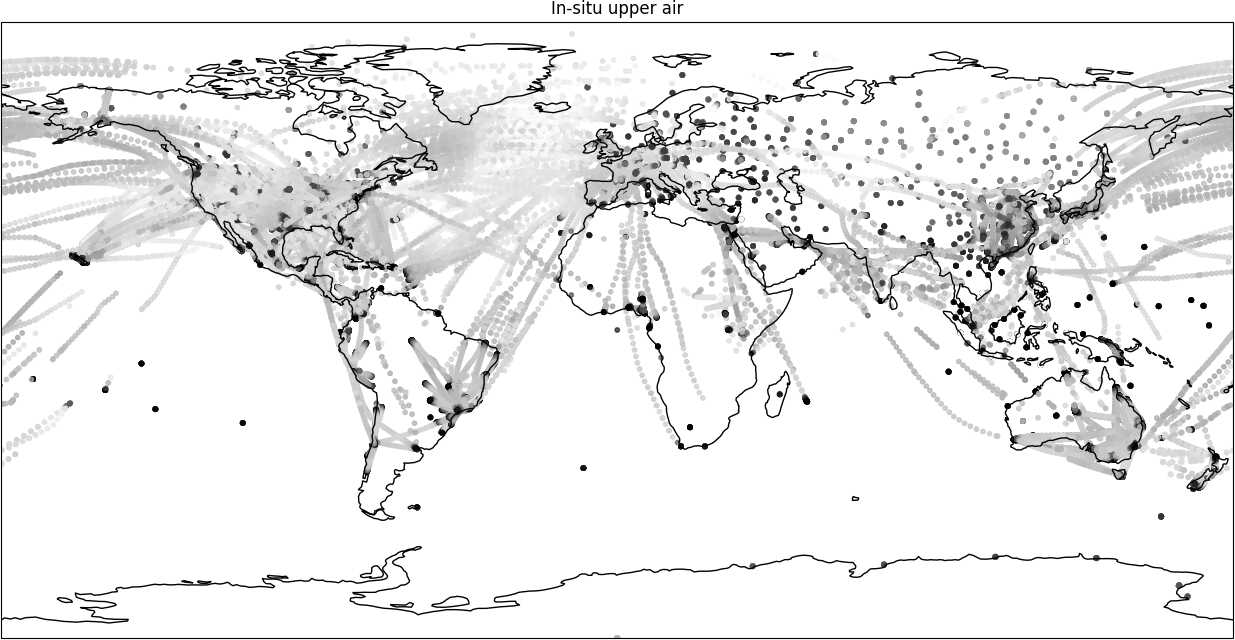}
}
\caption{Examples of data coverage from different observation types in a 12 hour window starting at 21 UTC on January 1st, 2021. NOAA-20 ATMS channel 5 (upper left), Meteosat 11 SEVIRI channel 5 (upper right), in-situ 2m temperature observations (lower left) and in-situ upper air observations (lower right).}
\label{fig:data-coverage}
\end{figure}

Observational data present distinct challenges compared to the gridded reanalysis datasets commonly used in previous studies. Observations are irregular in both space and time (see Figure \ref{fig:data-coverage}), contain both random and systematic errors that may evolve over time, and often measure quantities only indirectly related to the geophysical variables of interest. For example, infrared sounders on satellites do not measure atmospheric temperature, but rather the top-of-atmosphere radiance which is dependent on several geophysical variables including temperature. Furthermore, the coverage of observations changes throughout the training period as satellites are commissioned and decommissioned, and ground-based observation networks evolve.

The focus on medium-range forecasting shapes our data selection strategy. We prioritise observations that capture the large-scale thermodynamic structure of the atmosphere (such as from polar orbiting satellite sounding instruments), as these characteristics fundamentally govern atmospheric dynamics at the medium range.

Our dataset curation process selects observations based on three key criteria: their ability to capture atmospheric thermodynamic structure, the provision of physically meaningful output variables, and independence from reanalysis products or their derivatives. We maintain a strong preference for Level-1 observations in their native spatial and temporal resolution, deliberately avoiding retrieved or regridded Level-3 products that might introduce additional sources of error or hidden dependencies on NWP models. Level-1 observations are georeferenced and calibrated data that have been processed from raw instrument units into physically meaningful quantities such as brightness temperatures. One exception made in this study was in the use of Level 2 significant wave height retrievals derived from radar altimeter data as they provide valuable information about the ocean state. While our observation selection primarily draws from those ingested by ERA5, we specifically include several observation types that current operational data assimilation systems typically cannot fully utilise. These include surface-sensitive channels over land and cloud-affected infrared and microwave radiance data from various instruments, as well as visible spectrum measurements. We prioritise instruments currently in operation with a view towards real-time applications.

Observation quality control (QC) procedures are tailored to individual observation categories. A conservative ERA5 departure (observation minus forecast) QC check helps remove gross outliers for certain conventional observation types. While it is recognised that this introduces a partial dependence of the dataset generation pipeline on the reanalysis system, it is expected that this can be removed in a future iteration of the method. For satellite observations, we have built datasets both with and without the variational bias corrections (VarBC; \citep{dee2004VarBC}) applied to the brightness temperatures to allow further study of this aspect.

Although incomplete, our dataset encompasses most primary observation categories utilised in NWP systems; one notable exception are Atmospheric Motion Vectors (AMVs; \cite{forsythe2007amv}). We exclude AMVs due to their nature as derived products and their typical dependence on NWP-based background fields for height assignment, which would introduce implicit physical model dependencies which we seek to avoid. It is also noted that cross-track microwave humidity sounders such as the MHS onboard the MetOp and NOAA satellites and MWHS-2 onboard the Fengyun-3 series (which are known to have a significant impact in the physics-based assimilation system) have not yet been added.

In this study, the model was trained on 18 years of data between 2004 and 2021, with 2022 used for validation. The data sources used are shown in Figure \ref{fig:dop-obs-inventory}, with further details provided in Table \ref{table:dop-observations} in the Appendix.

\begin{figure}[!htpb]
\centering
\includegraphics[width=\linewidth]{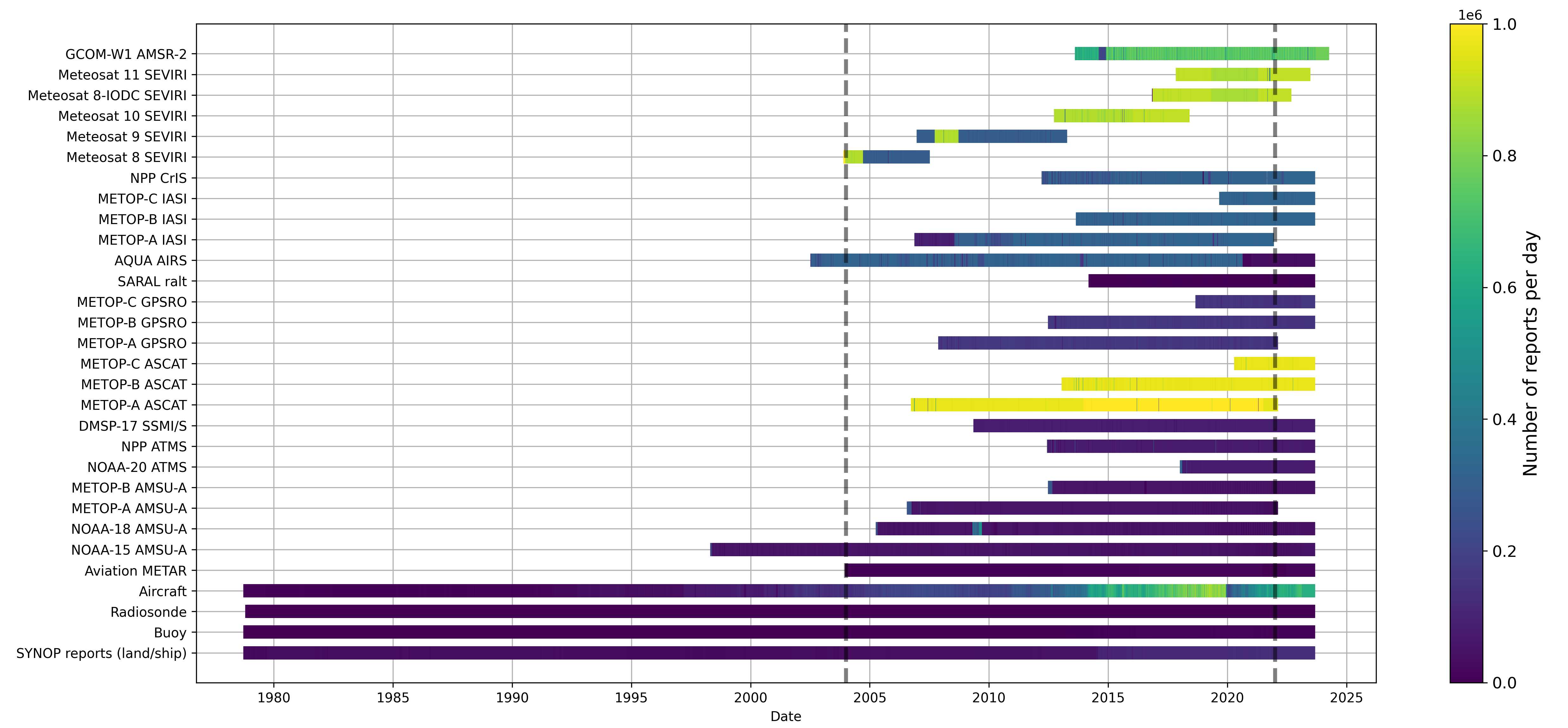}
\caption{A summary and timeline of the observation types currently included in the training dataset. This comprises both in-situ conventional data (from, e.g., surface stations and weather balloons) and Level-1 satellite observations from several instruments, including from geostationary and polar orbiters. Satellite observations are generally indicated by satellite names and instrument names; see the Appendix for a full list. Colours indicate the number of reports per day (each report may contain multiple observed variables or satellite channels). The period used for training the model described in this paper is marked by vertical dashed lines.}
\label{fig:dop-obs-inventory}
\end{figure}

\section{Model}
\label{ref:section-model}

GraphDOP was built around an encoder--processor--decoder network architecture similar to that used in ECMWF's data-driven forecast system AIFS \citep{lang2024aifs}. During training, the model is presented with data from a sequence of one or more non-overlapping observation time intervals (or "windows") as input, and produces a forecast of the observations in the subsequent window.

\begin{figure}[!htpb]
\centering
\includegraphics[width=\linewidth]{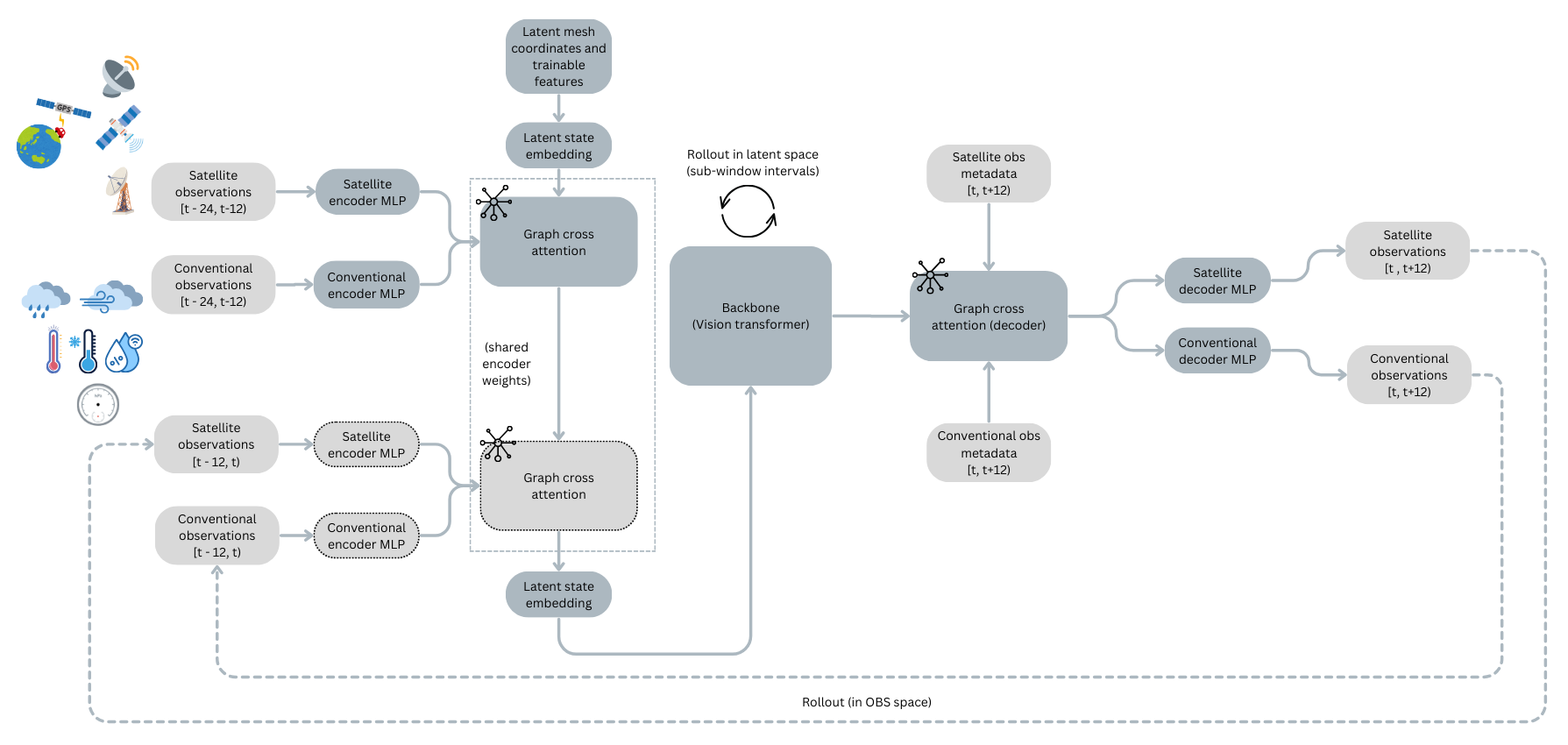}
\caption{A schematic representation of the GraphDOP model. In this illustration, the model receives two 12-hour observation data windows that are embedded sequentially into a latent space representation using graph attention \citep{lang2024aifs}. Multi-layer perceptrons (MLPs) are used to project multi-channel observations from individual instruments to a common feature dimension. The MLP weights are shared across all input windows. One or more invocations of the backbone module (rollout in latent space) advance the state throughout the 12-hour target observation window. A graph transformer decoder calculates intermediate representations of the observations at the target locations; finally, instrument-specific decoder MLPs produce the forecasts. Optionally, the forecasts can be fed back as the input to the next forecast window (rollout in observation space, dashed gray lines).}
\label{fig:dop-model-schematic}
\end{figure}

The high-level architecture of GraphDOP is illustrated in Figure \ref{fig:dop-model-schematic}. The encoder - and decoder - are GNNs that project the observations available inside each input window onto - and out of - a latent space representation of the atmospheric state. Since the location of the observations can vary between different input (and output) windows, the graphs used by the encoder and decoder are allowed to change from one batch of data to the next, and are built on the fly from the observation data, using GPU-optimised functions available in the PyTorch Geometric and PyTorch Cluster software packages \citep{fey2019pyg}. The encoder graph defines a graph edge between each observation inside the current output window and its nearest neighbour on the latent mesh. In the decoder, each target observation is connected by an edge to its three nearest neighbours on the latent mesh. The encoder and decoder edge features are the edge direction (forward bearing) and Haversine distance between the source and target node linked by the edge. We note that both the adjacency structure and edge features of these dynamic graph mappings are constructed exclusively from a subset of the observation \textit{metadata}, specifically the latitude and longitude coordinates of the observation. During inference, the decoder need only use the metadata to calculate a forecast - this means that (as demonstrated below) GraphDOP can produce forecasts at arbitrary locations and/or times inside the target window, including those where "real" observations from a particular instrument may not be available.

The processor module is a transformer with windowed attention \citep{lang2024aifs} and is responsible for advancing the latent atmospheric state representation forward in time throughout the target window, either in a single "step", or over several steps, each equal to a fraction of the total length of the output window. Furthermore, autoregressive rollout \citep{keisler2022forecasting,lam2022graphcast,lang2024aifs} allows the model to produce forecasts at longer lead times, by feeding back its current forecast window as the input for the next step.

The training objective is a weighted mean squared error (WMSE) accumulated over $T \ge 1$ target observation windows. GraphDOP weights the squared error contribution from each satellite channel and conventional observation by a fixed value; these empirical weights $w_{c,i}$ were chosen to balance the loss contributions of observations $o \in {\cal O}_{ic}$ from individual channels $c \in {\cal C}_i$ of a given instrument $i \in {\cal I}$. In addition, we support per-satellite (or conventional observation) weights $w_i$, that can be used to assign higher importance to one or more observation targets during model training or fine-tuning. If $y$ denotes the "true" observation and $\hat{y}$ is the model prediction, the GraphDOP WMSE objective can be written as follows:

\begin{equation}
{\cal L}_{\rm{DOP}} := \frac{1}{T \times | {\cal I} | \times |{\cal C}| \times |{\cal O}|} \sum_{t=1}^T \sum_{i \in {\cal I}} w_i \sum_{c \in {\cal C}_i} w_{c,i} \sum_{o \in {\cal O}_{ic}} \left(y_{tico} - \hat{y}_{tico} \right)^2 \,.
\label{eqn:dop-loss}
\end{equation}

Here $| \cdot |$ denotes the size of the respective set, with $|{\cal C}| = \sum_{i \in {\cal I}} |{\cal C}_i|$ and $|{\cal O}| = \sum_{i \in {\cal I}, c \in {\cal C}_i} |{\cal O}_{ic}|$.

The model version described herein was trained on 64 NVIDIA H100 64~GB GPU devices for 70,000 steps using mixed precision (\texttt{float16}; \cite{micikevicius2018mixed}). Sequence parallelism (see \cite{lang2024aifs} for a description) allows the input and output graphs to be sharded (i.e., split) across 8 GPU devices, resulting in an effective batch size of 8. The latent channel dimensions of the encoder, processor and decoder were all set to 1024. During training, the learning rate was annealed to $3 \times 10^{-7}$ from a starting value of $10^{-3}$ using a cosine scheduler with a linear warm-up interval of 1000 steps. During training, the model received one 12-hour window of observations as input and forecasted all observations within the next 12-hour window. To allow the model to optimise the forecast over a longer time horizon, starting from step 63000, we gradually increase $T$ by 1 every 1500 training steps. In contrast to traditional assimilation methods, GraphDOP does not use a background state. The windows seen by GraphDOP are aligned with those currently used in the operational long-window assimilation (LWDA) configuration of 4D-Var at ECMWF \citep{ecmwf4dvar48r1doc}, i.e., 09z - 21z and 21z - 09z. The observations used for training and validation are listed in Table \ref{table:dop-observations} in the Appendix.

The 40320 latent space (processor) nodes lie on an O96 reduced Gaussian grid \citep{Wedi2014}, with a spatial resolution of approximately 1 degree (111 km). To prevent overfitting during the training procedure, we randomly drop a quarter of the satellite observations and half of the conventional observations inside each input and output window.

\section{Qualitative evaluation}
\label{ref:section-qualitative-evaluation}

\subsection{IASI surface-sensitive channels} 

GraphDOP was trained to produce forecasts for several types of observations, including satellite brightness temperatures. While they are not geophysical parameters of direct user interest, raw satellite brightness temperatures are sensitive to various physical properties of the Earth System (atmosphere, ocean and land) and their correct forecast, particularly at medium-range lead times, is essential for the viability of an observation-driven forecast model. This is because polar orbiting satellites provide global coverage and are almost continuous in time, with very high spatial density. Together with geostationary instruments, polar orbiters produce valuable information over areas where conventional observations are sparse, for instance over oceans \citep{Bauer2015}. Observing System Experiments (OSEs) using conventional data assimilation confirm that accurate medium-range forecasting would not be possible without satellite data \citep{McNally2014}. Therefore, it is essential to understand how well an observation-driven forecast model is able to exploit and predict these observations.

\begin{figure}[!ht]
\centering
\includegraphics[width = .95\linewidth]{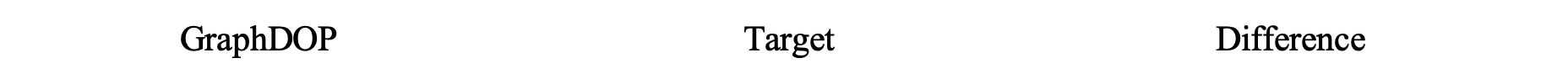}
\subfigure[Day one]{\includegraphics[trim=0 0 55 0, clip, height=.16\linewidth]{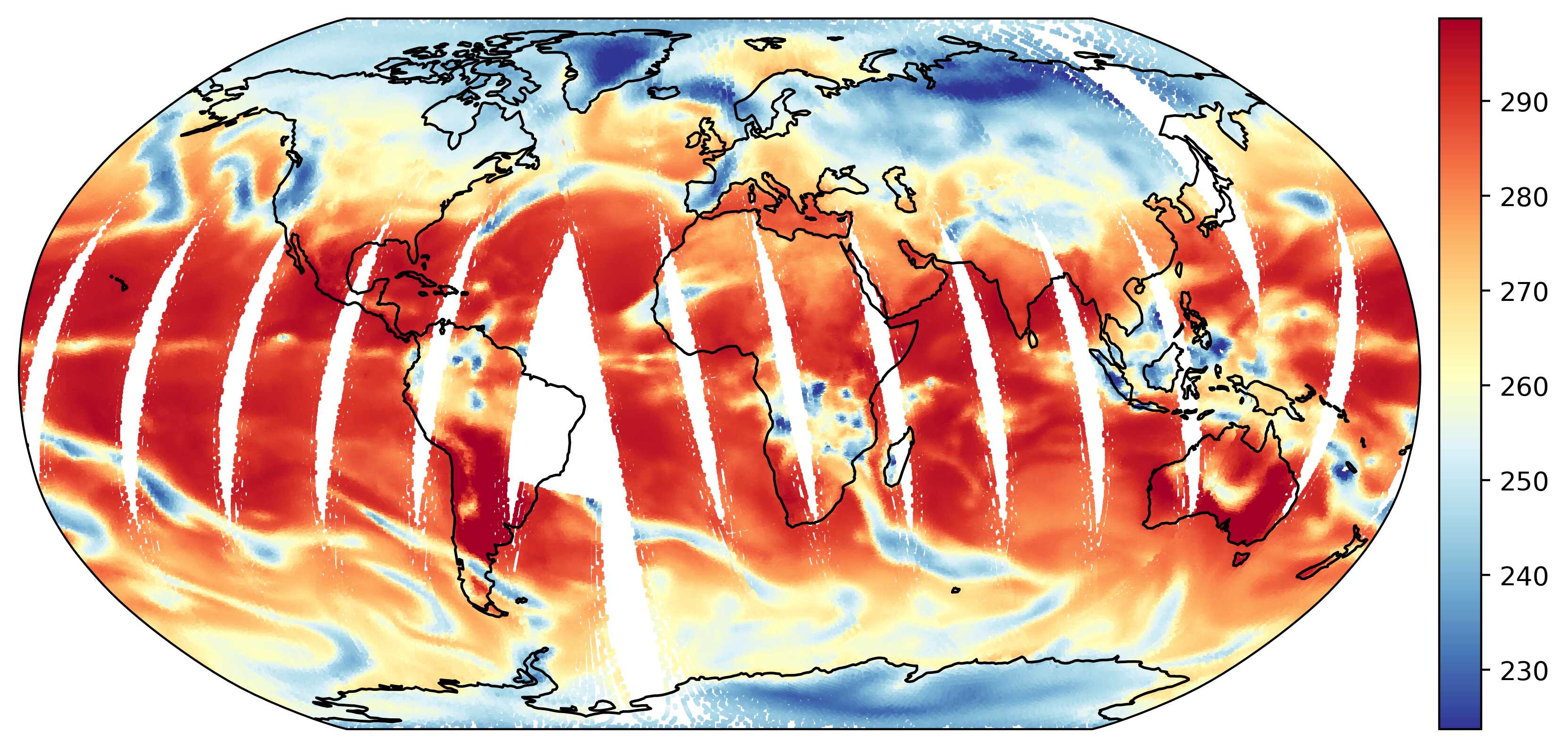} \includegraphics[height=.16\linewidth]{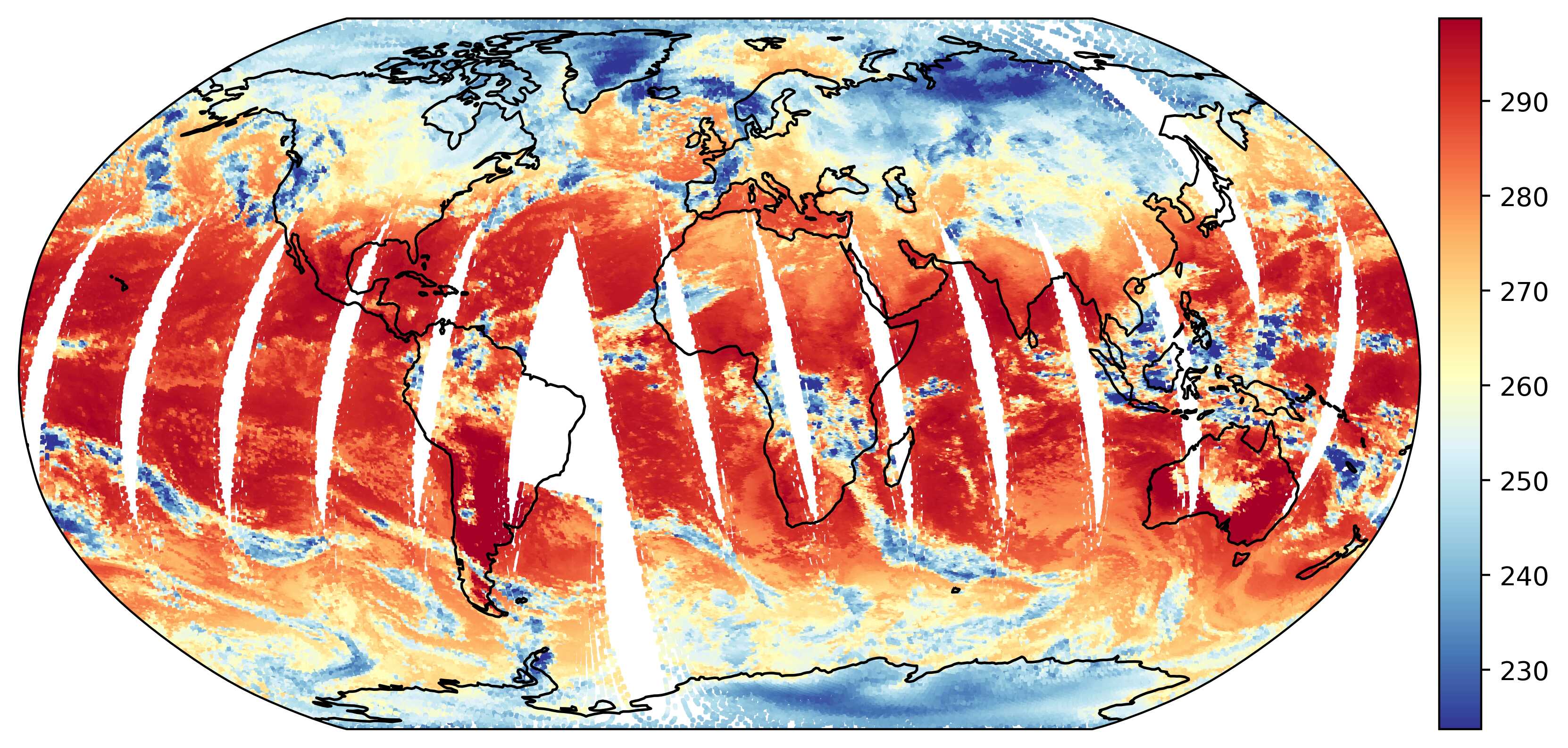}\includegraphics[height=.16\linewidth]{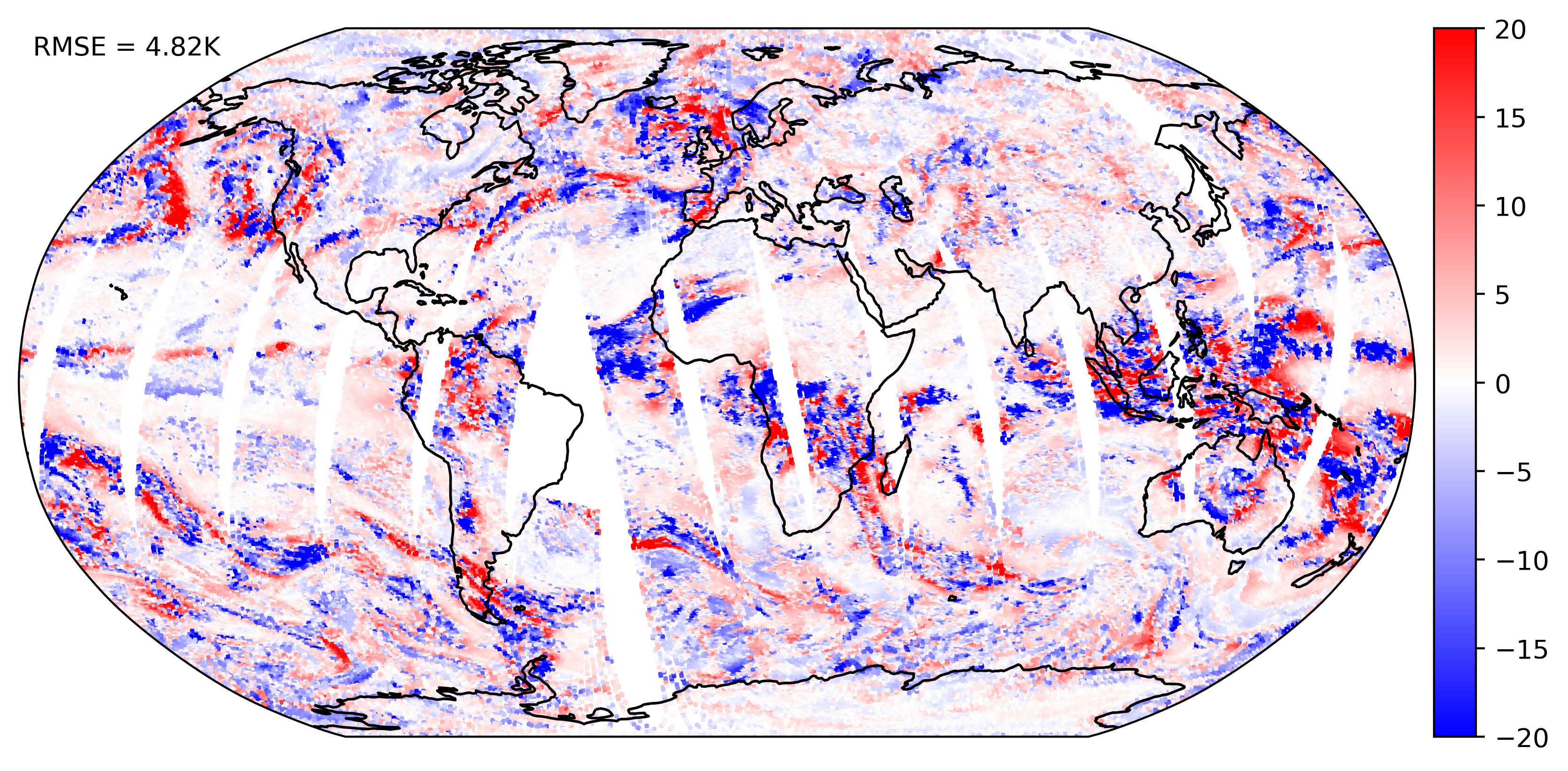}} \\
\subfigure[Day two]{\includegraphics[trim=0 0 55 0, clip, height=.16\linewidth]{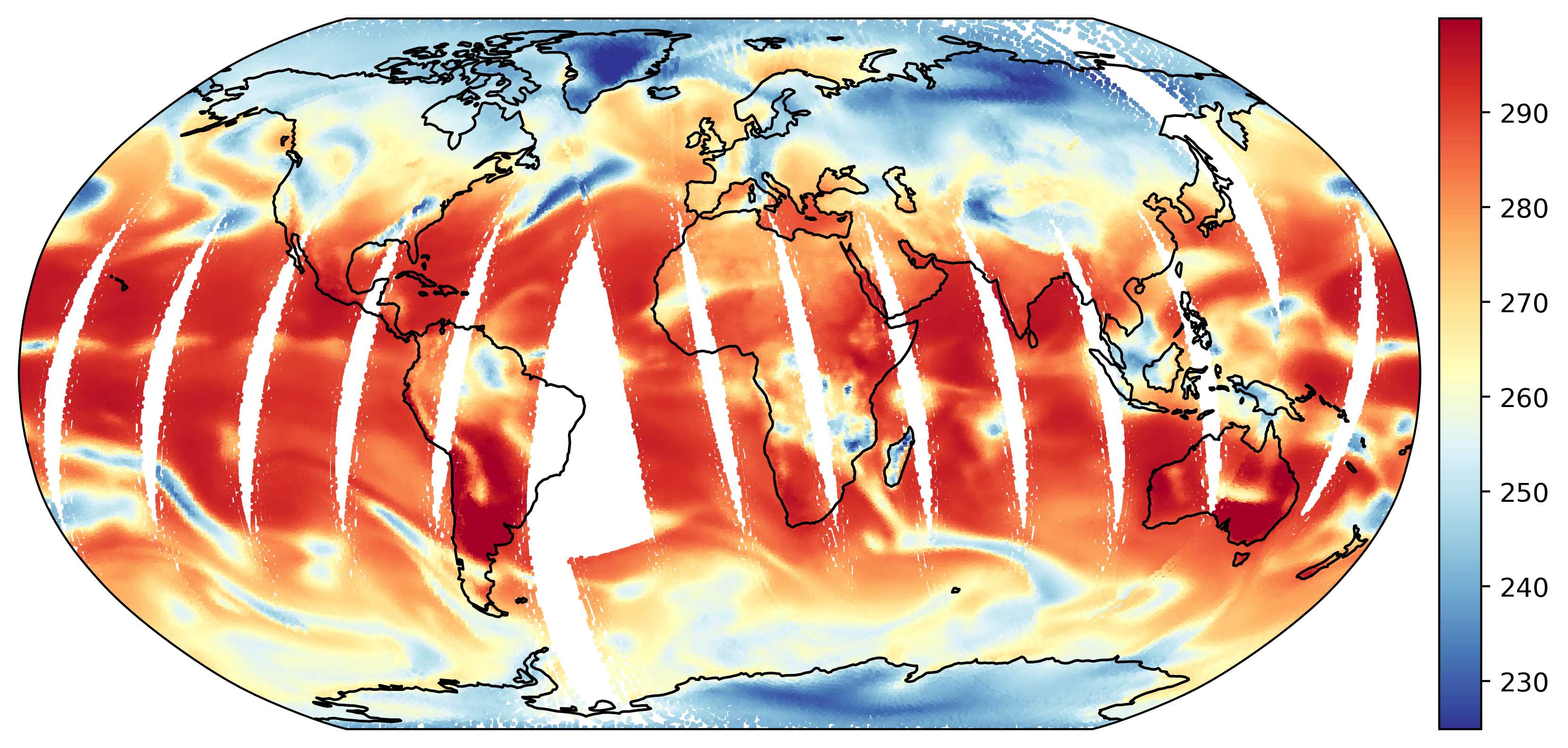} \includegraphics[height=.16\linewidth]{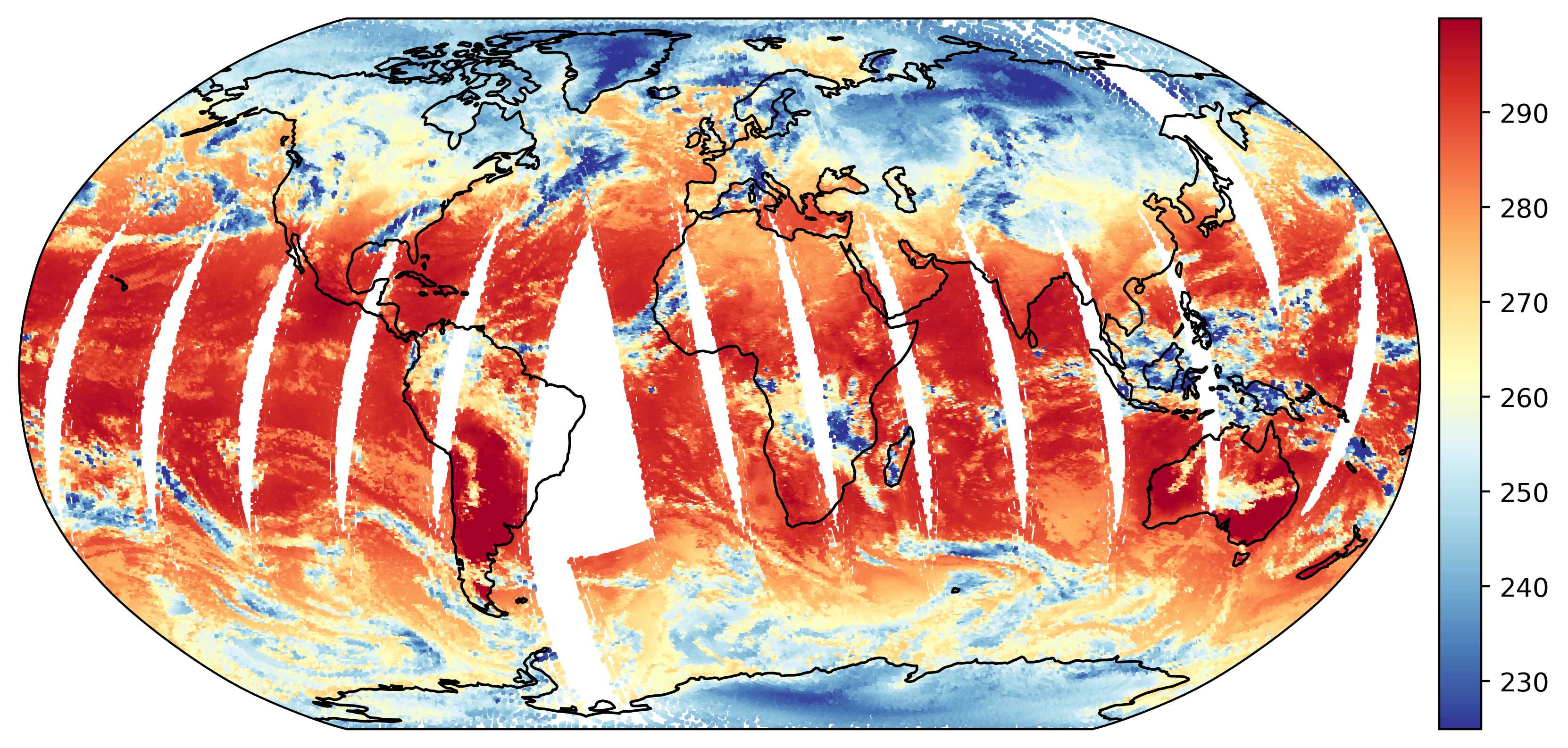}\includegraphics[height=.16\linewidth]{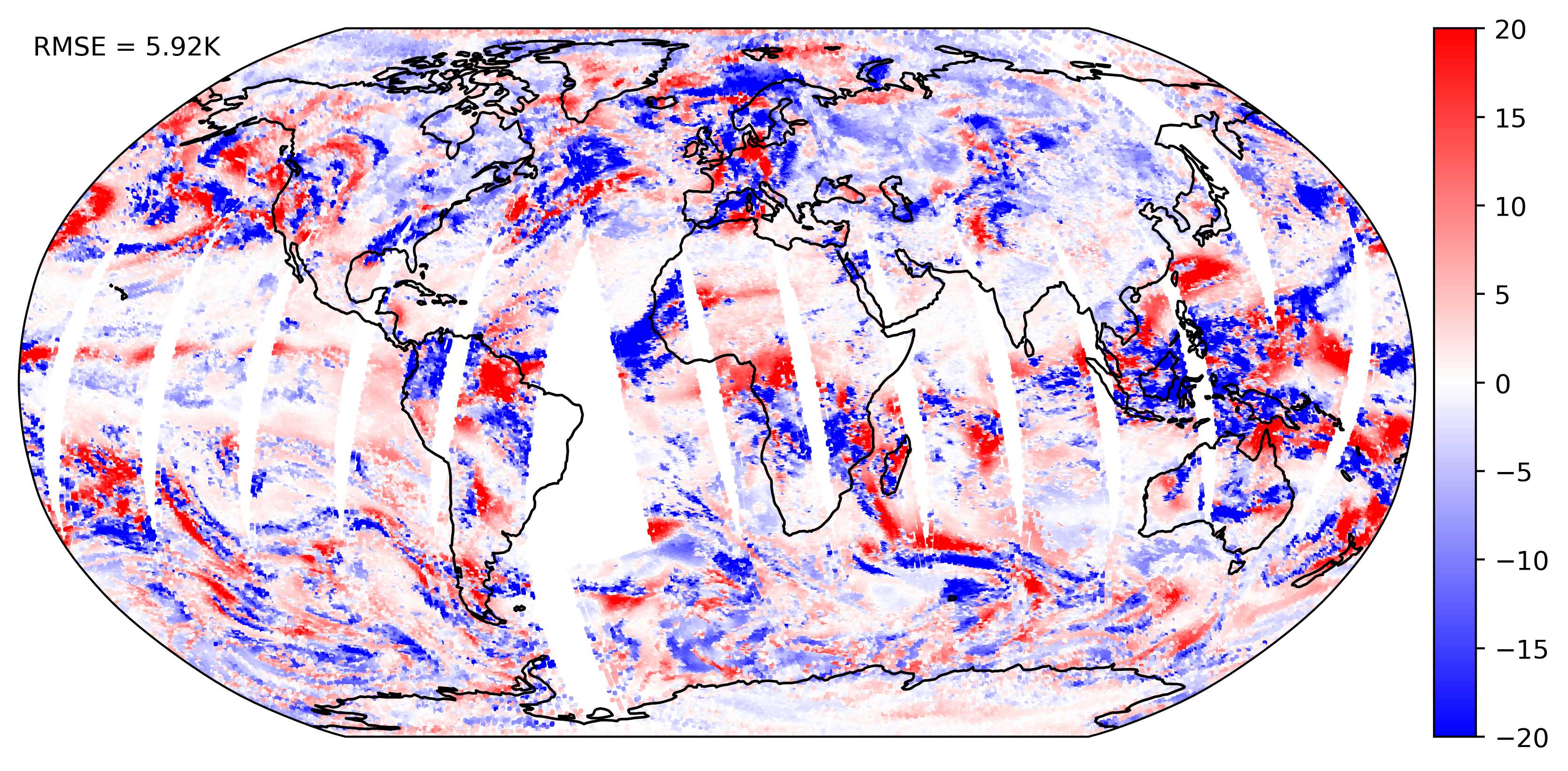}} \\
\subfigure[Day three]{\includegraphics[trim=0 0 55 0, clip, height=.16\linewidth]{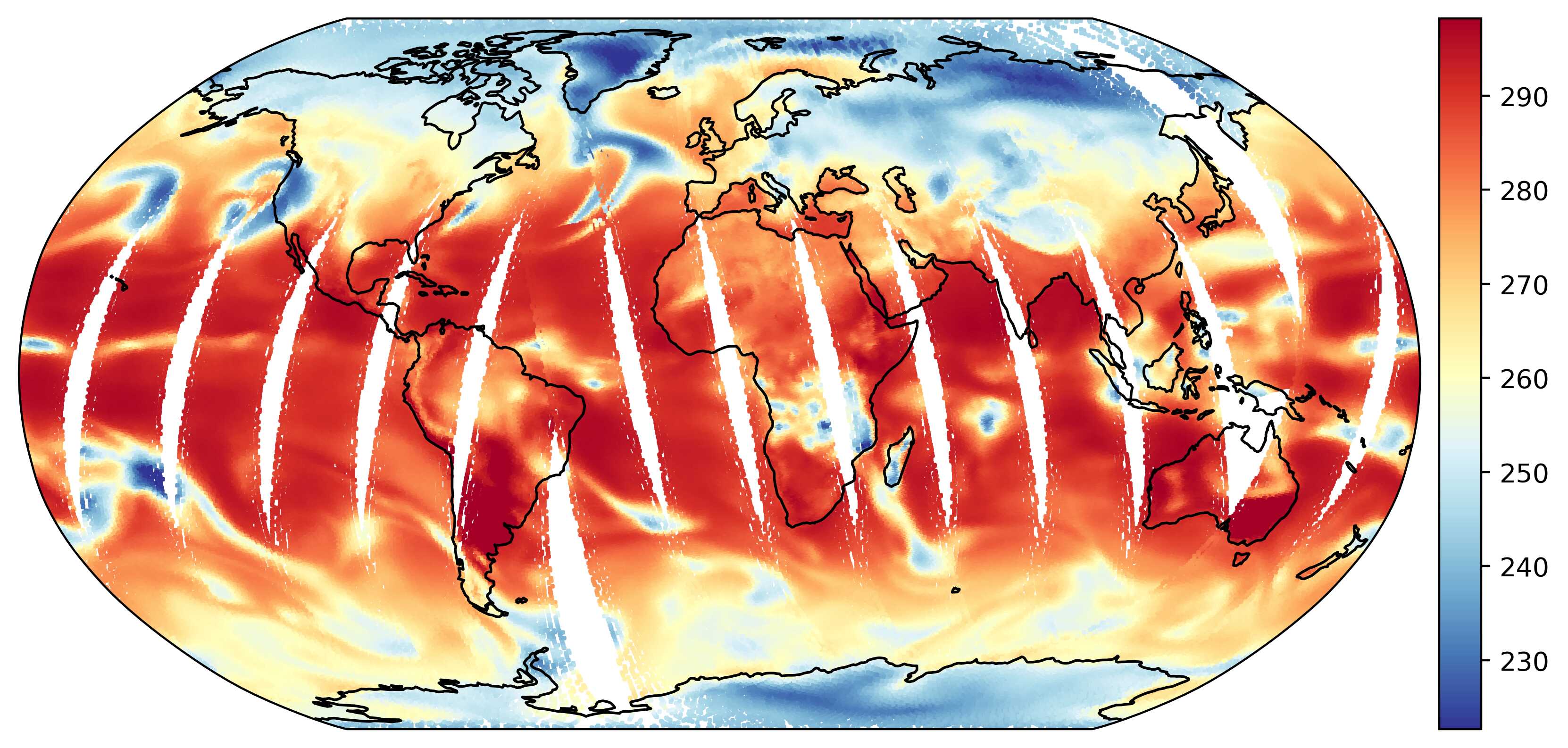} \includegraphics[height=.16\linewidth]{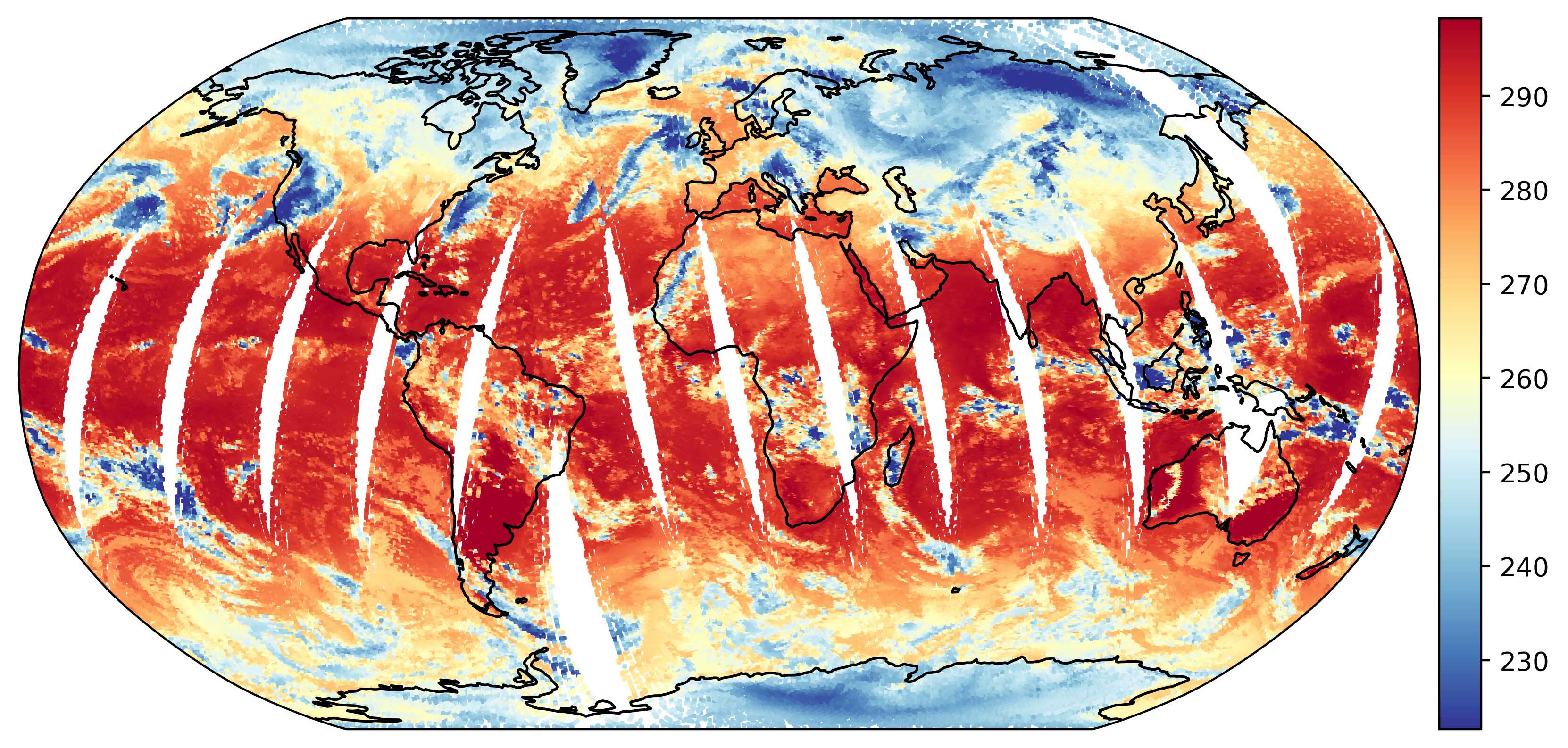}\includegraphics[height=.16\linewidth]{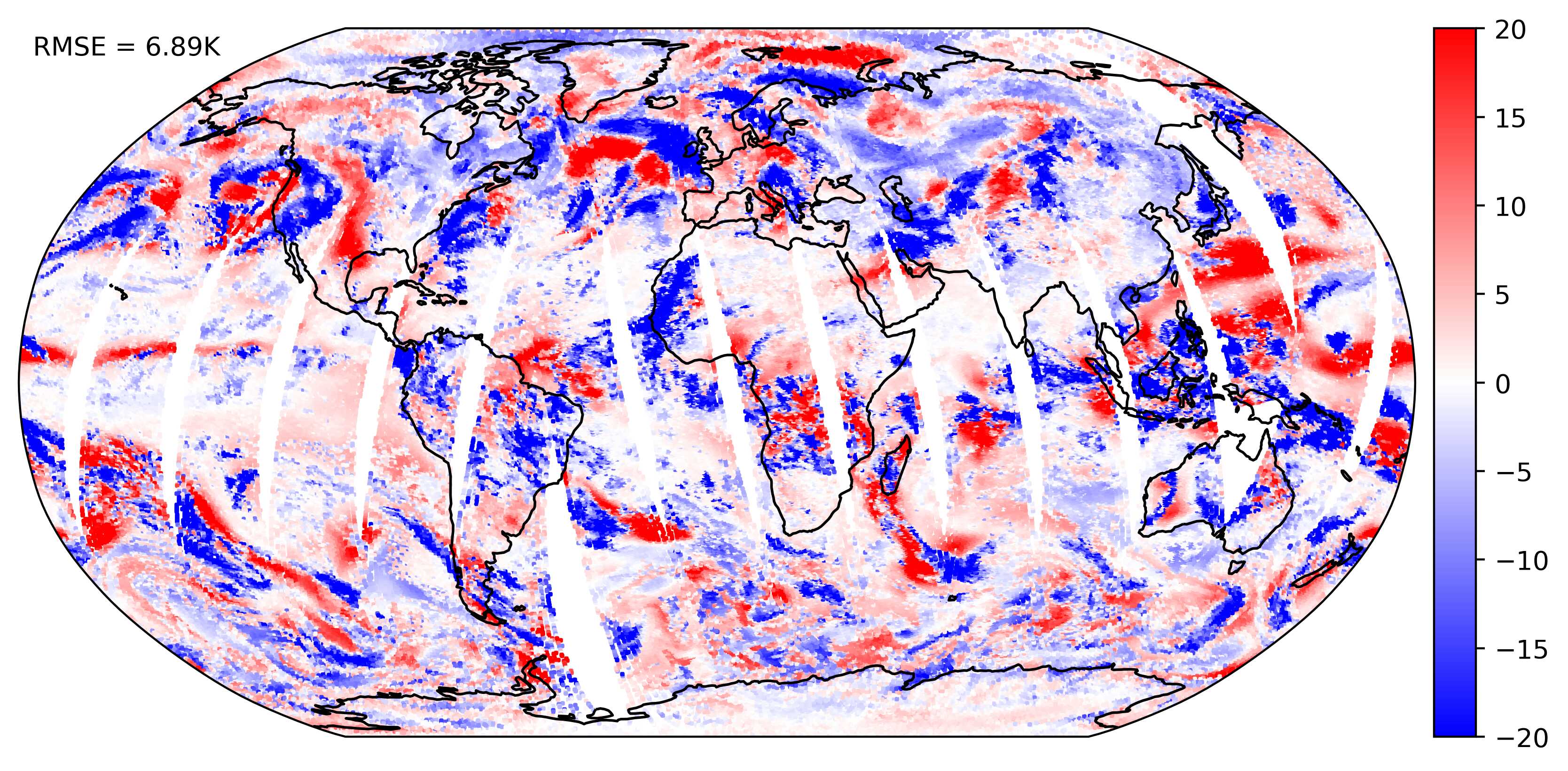}} \\
\subfigure[Day four]{\includegraphics[trim=0 0 55 0, clip, height=.16\linewidth]{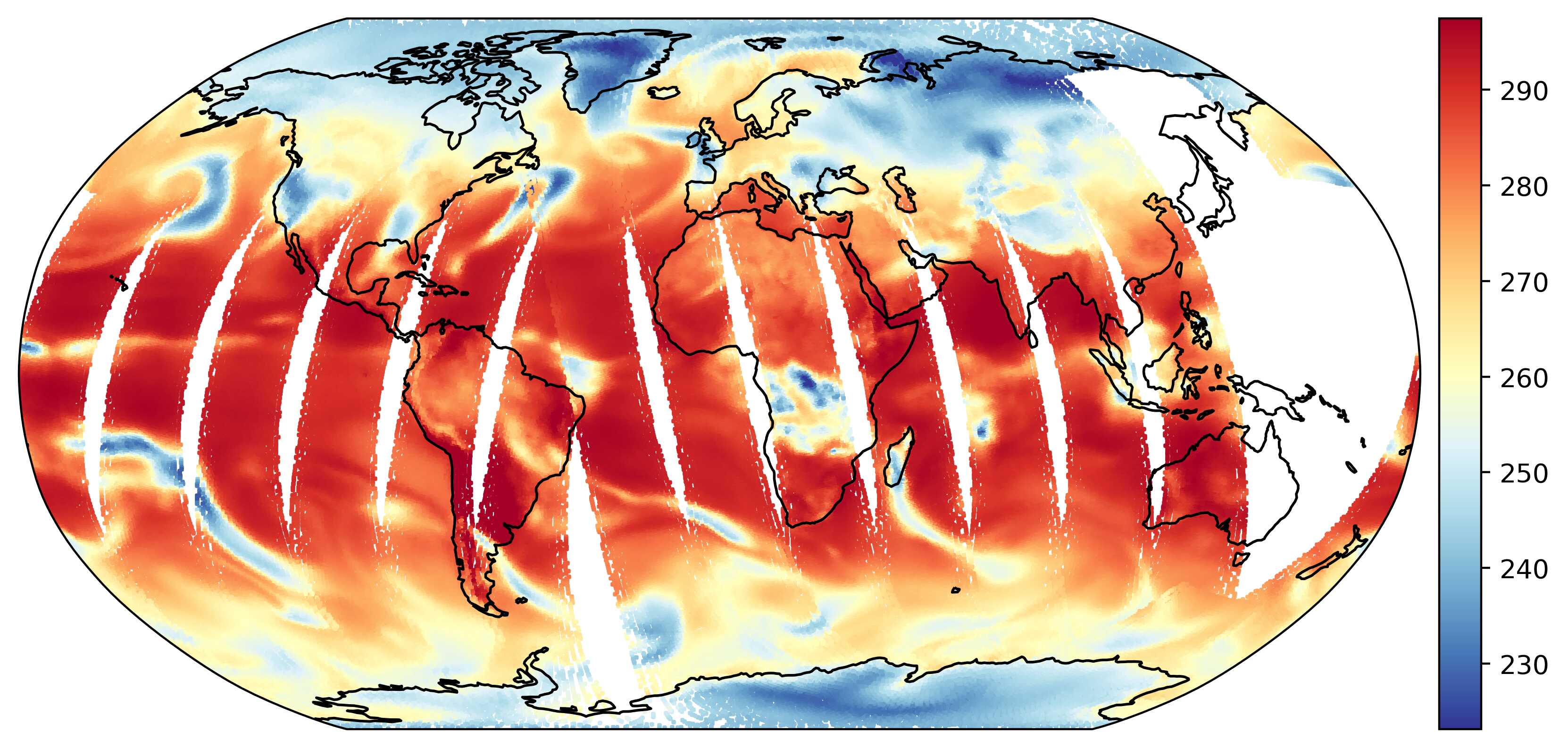} \includegraphics[height=.16\linewidth]{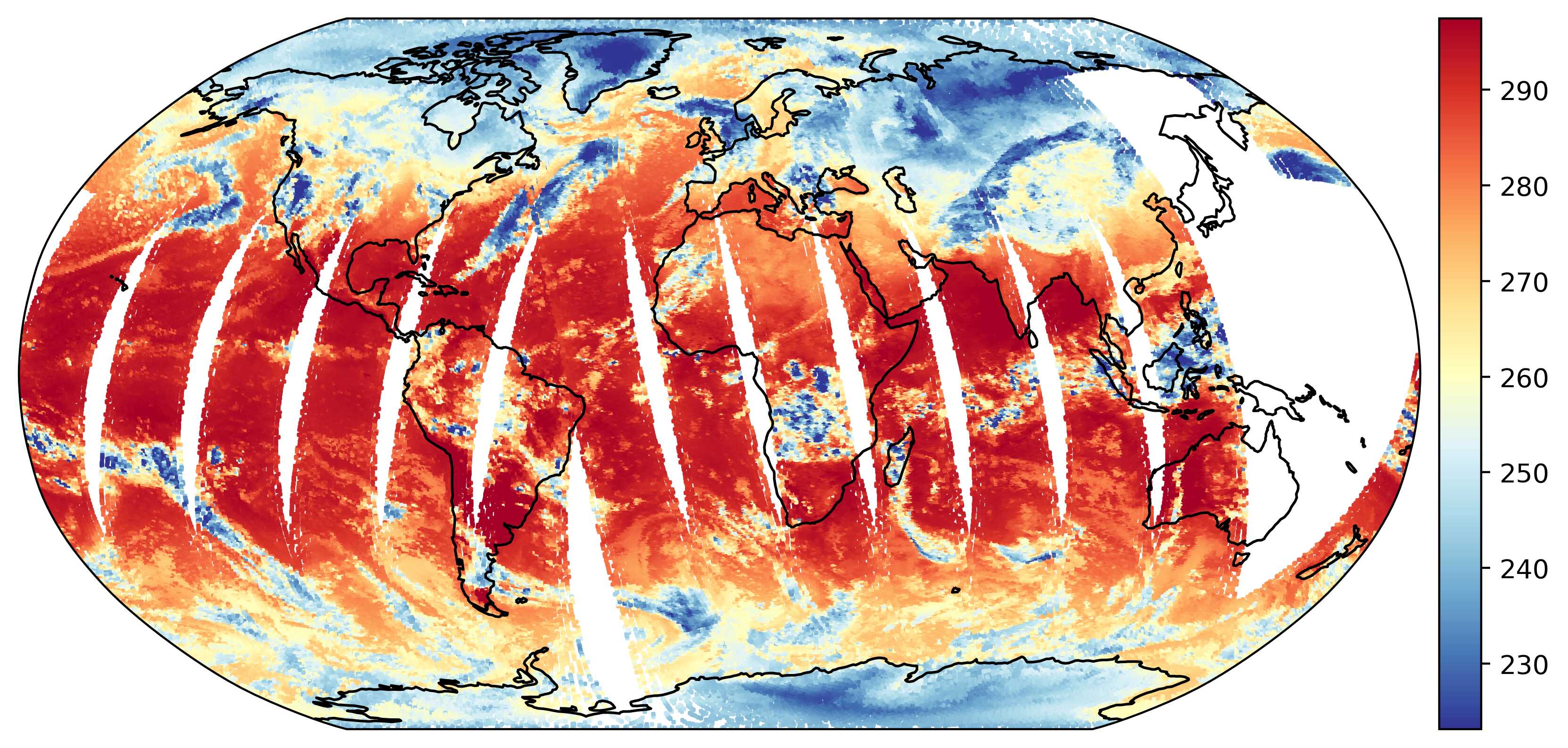}\includegraphics[height=.16\linewidth]{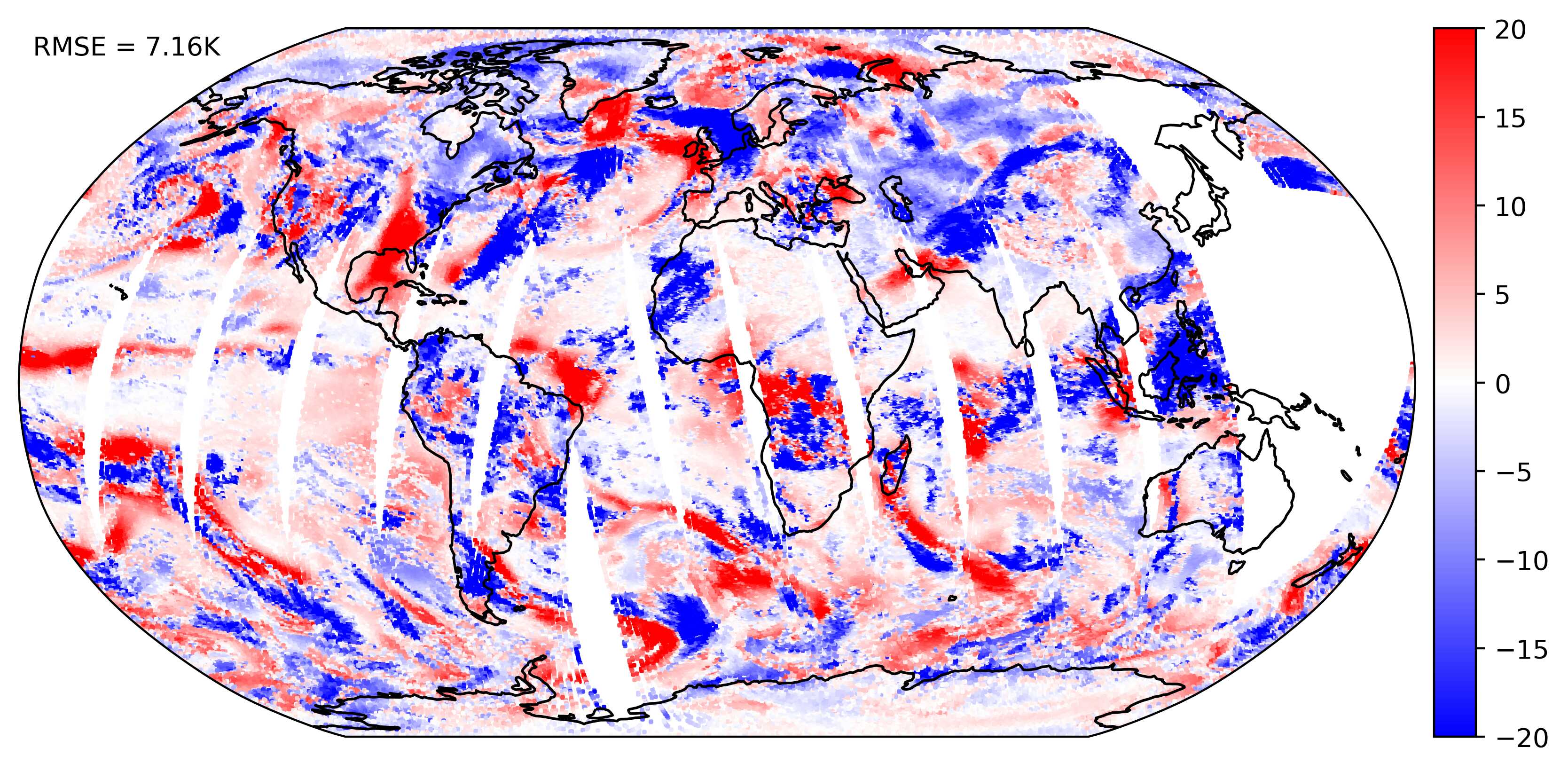}}\\
\caption{IASI channel 921 (wavenumber 875.0 $cm^{-1}$) brightness temperatures (K): forecasted (left), observed (middle), and difference
(observed minus forecast; right). We show 12-hour samples, starting from forecast day 1 (Jan 7, 2023; top row) through to day 4 (Jan 11, 2023; bottom row). The global forecast RMSE for the 12 hour sample is printed in the top left corner. Blue shades are indicative of "cold" features such as clouds, while red shades correspond to "warm" features, e.g., warm surface areas unobscured by clouds.}
\label{fig:IASI_4days}
\end{figure}

Figure \ref{fig:IASI_4days} shows the four-day evolution of brightness temperature forecasts (left) compared to observations (middle) for the Infrared Atmospheric Sounding Interferometer (IASI) channel 921 (wavenumber 875.0 cm$^{-1}$). The forecasts demonstrate that GraphDOP is able to represent the evolution of synoptic-scale weather systems. In particular, when looking at Europe, we identify a frontal cloud feature over Spain that is present in the day-one forecast. This front moves eastward, ending up over south-eastern Europe at day four. The movement of the feature is captured very well by the GraphDOP model. Day-three forecasts also show two hook-shaped cloud features in the North Pacific.

It is apparent that the GraphDOP forecast becomes smoother at longer lead times, as higher spatial frequency features are progressively dampened. This can be attributed in large part to the deterministic WMSE training objective used for GraphDOP \citep{blogaifs2,lang2024aifs}. The WMSE promotes spectral smoothing in the forecast fields - to avoid so-called "double-penalties" \citep{Hoffman1995,ebertdouble} - particularly over long training rollout sequences. We plan to address this in a future model version trained to optimise a probabilistic objective \citep{karras2022elucidating,alexe2024ecmwfnl,lang2024ensscore}. Future work will also investigate the use of principal component (PC) scores derived from IASI spectra \citep{Matricardi2014}, that would allow finer control over specific features deemed important in the optimization, e.g., cloud information (that is largely contained in the first PC eigenvector).

\subsection{Gridded forecasts of weather parameters evaluated against ERA5}

The model is capable of producing forecasts of observations at arbitrary time and space locations. The forecasts shown here were produced on an o96 reduced Gaussian grid. We note that any observation used during training can be forecasted on such a grid. As in the training, our choices of input/output windows are aligned with those used by traditional 4D-Var to produce the ERA5 analysis, namely 21z - 09z and 09z - 21z. 

\begin{figure}[!ht]
\centering
\includegraphics[trim=0 0 50 0, clip, height=.25\linewidth]{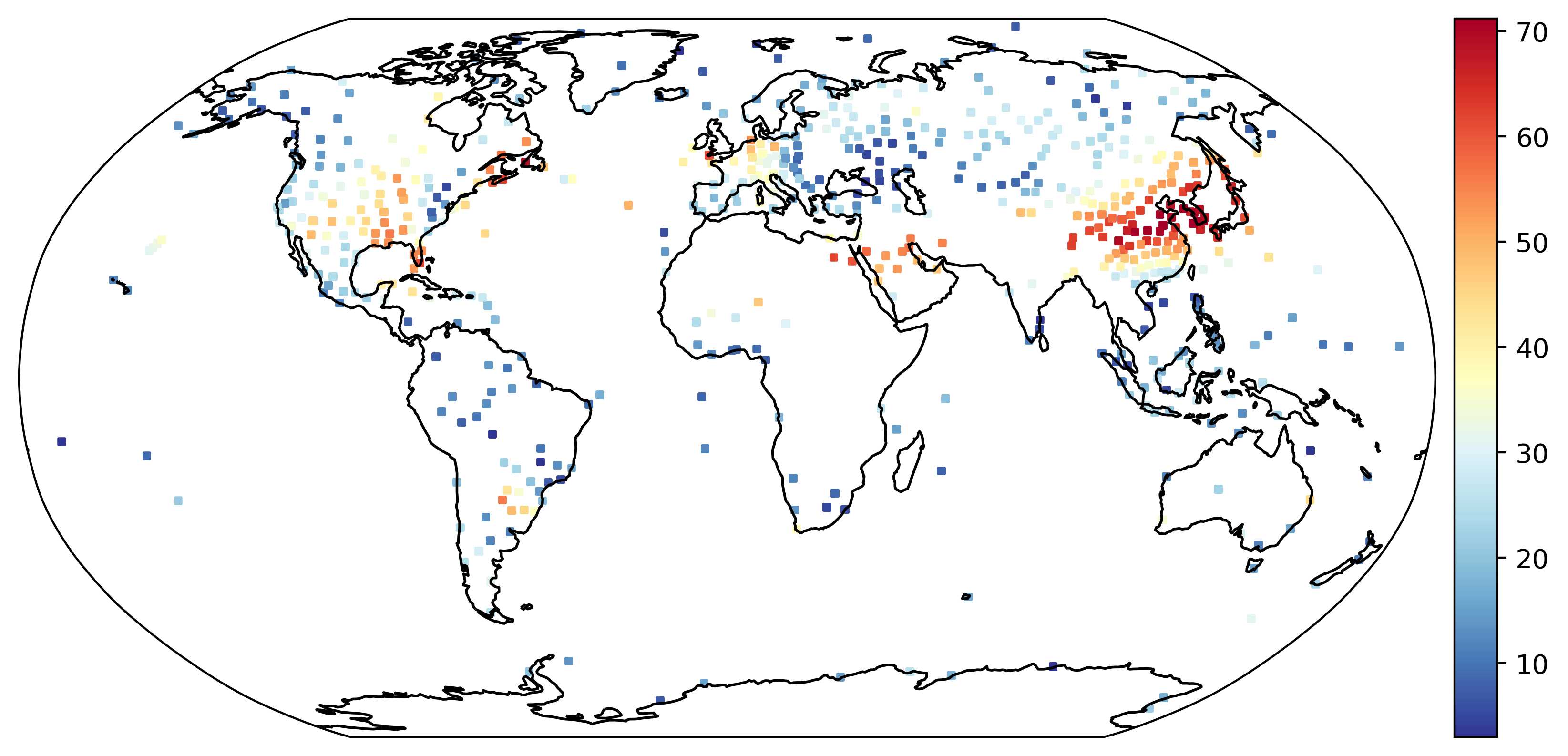}
\hfill
\includegraphics[height=.25\linewidth]{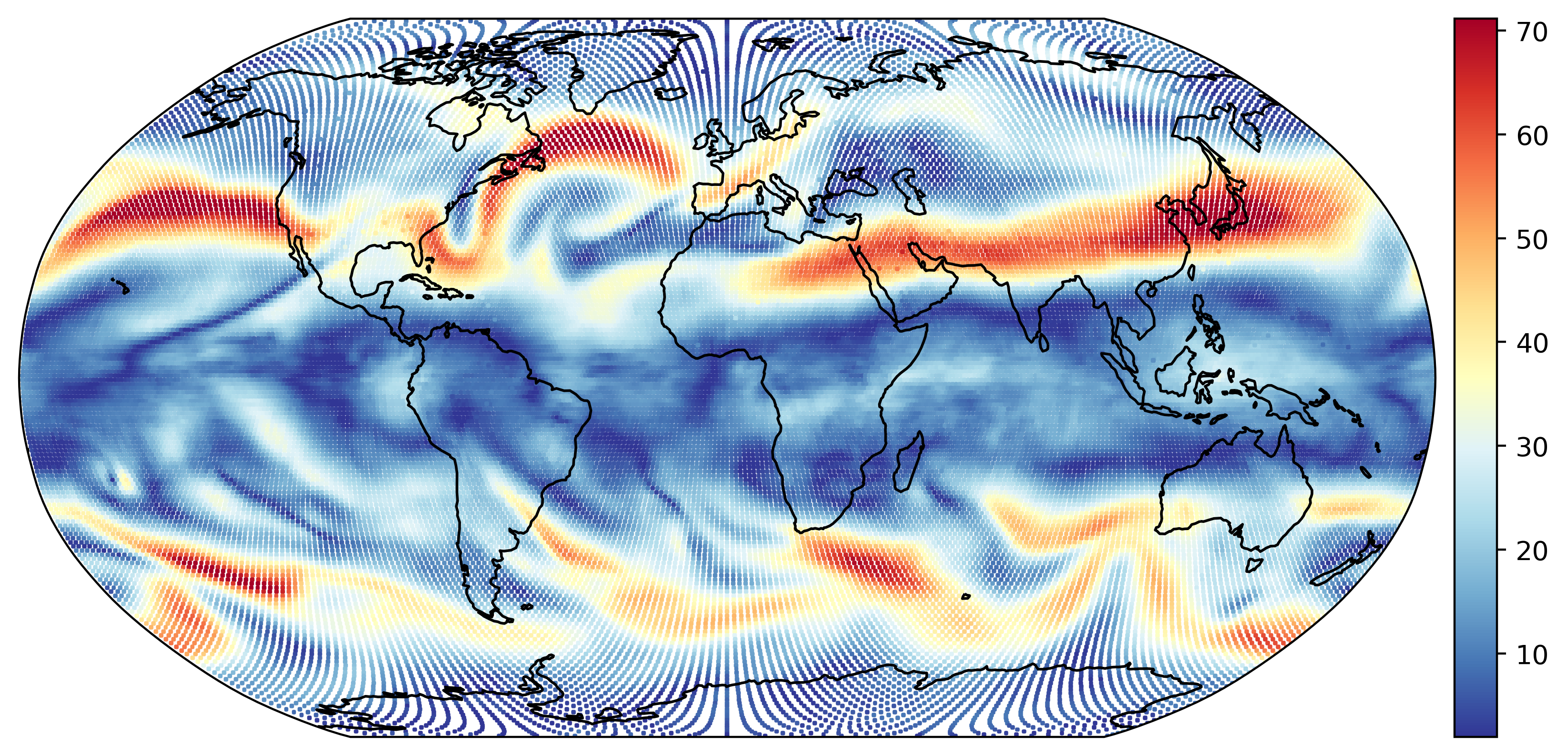}
\caption{Observations of wind speed at 200~hPa used as input to the network (left) and the gridded 200~hPa wind speed from a 24-hour GraphDOP forecast (right) valid on Jan 15, 2023, 12z.}
\label{fig:200hpa_winds_input_grid}
\end{figure}

Figure \ref{fig:gridded_examples_day1} and figure \ref{fig:gridded_examples_day5} in the Appendix show day-one and day-five gridded forecasts of sea surface temperature (SST), 2-meter temperature, 10-meter wind speed, wind speed at 200~hPa and temperature at 850~hPa. The network is able to generate predictions on a dense grid even for observation types with sparse coverage (we recall that there are absolutely no gridded ERA5 fields used during the training of the network). Figure \ref{fig:200hpa_winds_input_grid} contrasts the input observations of 200~hPa wind speed (primarily from radiosondes) with the network's predictions 24 hours later on a regular grid. Even in regions with very few observations of this variable, such as over the Pacific Ocean, the network captures detailed synoptic and large-scale structures associated with jet streams and other meteorological features. This conclusion is supported by, e.g., the sea-surface temperature forecasts shown in Figure \ref{fig:gridded_examples_day1}(a), where eddies are clearly visible. We also notice a cooler current moving up on the East coast of South America that is well captured by GraphDOP. The largest errors in the SST forecast occur in the Arctic; this may be caused by biases in buoy measurements when sea ice is present. We hypothesise that the network has learned relationships between satellite brightness temperatures and upper-level winds that generalise well to areas without direct wind observations. This provides encouraging evidence that the network can effectively combine information from different observation types to enhance its predictions.

\begin{figure}[!ht]
 \centering
 \includegraphics[width = \linewidth]{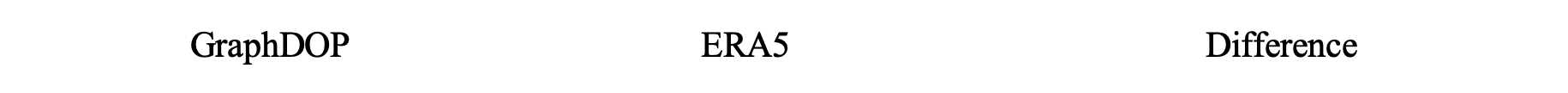}
\subfigure[Sea surface temperature]{\includegraphics[trim=0 0 55 0, clip, height=.16\linewidth]{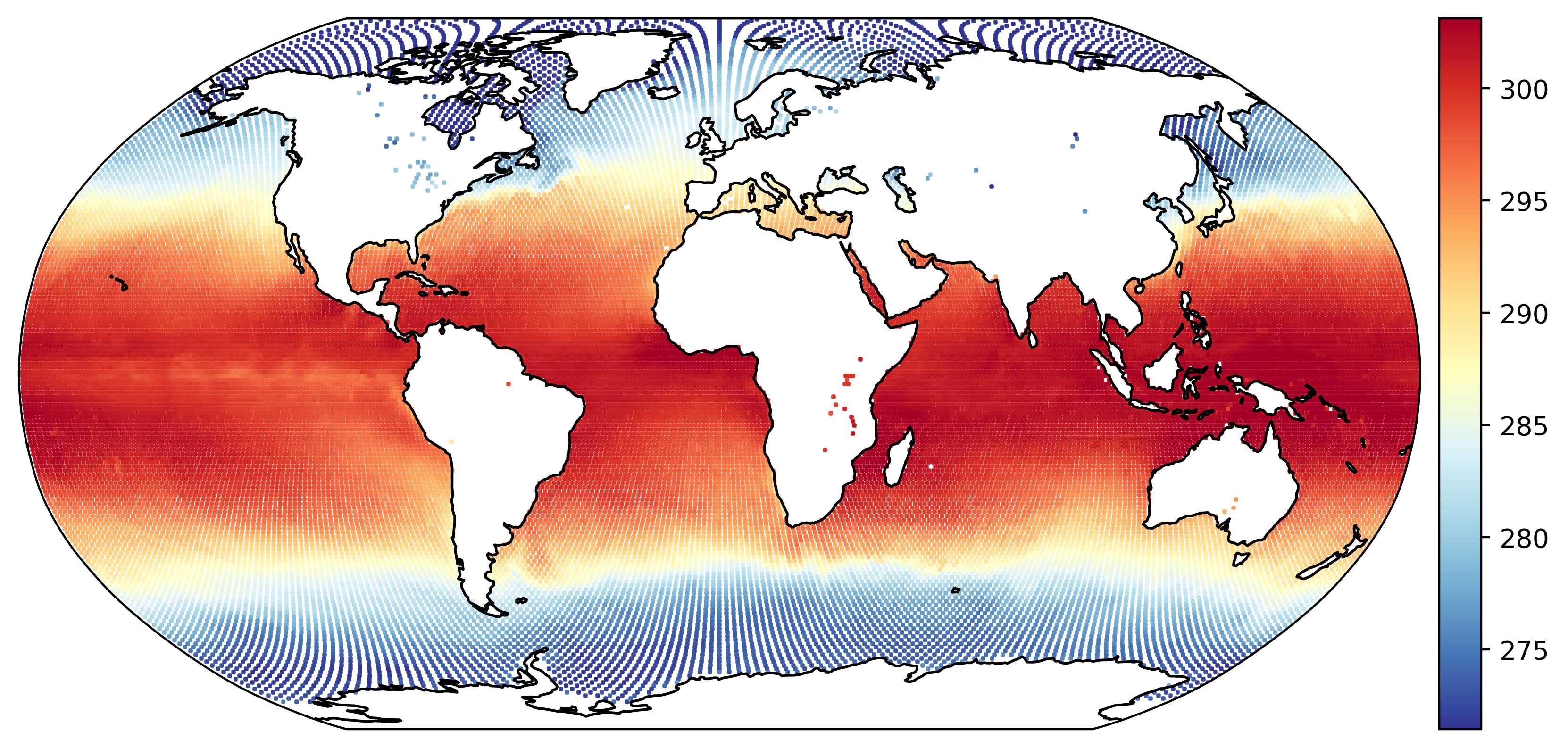} \includegraphics[height=.16\linewidth]
{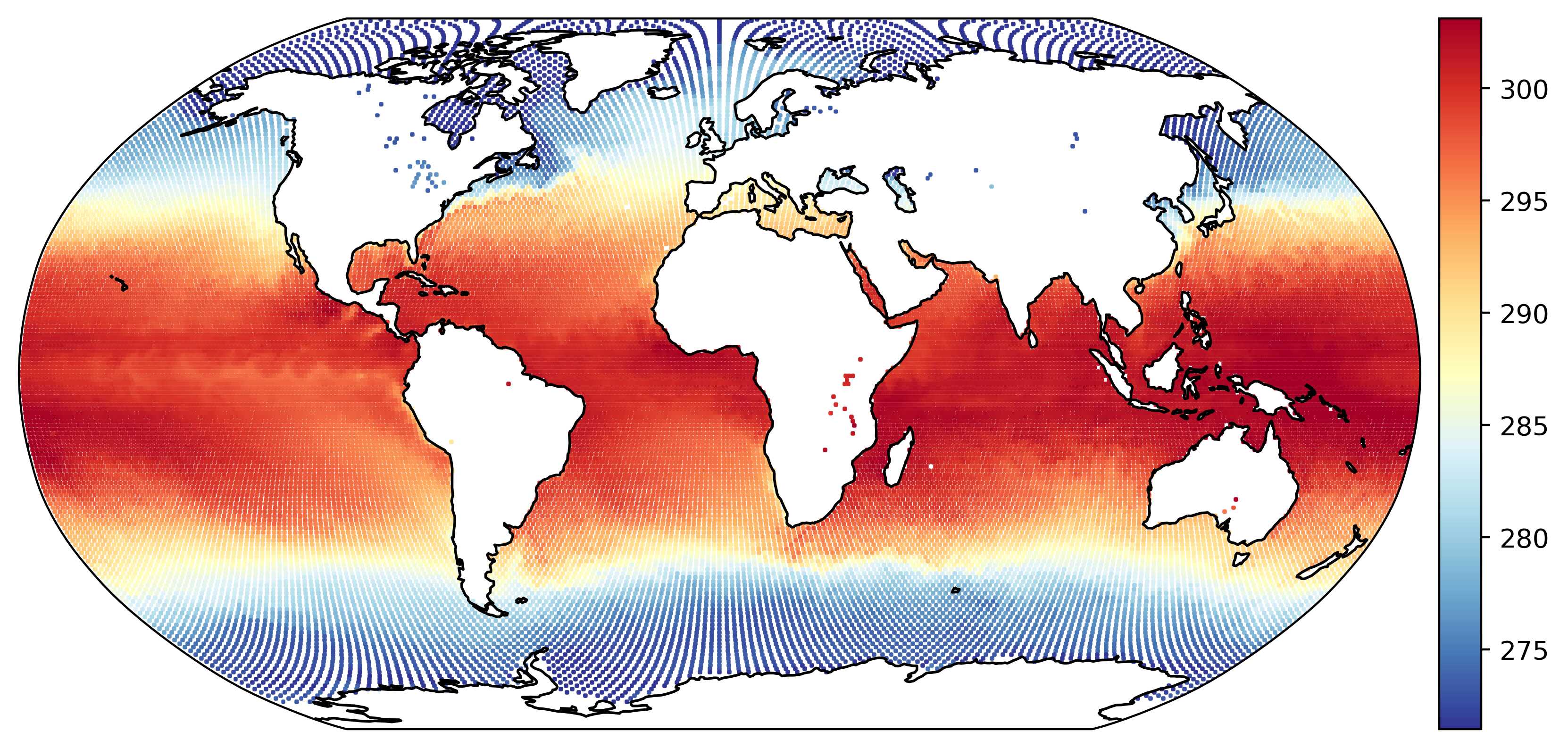}\includegraphics[height=.16\linewidth]{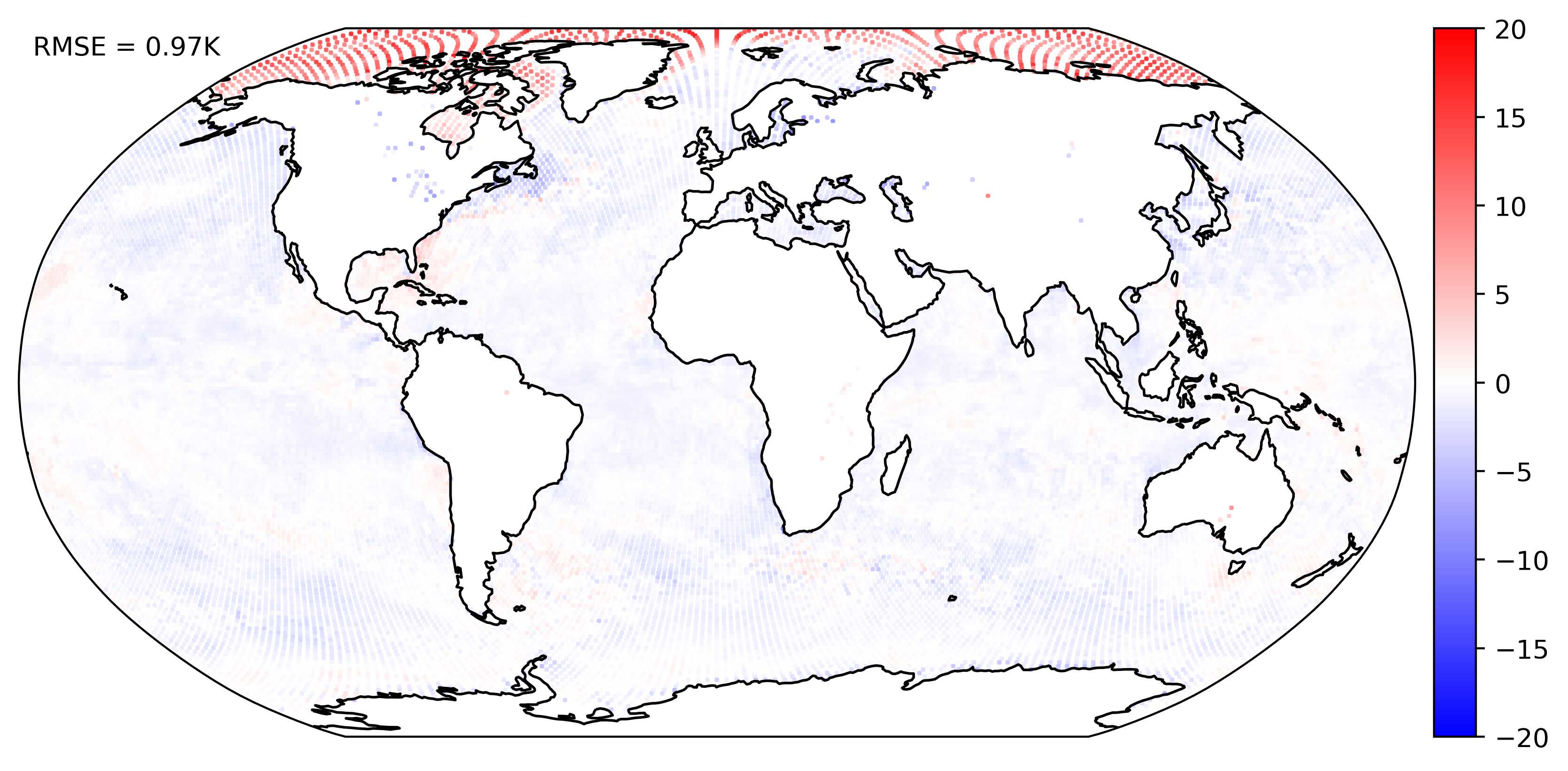}} \\
\subfigure[2-meter temperature]{\includegraphics[trim=0 0 55 0, clip, height=.16\linewidth]{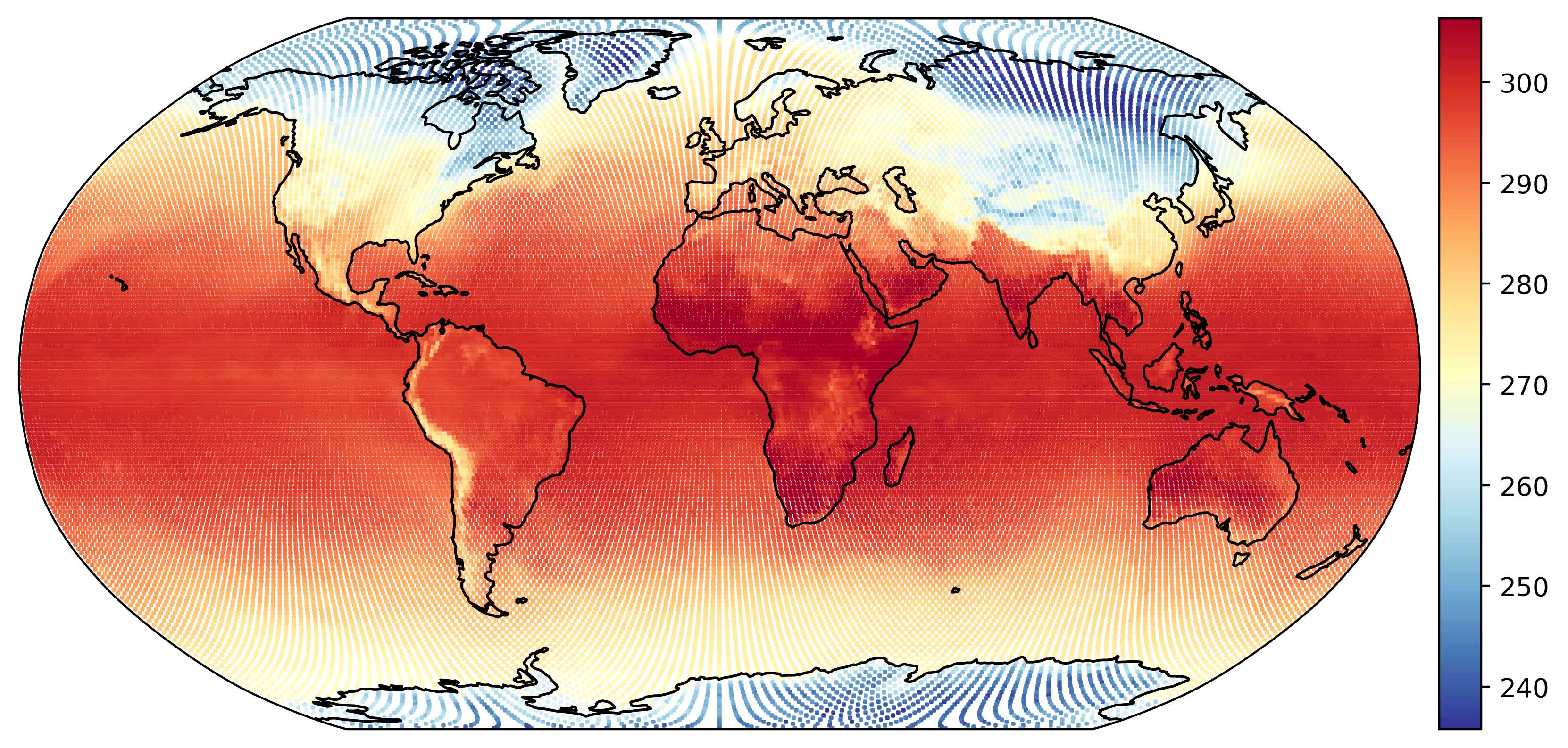} \includegraphics[height=.16\linewidth]{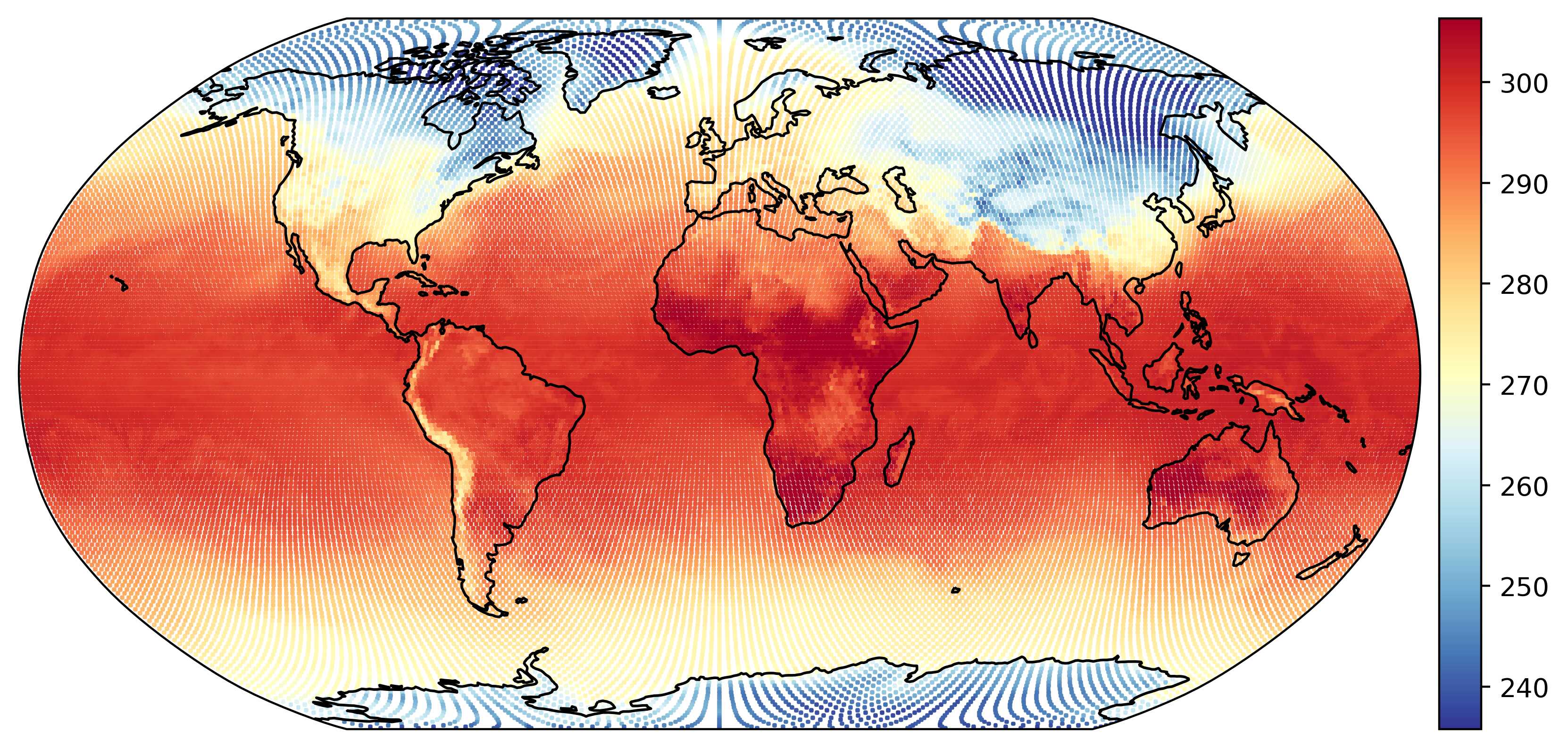}\includegraphics[height=.16\linewidth]{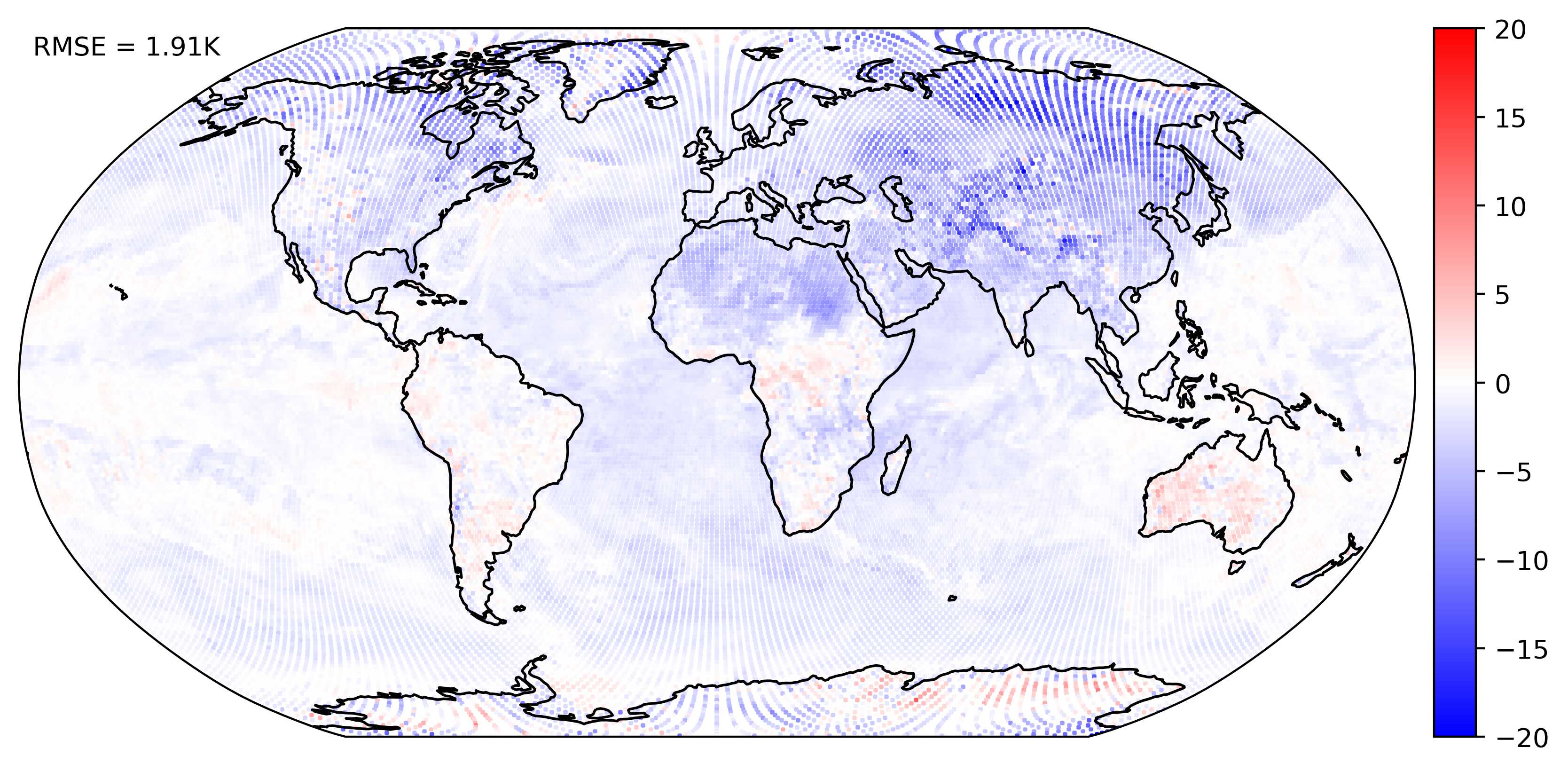}}\\
\subfigure[10-meter wind speed]{\includegraphics[trim=0 0 55 0, clip, height=.16\linewidth]{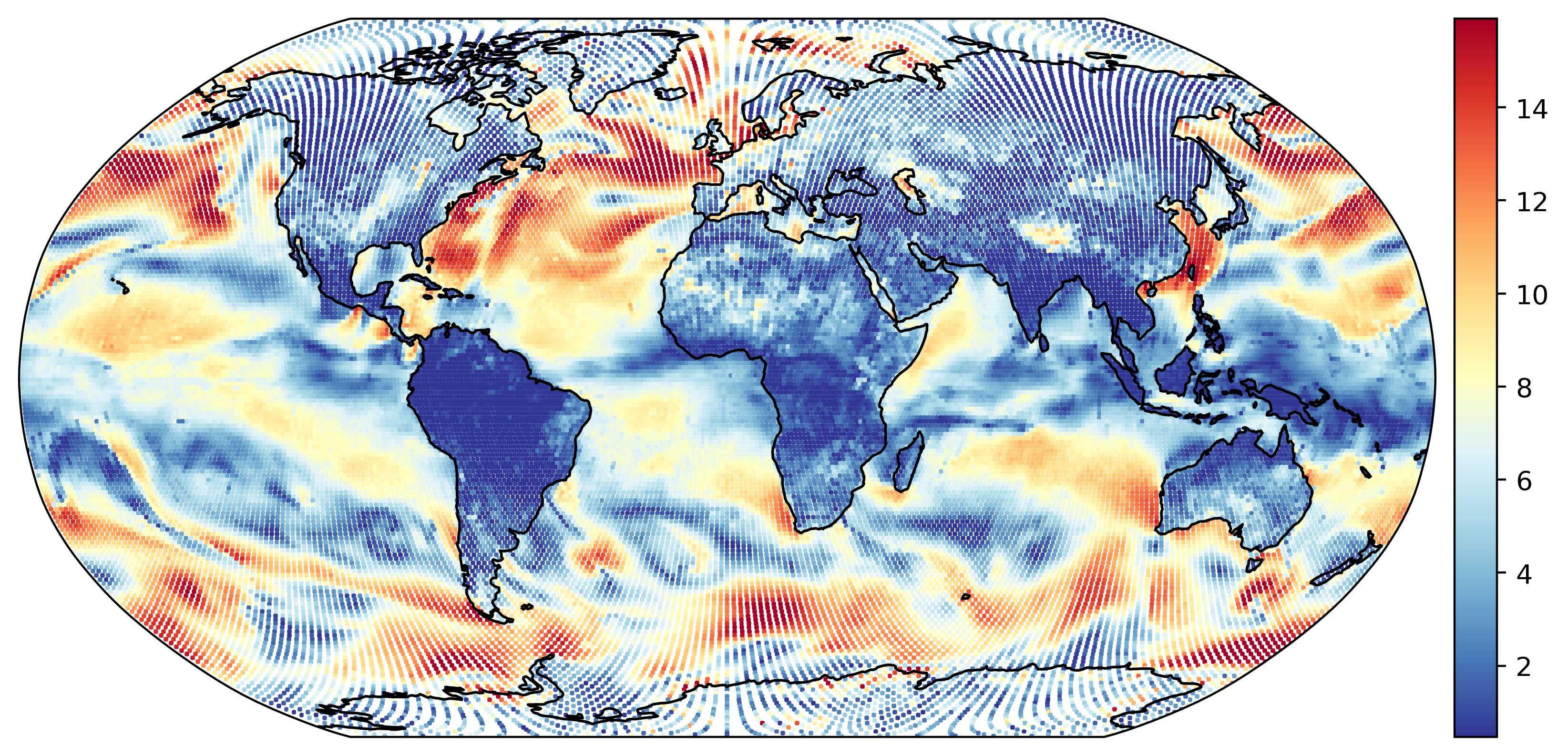}
\includegraphics[height=.16\linewidth]{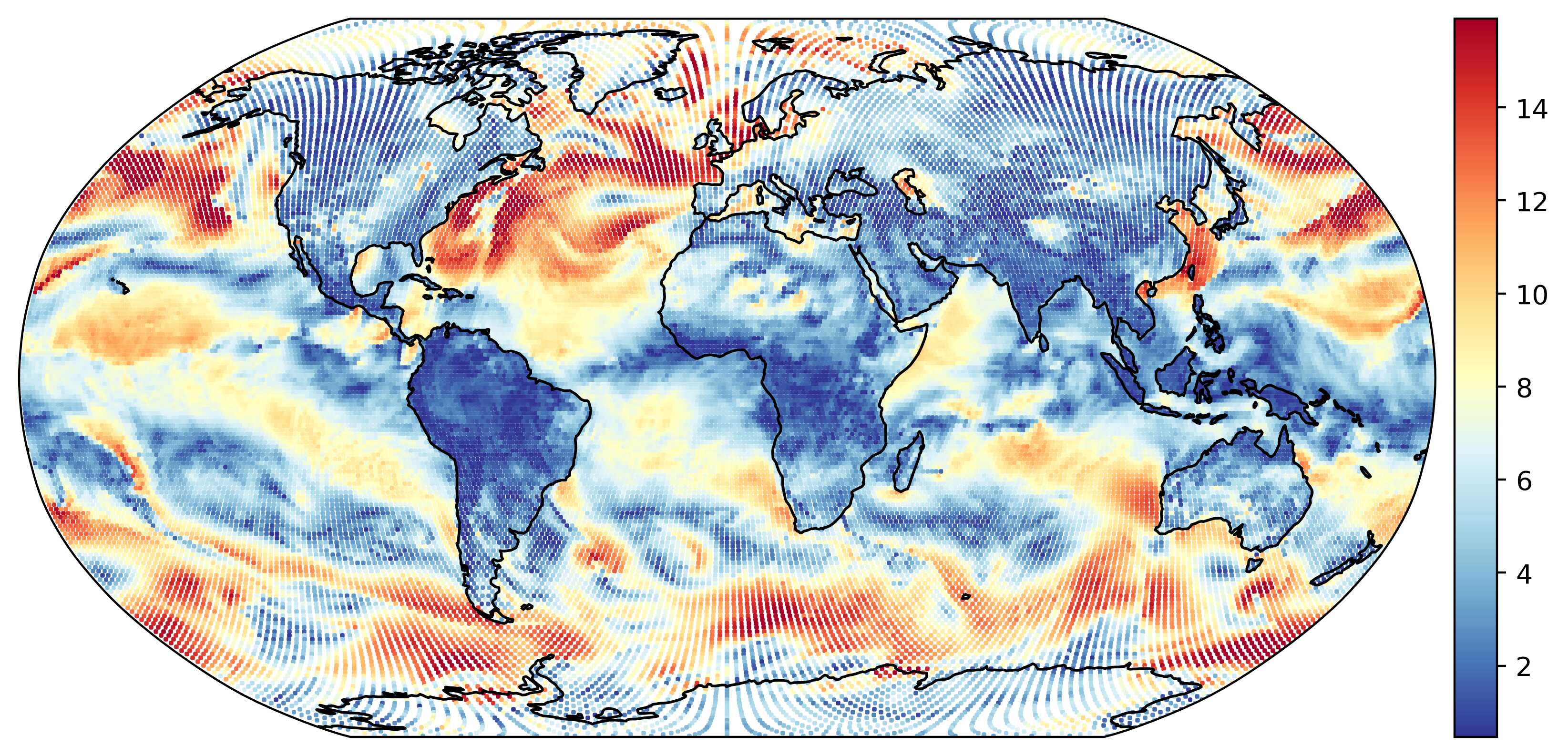}\includegraphics[height=.16\linewidth]{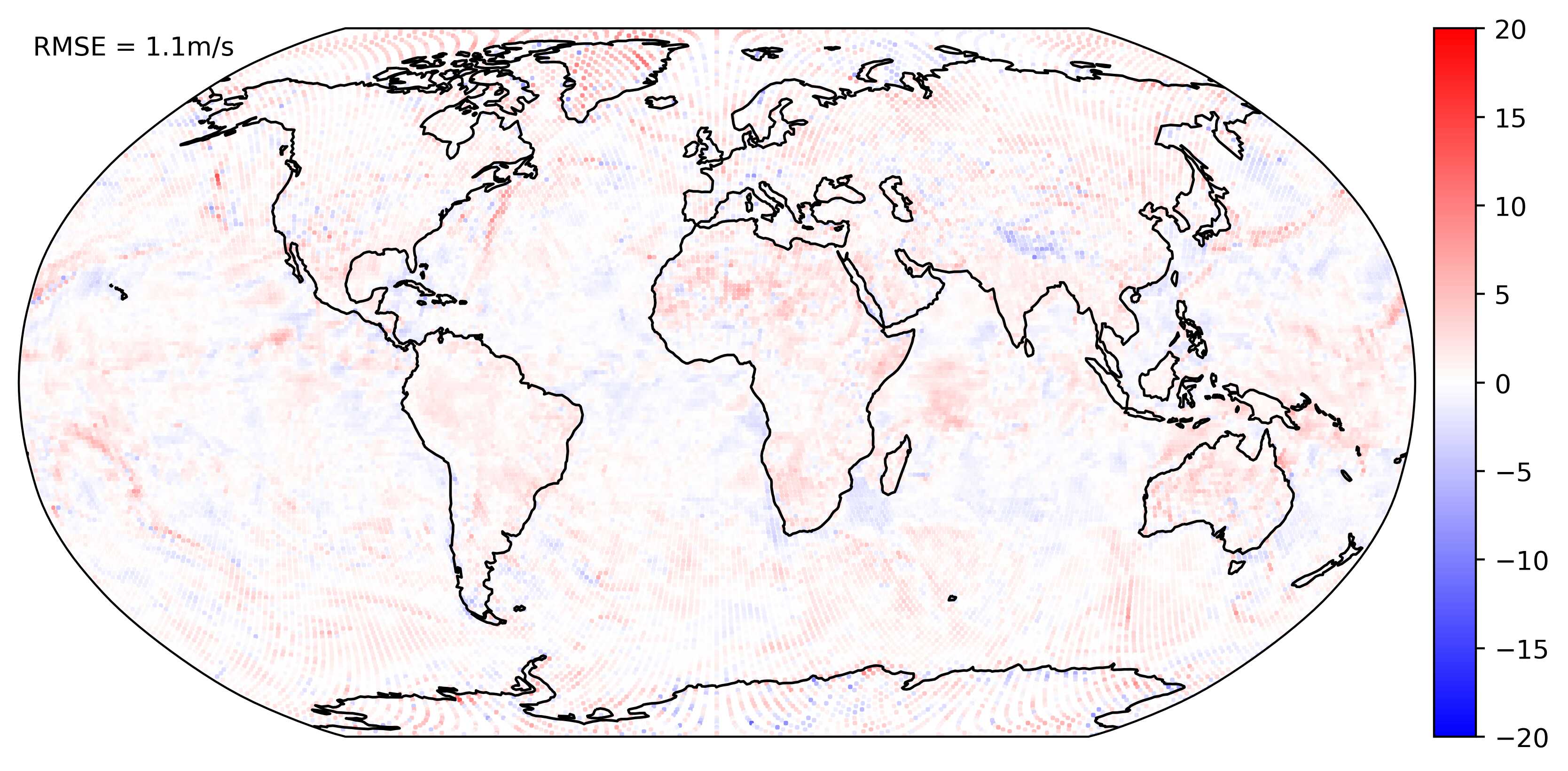}}\\
 \subfigure[Wind speed at 200hPa]{\includegraphics[trim=0 0 55 0, clip, height=.16\linewidth]{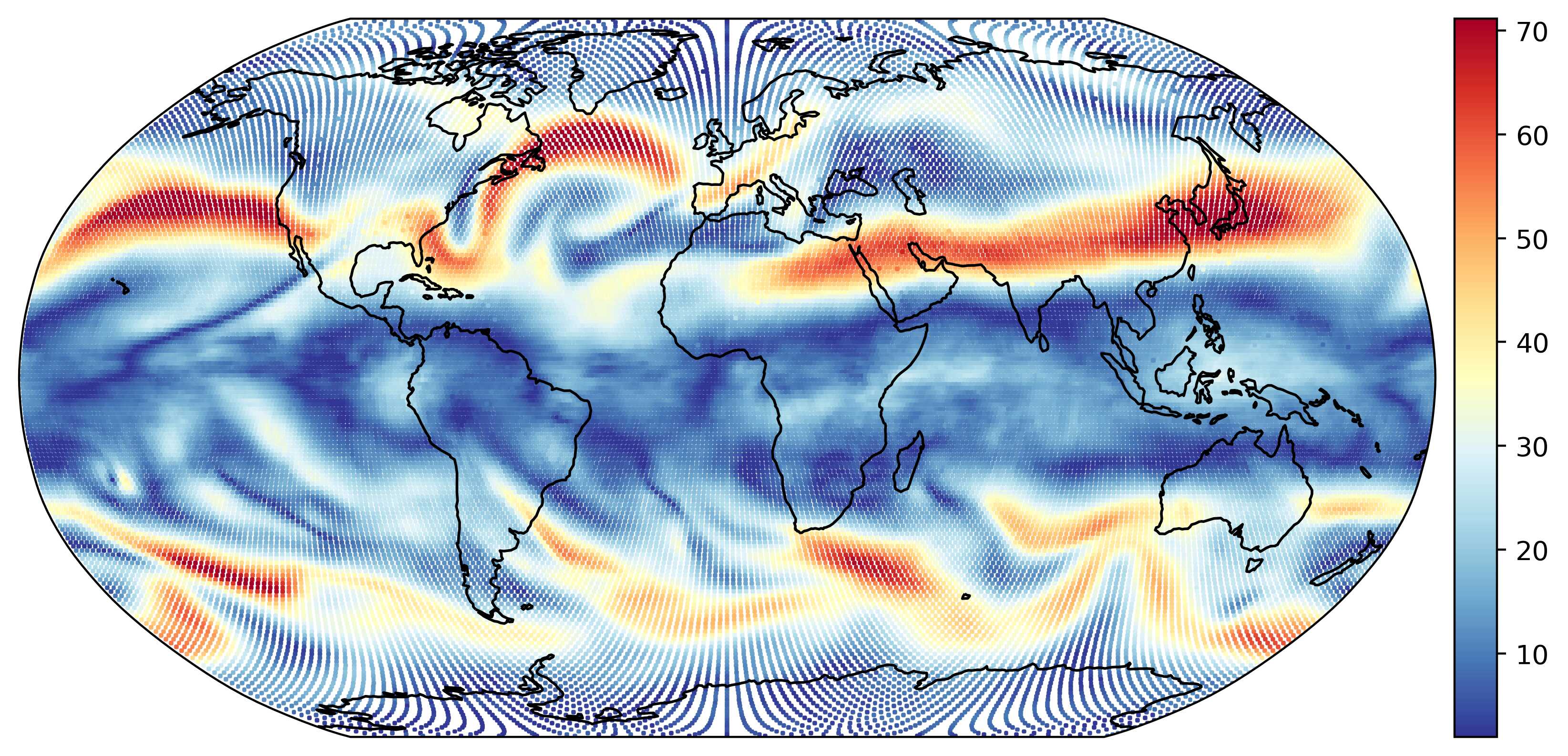}
 \includegraphics[height=.16\linewidth]{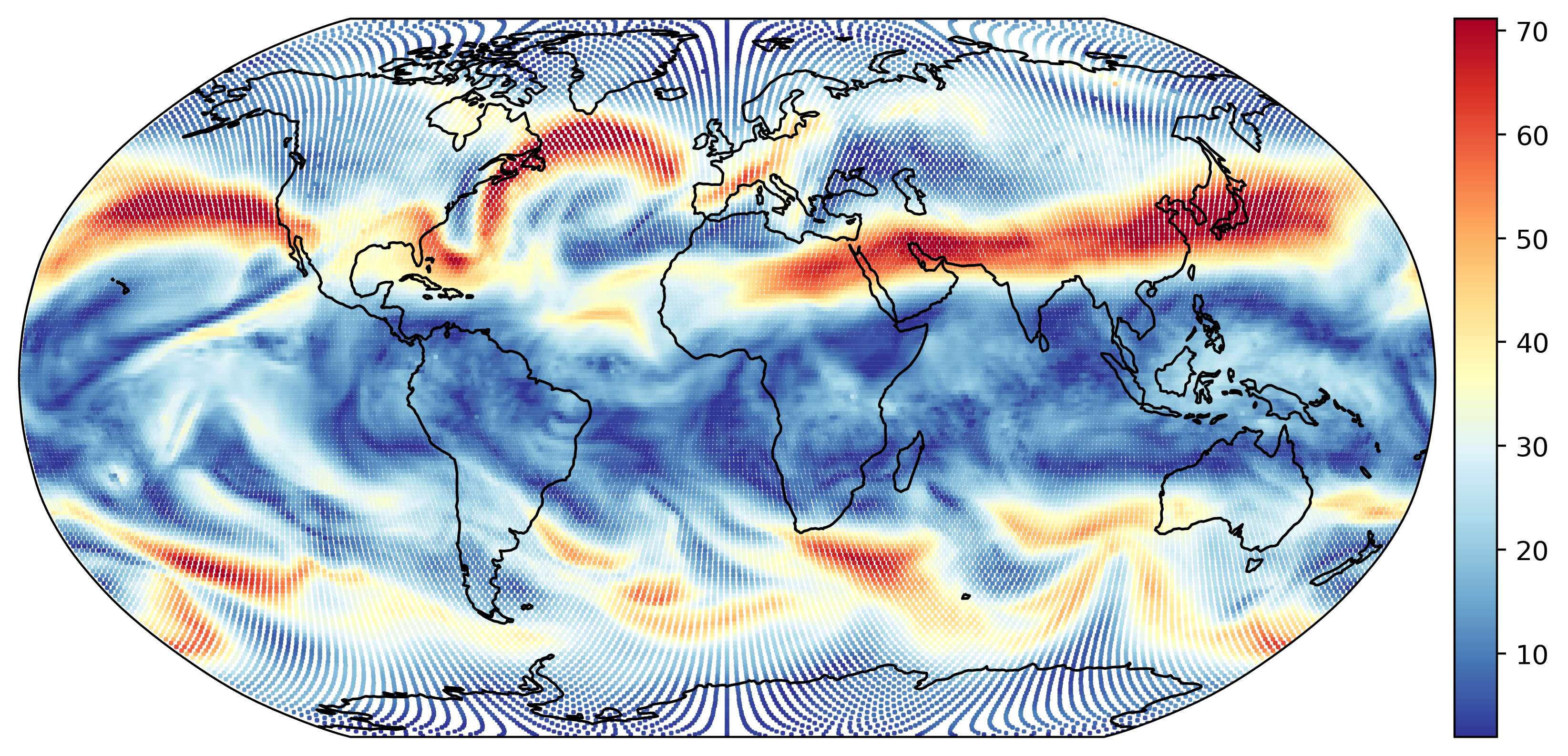}\includegraphics[height=.16\linewidth]{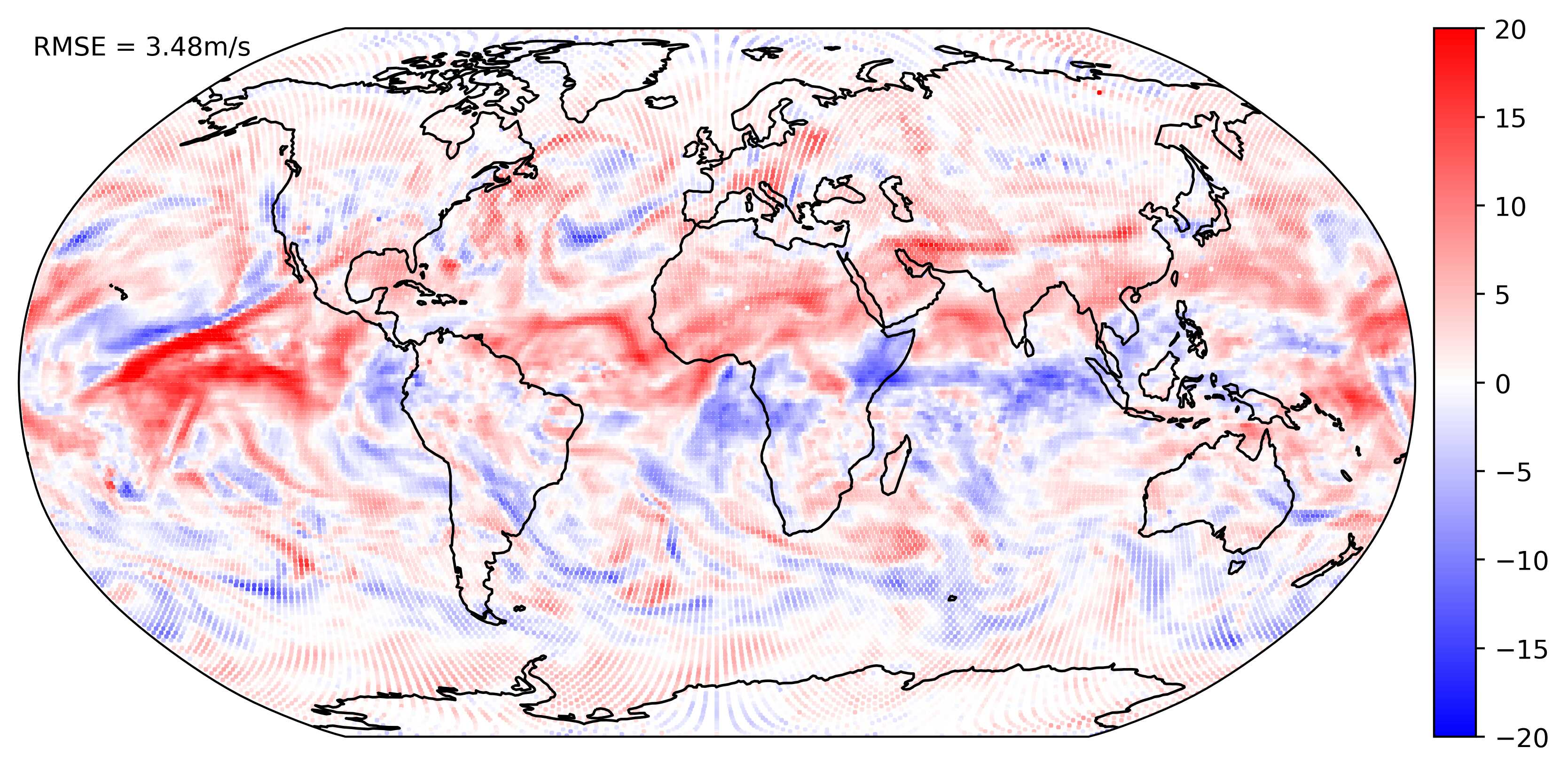}}\\
 \subfigure[Temperature at 850hPa]{\includegraphics[trim=0 0 55 0, clip, height=.16\linewidth]{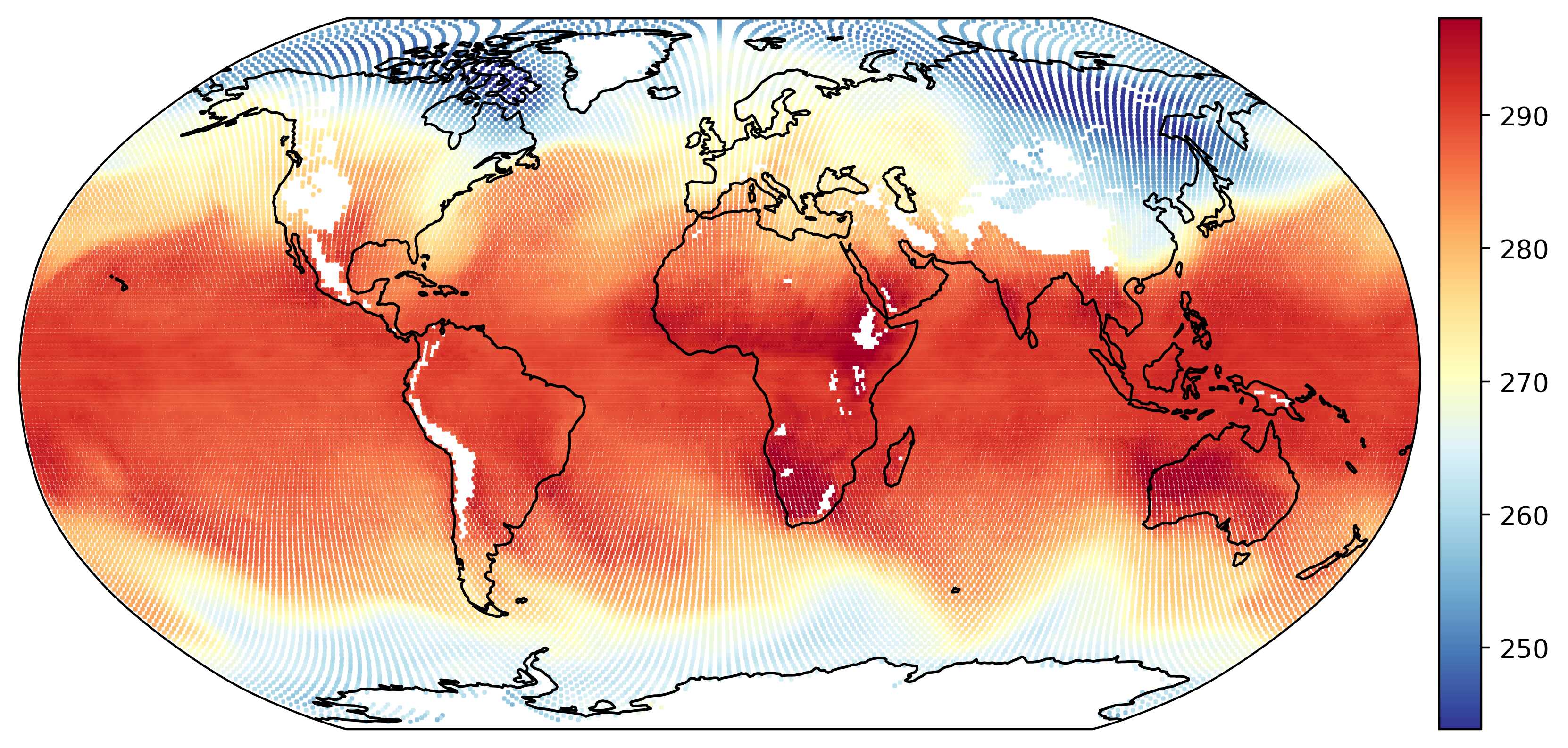} \includegraphics[height=.16\linewidth]{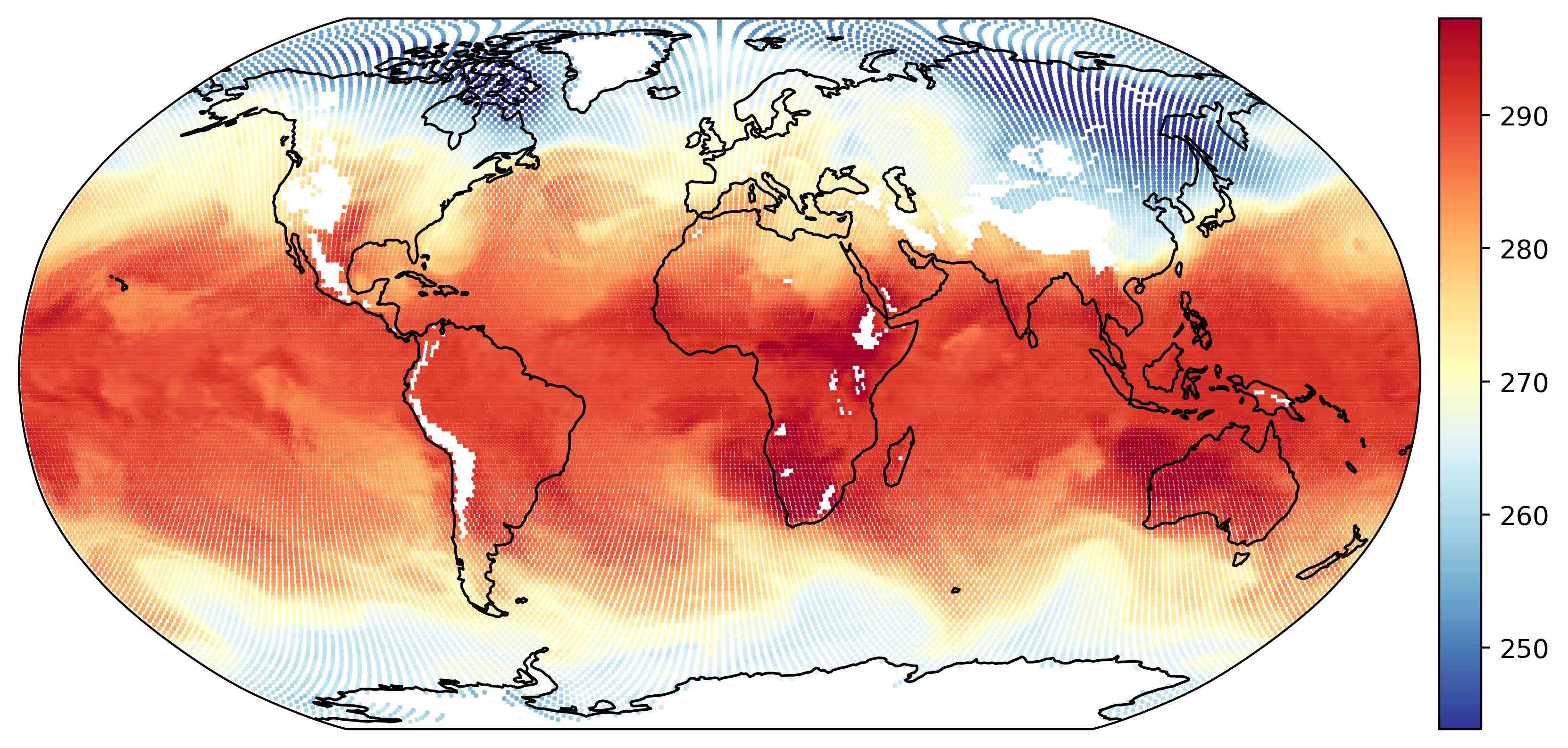}\includegraphics[height=.16\linewidth]{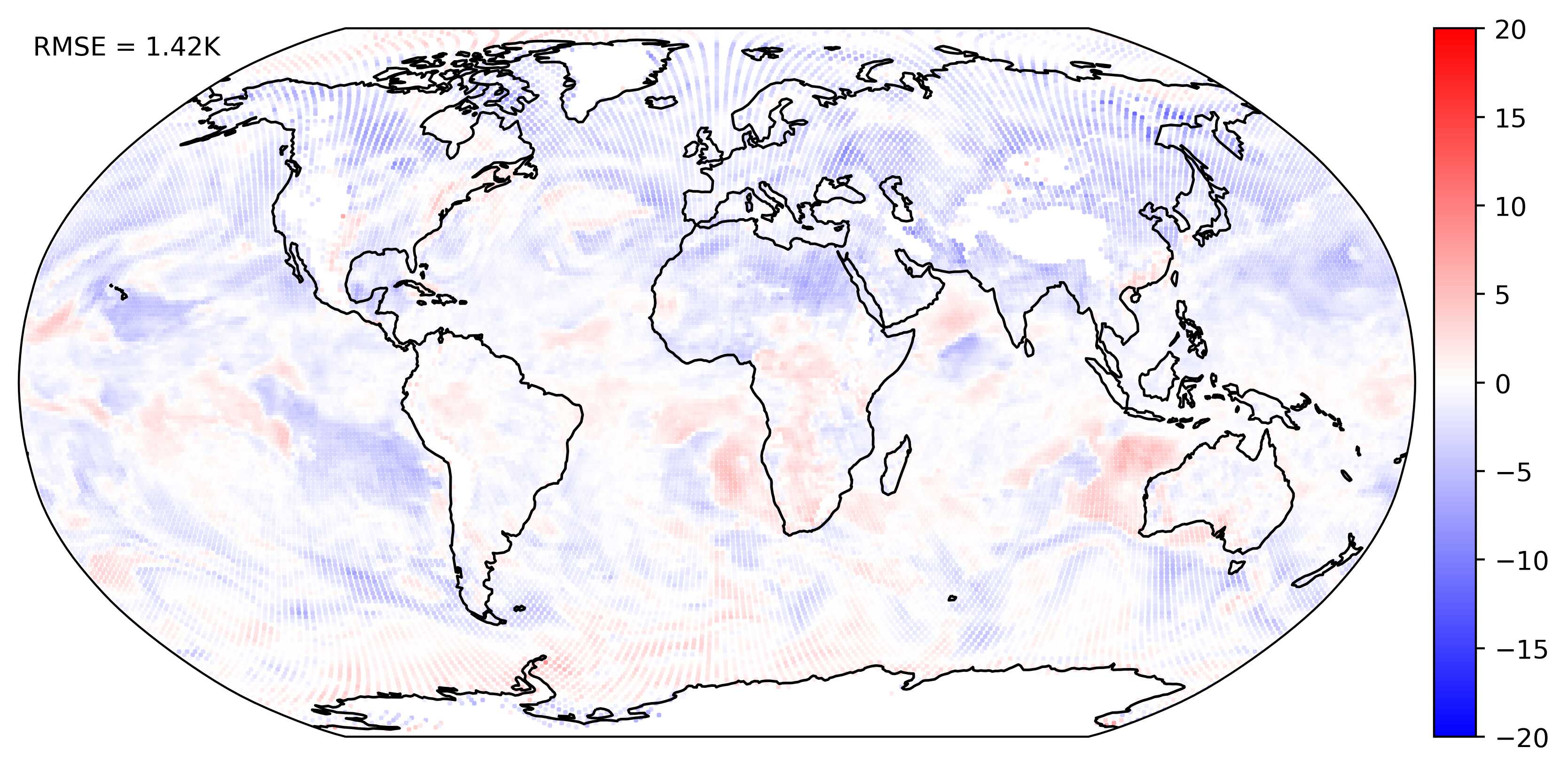}}
 \caption{Gridded forecasts at a lead time of 24 hours, valid on Jan 15, 2023, 12z (right) compared to the ERA5 reanalysis (middle). Right panels show the difference (reanalysis - forecast), with the global forecast RMSE printed at the top right corner. In the bottom left and centre panels, the pixels where surface pressure is below 850~hPa are masked out (coloured white).}
 \label{fig:gridded_examples_day1}
 \end{figure} 

Overall, larger errors are present at longer lead-times, although the forecast RMSE remains reasonable. For instance, Figure \ref{fig:gridded_examples_day5}(d) shows a slight misplacement of the jet-stream, which is to be expected at a lead time of five days, even with a state-of-the-art physical model. At a lead time of 5 days, GraphDOP is able to capture smaller-scale details, for instance, a cold streak over the Alps.

\section{Quantitative verification}
\label{ref:section-quantitative-evaluation}

\subsection{Verification in observation space: a comparison between operational IFS and GraphDOP}
\label{ref:section-comparison-ifs}

This section presents a quantitative, observation-space evaluation of the GraphDOP forecasts. We compare these forecasts against the operational version (CY47R3 in late 2022 - early 2023) of ECMWF's physics-based IFS (\cite{Owens18}) that has a spatial resolution of ca. 9 km (Tco1279). We focus on conventional (SYNOP) observations of 2-meter temperature and on brightness temperatures from the Advanced Microwave Sounding Unit-A (AMSU-A) sounder and the Special Sensor Microwave Imager / Sounder (SSMIS) satellite sensors \citep{Baordo15,Duncan21}.

ECMWF uses a set of headline scores to monitor the evolution of the IFS forecast skill over time which include verification against radiosondes and weather station observations \citep{Haiden24}. These high-quality observations are very valuable as they provide to a large extent independent verification, but they lack temporal and spatial coverage which can lead to sampling issues. To compare the quality of the IFS and GraphDOP forecasts, we compute observation equivalents from IFS forecasts for all observation types and instruments routinely processed at ECMWF, following \cite{Dahoui16}. Observation-minus-forecast departures help us understand the sources and spatial characteristics of forecast errors. 

Two-metre temperature is a diagnostic variable in the IFS; it is calculated as a weighted average of surface (skin) temperature and the lowest model level temperature. Historically, t2m has proved extremely challenging to assimilate in a global physics-based weather model, with the ECMWF starting operational assimilation of t2m observations only in November 2024 \citep{Ingleby24}. This led to significant improvements in short-range t2m forecasts. For benchmarking against GraphDOP, the IFS operational 10-day forecasts have been reprocessed to output 2-meter temperature at the actual observation times and locations. As the observations assimilated in the operational IFS are different to those used to train and initialise GraphDOP, a strict match-up procedure has been implemented to ensure that only predicted observations that can be found in both forecasting systems are used to compute the forecast skill scores. The differences between these observation datasets are generally minor with over 95\% of observations successfully matched between the two forecasting systems. For technical reasons, the IFS operational forecasts initialised from the early-delivery analysis had to be used in this study which means that the GraphDOP forecasts have an approximately 5-hour advantage \citep{Lean21}. This is because the early delivery is produced from shorter assimilation windows of 9z - 16z and 21z - 4z. It is important to mention that GraphDOP is not tied to a given production schedule and it can run as early - or, indeed, as late - as the user requires. As it requires only a few minutes of GPU compute time to generate a 10-day forecast, GraphDOP can be seen as an on-demand forecasting tool with no constraints other than near real-time observation arrival.

\begin{figure}[htpb]
\centering
\includegraphics[width=\linewidth]{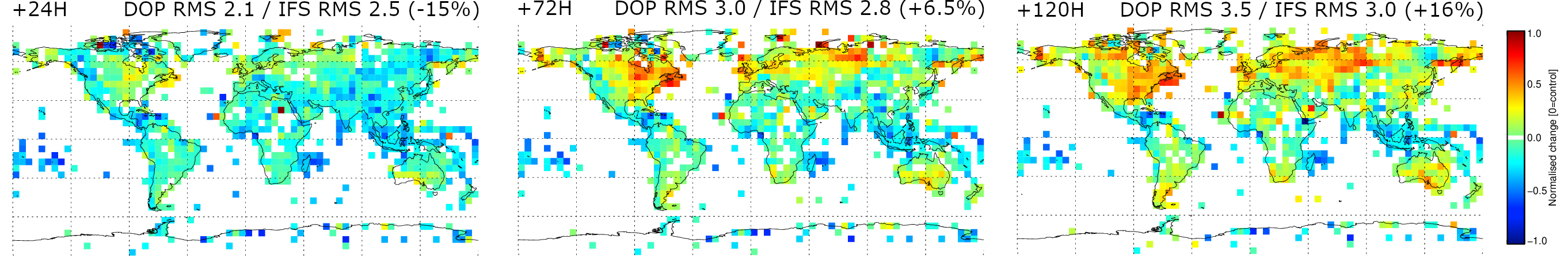}
\caption{Normalised RMS difference between IFS (CY47R3) and GraphDOP forecast departures for 2-meter temperature SYNOP observations at lead time 24h (left), 72h (middle) and 120h (right). Negative (positive) values represent an improvement (degradation) in GraphDOP forecasting skill compared to IFS. Statistics have been computed between December 10, 2022 and February 28, 2023 using all observations that have been processed by both forecasting systems.}
\label{fig:dop-ifs-t2m}
\end{figure}

Figure \ref{fig:dop-ifs-t2m} shows the normalised root-mean-square (RMS) differences between IFS and GraphDOP forecast departures for 2-meter temperature SYNOP observations at lead time 24h (left), 72h (middle) and 120h (right). Negative (positive) values represent an improvement (degradation) in GraphDOP forecast skill compared to IFS. 

At day 1, GraphDOP outperforms IFS, with a global improvement of 15\%. The results are more mixed at 3 to 5 days into the forecast: GraphDOP departures improve upon IFS over the Tropics but are degraded in the Northern hemisphere. It is remarkable that the GraphDOP forecasts have very small systematic errors, especially over the Tropics where the GraphDOP bias is -0.1~K at day five versus 0.9~K in the IFS system. More work is required to study these patterns. For example, we know that the IFS forecasts over winter present a night-time cold bias of 0.5–1~K in large parts of Europe, and a warm bias of several Kelvin throughout the day in parts of Scandinavia. As such, we plan to evaluate GraphDOP forecasts in specific weather conditions to get a comprehensive view of diurnal, seasonal and regional error dependence. Ongoing research work combining ML and data assimilation also aims to improve forecast bias at the surface in the IFS system \citep{Bonavita20,farchi24}. We also note that GraphDOP is evaluated here against "raw" IFS forecasts, and it is known \citep{ziedT2Mpproc2023} that post-processing can improve the RMSE of IFS forecasts at surface stations by 10 - 15\%. 

The assimilation of all-sky and all-surface brightness temperatures in NWP \citep{Geer22} is a challenging research topic. Since the beginning of satellite data assimilation in the 1980s, most cloud-affected observations have been rejected during 4D-Var quality control, following a "clear‑sky" approach. This is because of known shortcomings of the forecast model and observation operators in areas of cloud and precipitation. Over the years, ECMWF has expanded the coverage of all-sky assimilation to, as of December 2024, nine microwave sensors that form a major part of the current observing system \citep{Duncan21}. In contrast, GraphDOP is using brightness temperatures in a very different manner to traditional data assimilation methods. GraphDOP does not rely on sophisticated cloud physics parametrisations and radiative transfer models as the brightness temperatures are directly ingested by the GNN during the training process. Because of this, future versions of GraphDOP (and other AI-DOP models) may be able to exploit more high-resolution, and more complex all-sky observations than are currently being assimilated in physics-based NWP systems.

\begin{figure}[htpb]
\centering
\includegraphics[width=.6\linewidth]{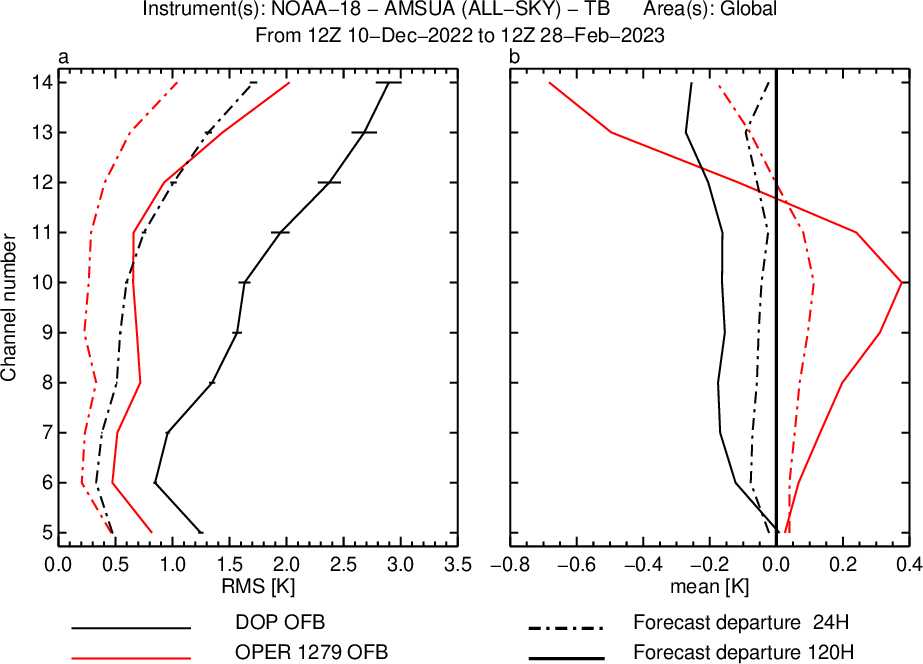}
\caption{RMS forecast departures (left panel) and mean forecast departures (right panel) with respect to AMSU-A all-sky (NOAA-18) for GraphDOP (black) and IFS (red) at forecast lead time +24h (dashed lines) and +120h (solid lines). Statistics have been computed between December 10, 2022 and February 28, 2023 using all observations that have been processed by both systems.}
\label{fig:dop-ifs-amsua}
\end{figure}

Although brightness temperatures are not geophysical parameters required by forecast users, they are sensitive to temperature, water vapour, cloud, and precipitation. An accurate radiance forecast reflects an accurate representation of the atmospheric state and its physical processes. The IFS simulated brightness temperatures at their valid time and location have been saved at different lead times and are compared to the GraphDOP forecasts. The left panel of Figure~\ref{fig:dop-ifs-amsua} shows the RMS forecast departures with respect to AMSU-A all-sky for GraphDOP (black) and IFS (red) at forecast lead time +24h (dashed lines) and +120h (solid lines).

\begin{figure}[htpb]
\centering
\includegraphics[width=.6\linewidth]{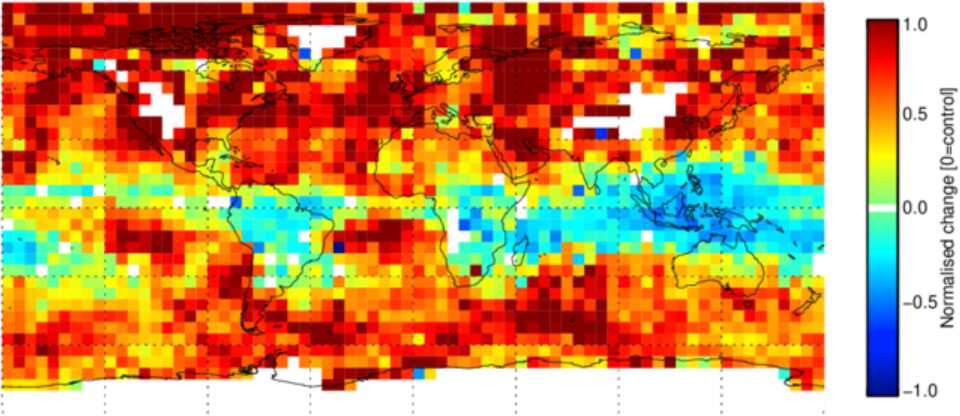}
\caption{Normalised RMS difference between IFS (CY47R3) and GraphDOP forecast departures for AMSU-A channel 5 after five days. Negative (positive) values represent an improvement (degradation) in GraphDOP forecasting skill compared to IFS. Statistics have been computed between December 10, 2022 and February 28, 2023 using all observations that have been processed by both forecasting systems.}
\label{fig:dop-ifs-amsua5}
\end{figure}

While the IFS forecasts are significantly better globally for all the channels, it is interesting to highlight the good performance of GraphDOP at predicting channel 5 which is sensitive to the lower tropospheric temperature. Over the Tropics, the GraphDOP forecasts of AMSU-A channel 5 brightness temperatures have smaller RMS departures than IFS at day 1 (by 25\%) and day 5 (by 8\%, see Figure~\ref{fig:dop-ifs-amsua5}). This is a very promising result as it demonstrates the ability of GraphDOP to combine the information from different instruments to produce a skilful joint forecast of surface and lower tropospheric temperature over the Tropics.

The right panel of Figure~\ref{fig:dop-ifs-amsua} shows the mean forecast departures with respect to AMSU-A all-sky. GraphDOP presents smaller biases in the predicted brightness temperatures, especially for the upper channels that are sensitive to the stratospheric temperature. It is known that the IFS model develops larger systematic errors in the stratosphere and the development of model bias correction methods is an ongoing area of active research \citep{Sheperd18,Laloyaux20}. 

\begin{figure}[htpb]
\centering
\includegraphics[width=.6\linewidth]{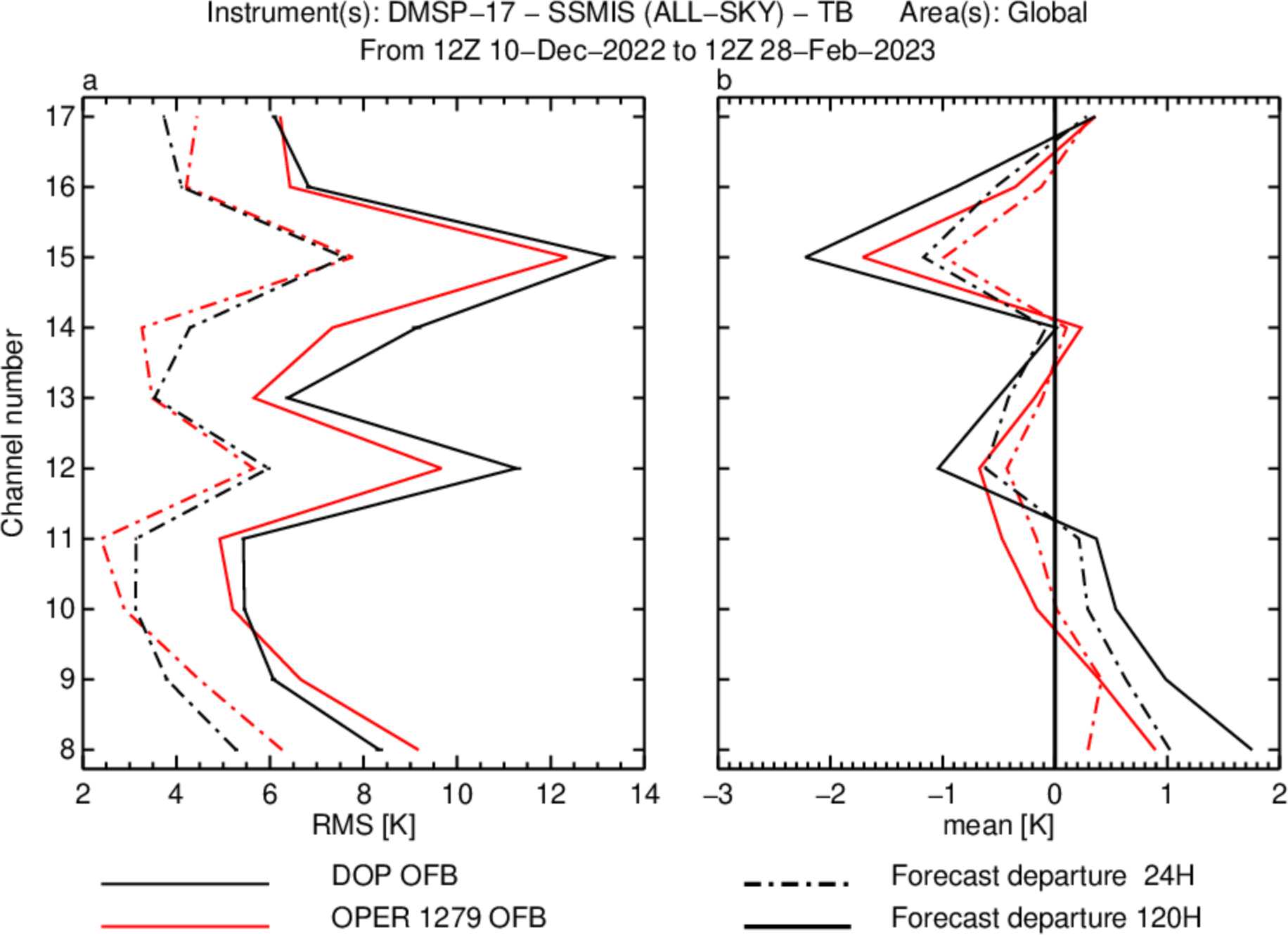}
\caption{RMS forecast departures (left panel) and mean forecast departures (right panel) for SSMIS all-sky (DMSP-17) brightness temperature forecasts from GraphDOP (black) and IFS (red) at +24h (dashed lines) and +120h (solid lines). Statistics have been computed between December 10, 2022 and February 28, 2023 using all observations that have been processed by both systems.}
\label{fig:dop-ifs-ssmis}
\end{figure}

The Special Sensor Microwave Imager/Sounder (SSMIS) is another passive microwave instrument that has several channels sensitive to temperature, humidity and surface parameters \citep{Baordo15}. Figure~\ref{fig:dop-ifs-ssmis} shows the RMS (left panel) and mean (right panel) forecast departures with respect to SSMIS all-sky in a similar way to what is presented for AMSU-A in Figure~\ref{fig:dop-ifs-amsua}. The humidity-sensitive channels examined here show a large sensitivity to clouds, particularly in the case of the window channels 12-17.

SSMIS confirms the good performance of GraphDOP at predicting tropospheric humidity and clouds as the RMS and mean forecast departures show a similar general behaviour. The improvements in SSMIS channels 8 and 9 suggests lower tropospheric moisture (cloud or column water vapour) or convection. Another set of IFS forecasts initialised from the early-delivery analysis and produced at a lower horizontal resolution of ca. 110~km (T159) was produced. These forecasts have approximately the same spatial resolution as the o96 GraphDOP output grid, but they do not reduce the errors for the lower channels (not shown). This suggests that the GraphDOP improvement is genuine and not caused by double penalty effects where smoothing or eliminating cloud and precipitation usually lead to a lower RMS. Forecast activity diagnostics (under development at present) are expected to provide better insight into the reasons for the reduced RMS forecast departures for channel 8 and 9.

In the IFS system, knowledge of the ocean surface skin temperature (SKT) is vital to the accurate use of satellite brightness temperatures and it is currently derived from a combination of external sources with a latency of up to 69 hours. Among other things, this can produce phase errors in Tropical Instability Waves in the Eastern Tropical Pacific (ETP) and motivates the ongoing work to allow the ocean SKT to update as part of 4D-Var \citep{Scanlon24}. GraphDOP is not affected by these latency issues as it uses only observations at their actual time and location without relying on external data products. In this context, it is promising that GraphDOP performs better in the ETP for SSMIS channels 12 and 13 that are sensitive to SKT.  

Clearly more work is needed to better understand the spatio-temporal correlations estimated by GraphDOP using, e.g., adjoint sensitivity analysis, as this may provide more insight into the atmospheric dynamics and physical processes that are successfully learnt by the model during training.

\subsection{Verification in grid space: GraphDOP forecasts generalised to arbitrary locations and grids}

\begin{figure}[!h]
\centering
\subfigure{\includegraphics[width=.3\linewidth]{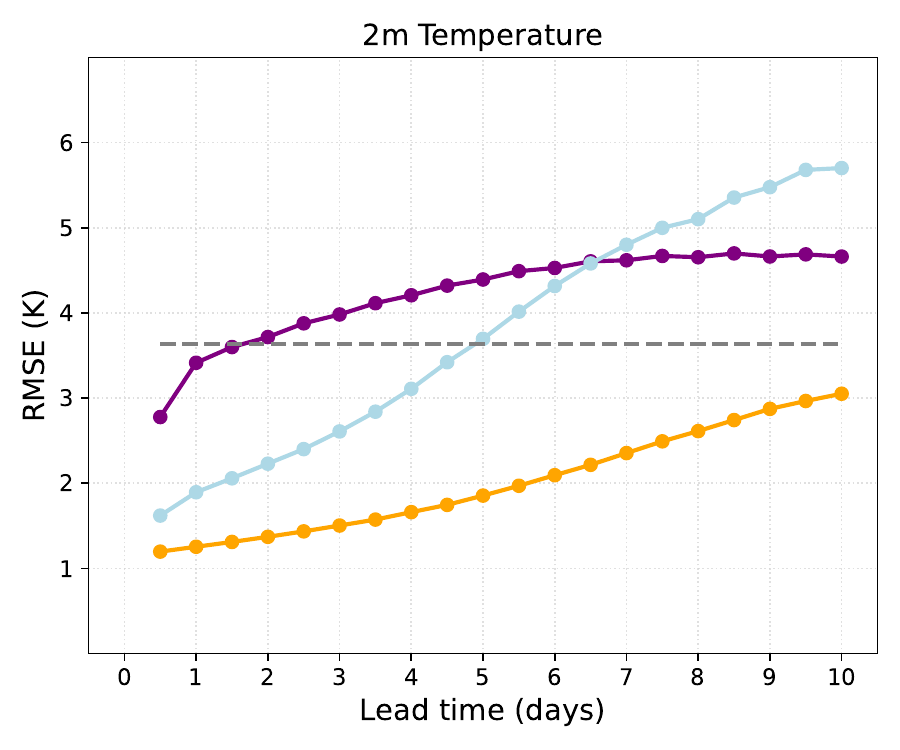} }
\subfigure{\includegraphics[width=.3\linewidth]{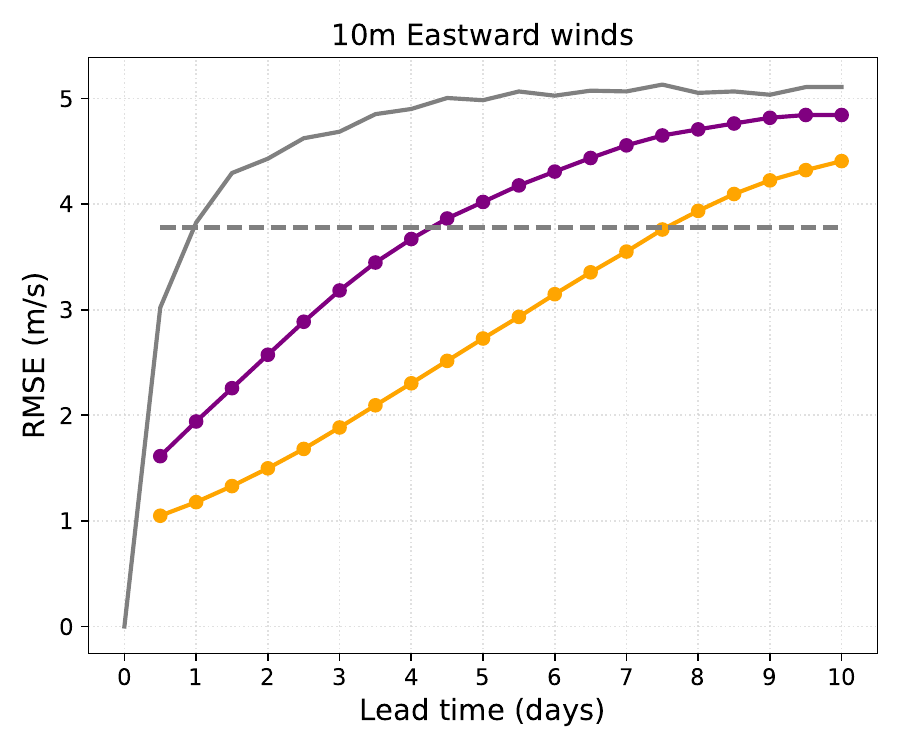} }
\subfigure{\includegraphics[width=.3\linewidth]{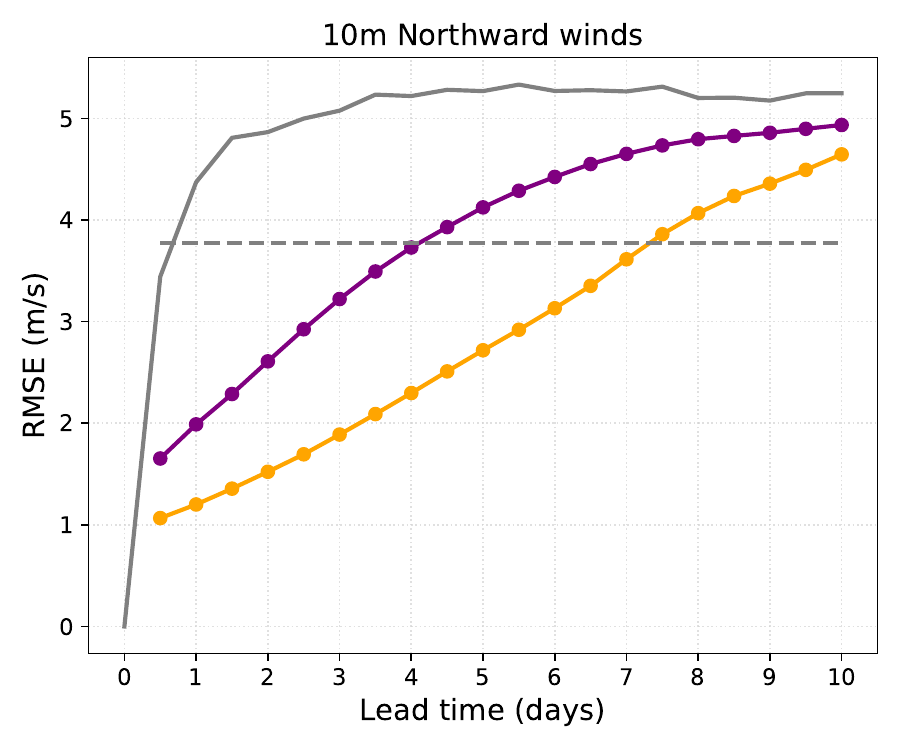} } \\
\subfigure{\includegraphics[width=.3\linewidth]{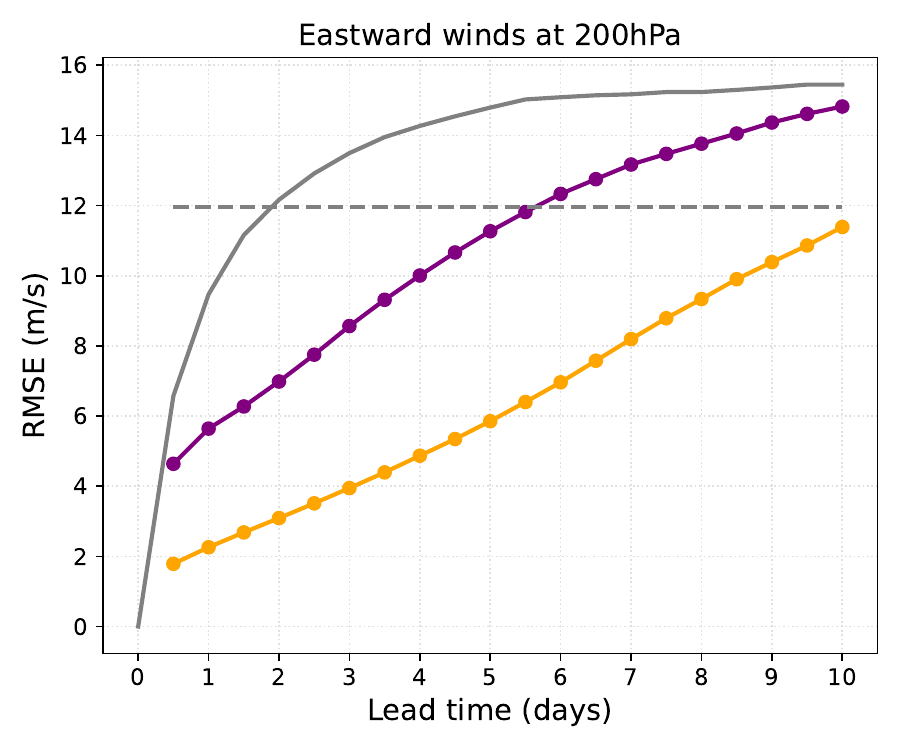} }
\subfigure{\includegraphics[width=.3\linewidth]{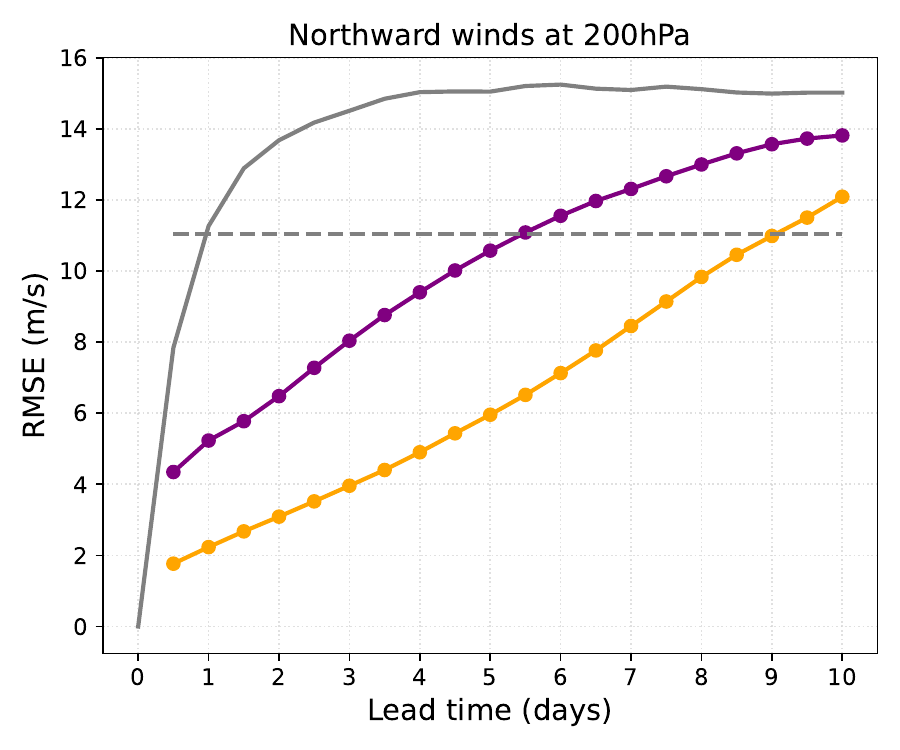} }
\subfigure{\includegraphics[width=.3\linewidth]{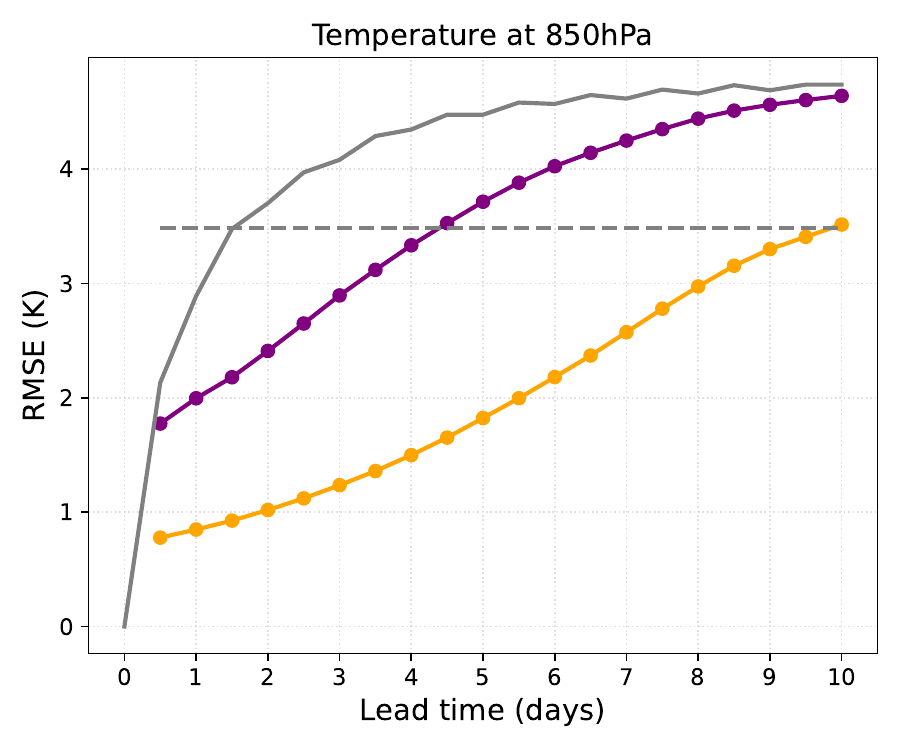}} \\
\subfigure{\includegraphics[width=.6\linewidth]{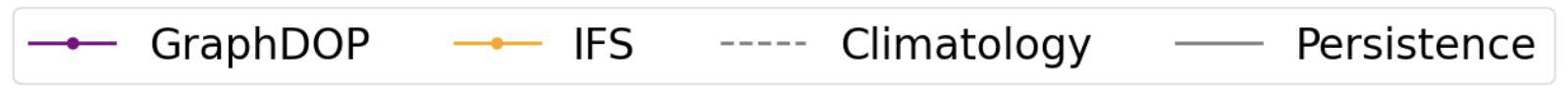}} \\
\caption{RMSE of gridded global GraphDOP forecasts, evaluated against ERA5 reanalysis, as a function of lead time. Statistics are computed for six variables: three surface variables (2-meter temperature and 10-meter zonal and meridional winds), and three upper-atmosphere variables (zonal and meridional 200~hPa winds and temperature at 850~hPa), and were computed on January 2023 forecasts. Persistence was left out for 2-meter temperature because of the diurnal cycle.}
\label{fig:rmse_era5}
\end{figure}

Figure \ref{fig:rmse_era5} shows the performance of GraphDOP by showing RMSE (evaluated against the ERA5 reanalysis, interpolated onto an o96 grid) for six surface and upper-air variables. We stress that the ERA5 fields are employed exclusively for verification, and were not used during the training of the network. We also compare GraphDOP forecast skill against that of two simple baselines \citep{rasp2024weatherbench2benchmarkgeneration}: persistence and climatology. The climatology used here is a six-hourly ERA5 climatology based on \citep{Jung2008}. The IFS operational forecasts initialised from the early-delivery analysis (described in Section \ref{ref:section-comparison-ifs}) and interpolated on an o96 grid are also evaluated against ERA5 and plotted in orange. Scores are computed on January 2023 forecasts, with input/output windows aligned with those used by traditional 4D-Var to produce the IFS forecasts and ERA5 analysis, namely 21z - 09z and 09z - 21z. We plot the RMSEs as a function of lead time, for every 12 hours up to day 10. The lead time is defined relative to the nominal analysis time, e.g., 00z for a 21z - 09z window, same as in 4D-Var. 

Across most variables, GraphDOP outperforms climatology up to day five. Skilful forecasting of t2m is more challenging owing to its strong dependency on orography (hence output resolution), the relatively coarse resolution of surface-sensitive satellite channels, and the sparsity of conventional t2m observations. Consequently, the t2m RMSE of the GraphDOP forecast grows rapidly from day 3 onwards. The RMSE at the observation times and locations - blue curve, calculated against the "true" \textit{in situ} observations - is, however, much lower and in line with results presented in Section \ref{ref:section-comparison-ifs}. This discrepancy is currently being investigated.

\section{Case studies}
\label{ref:section-case-studies}

\subsection{A 10-day forecast of microwave brightness temperatures over sea ice}
\label{ref:subsection-sea-ice}

Over its multi-decadal history, the atmospheric model at the heart of IFS has been progressively coupled with other Earth System components, such as land, ocean and sea ice. This Earth System modelling (ESM) approach aims to better represent physical processes and make optimal use of interface observations - i.e., measurements that are sensitive to several components of the Earth system - during coupled 4D-Var \citep{Mogensen18,Rosnay22}. The development of a global ESM requires significant research and development efforts, to ensure that model errors in one component do not amplify errors in another, that initial conditions are consistent across different model components, the spatial and temporal resolution of the model components are coherent, and that interface observations are optimally exploited during data assimilation \citep{Rosnay22,browne2024interface}. In contrast, GraphDOP processes observations directly at their valid time and location without relying on physics-based models. In this section, we look at how GraphDOP uses the information in microwave brightness temperatures sensitive to sea ice. Sea-ice extent is a key indicator of climate variability and plays a critical role in regulating the Earth's energy balance and ocean circulation.

To produce a good forecast of sea ice extent, an observation-driven system such as GraphDOP must learn a coherent latent representation of the complex correlations between the sea-ice state, conventional observations of relevant geophysical parameters - sea surface temperature, winds, etc. - and (microwave or infrared) brightness temperature observations available over the area of interest. In addition, the initialising observations play a crucial role in the development of the forecast.

\begin{figure}[!ht]
\centering
\includegraphics[width = \linewidth]{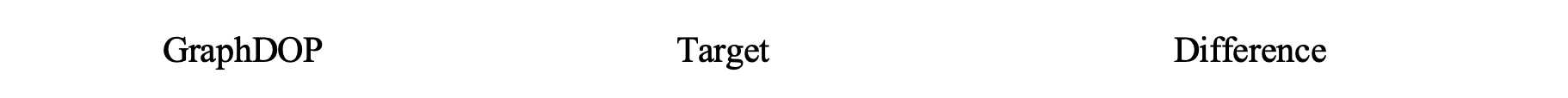}
\subfigure{\includegraphics[trim=0 0 50 0, clip, height=.3\linewidth]{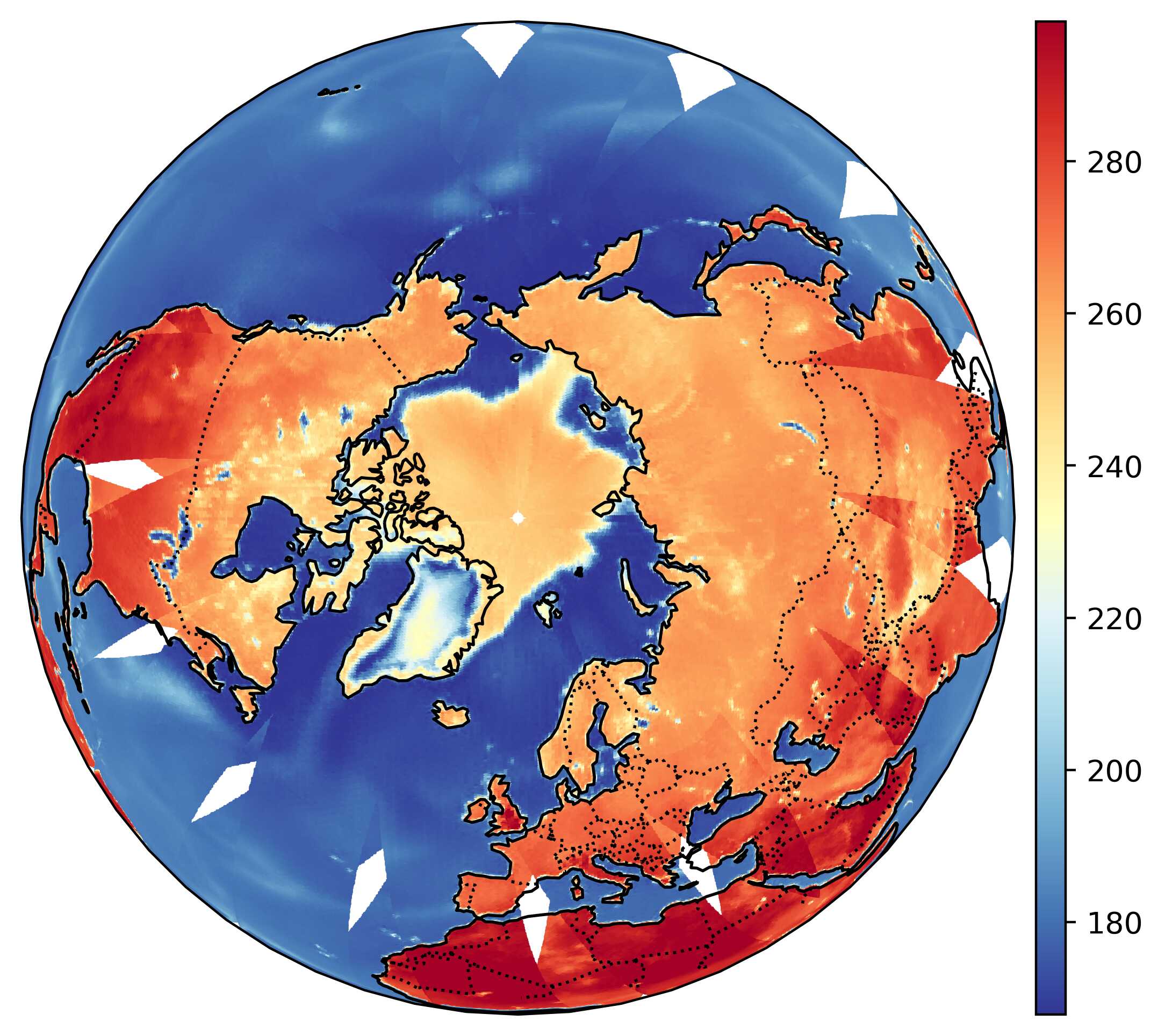} \includegraphics[height=.3\linewidth]{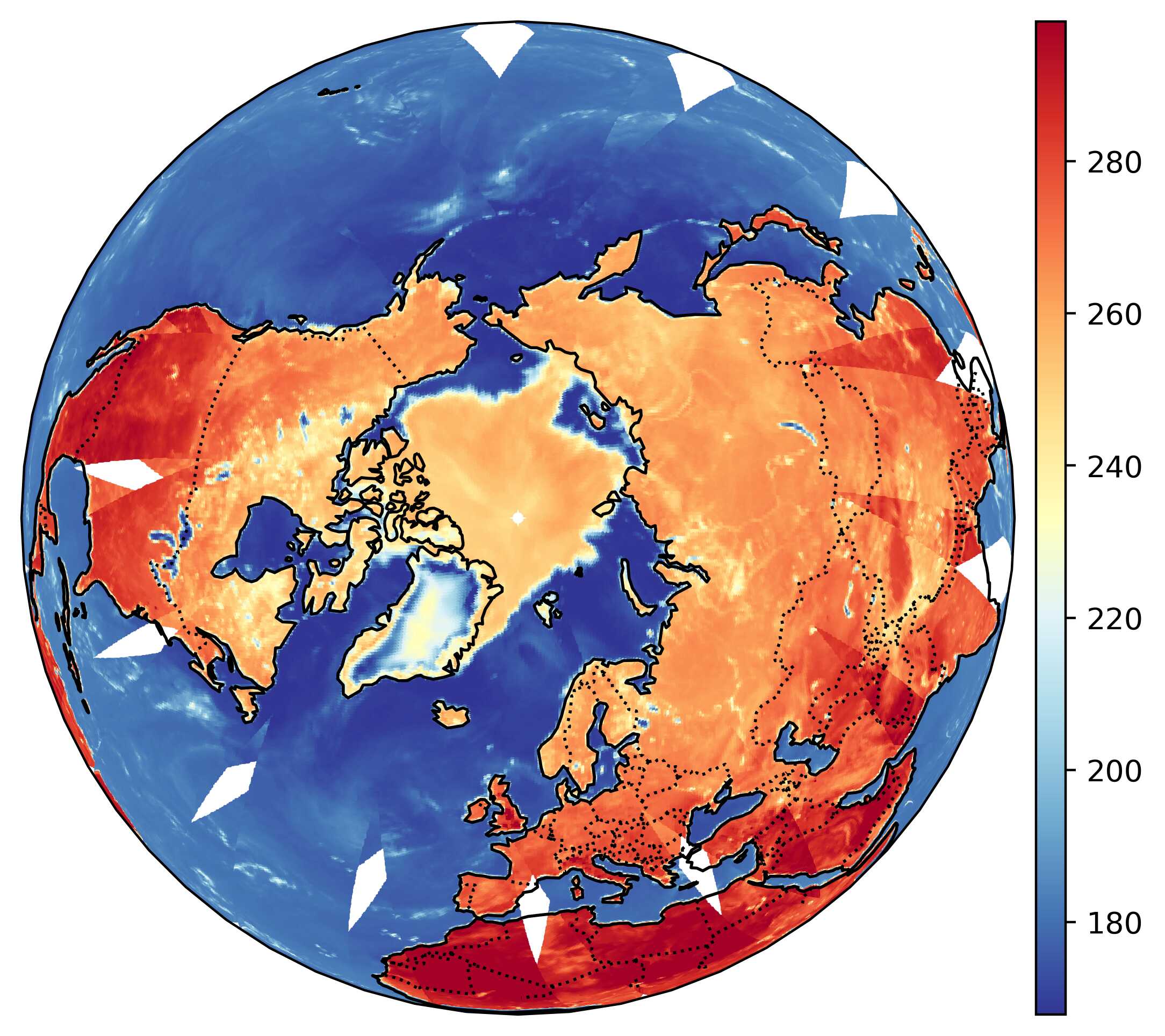}\includegraphics[height=.3\linewidth]{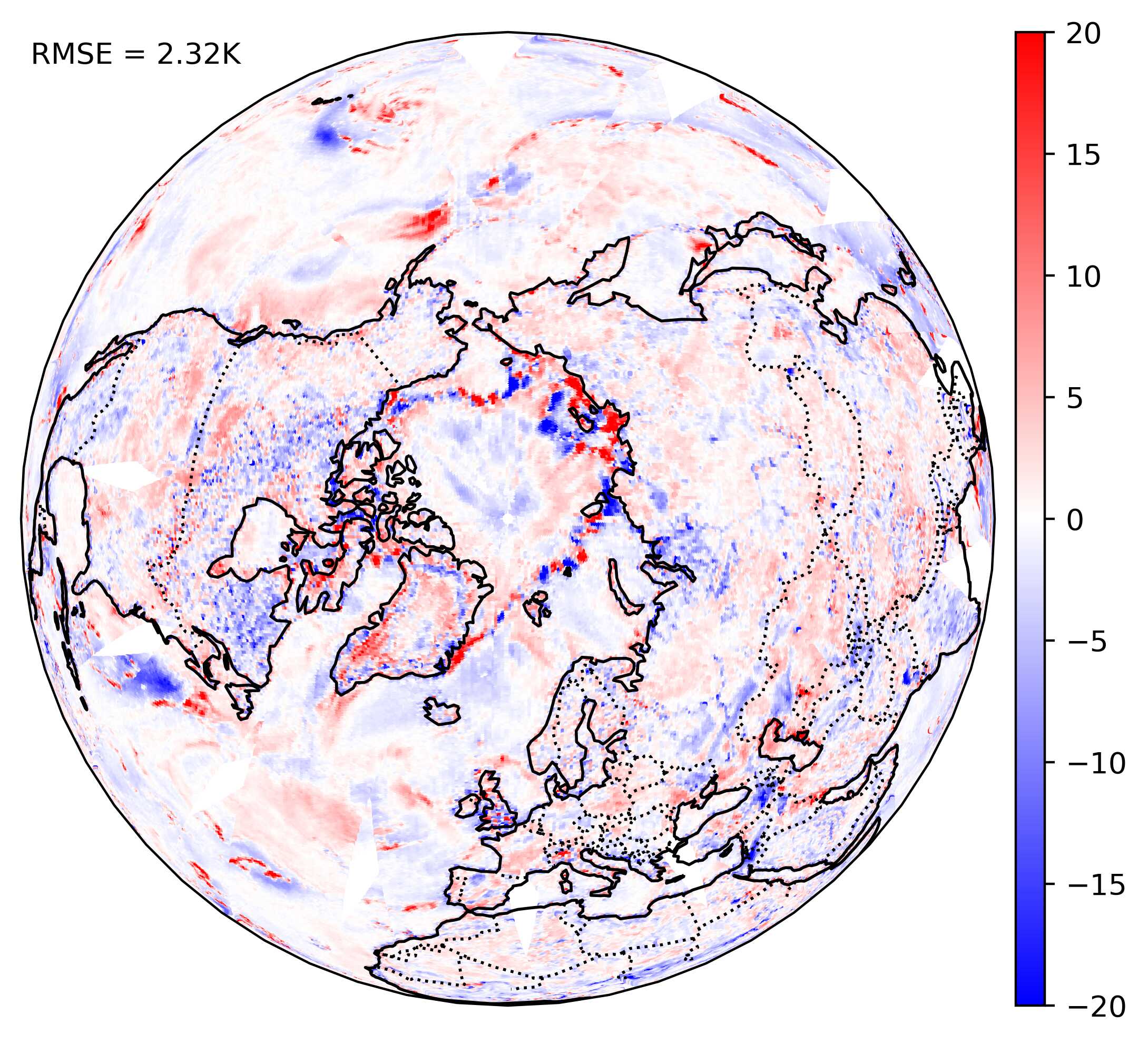}}
\subfigure{\includegraphics[trim=0 0 50 0, clip, height=.3\linewidth]{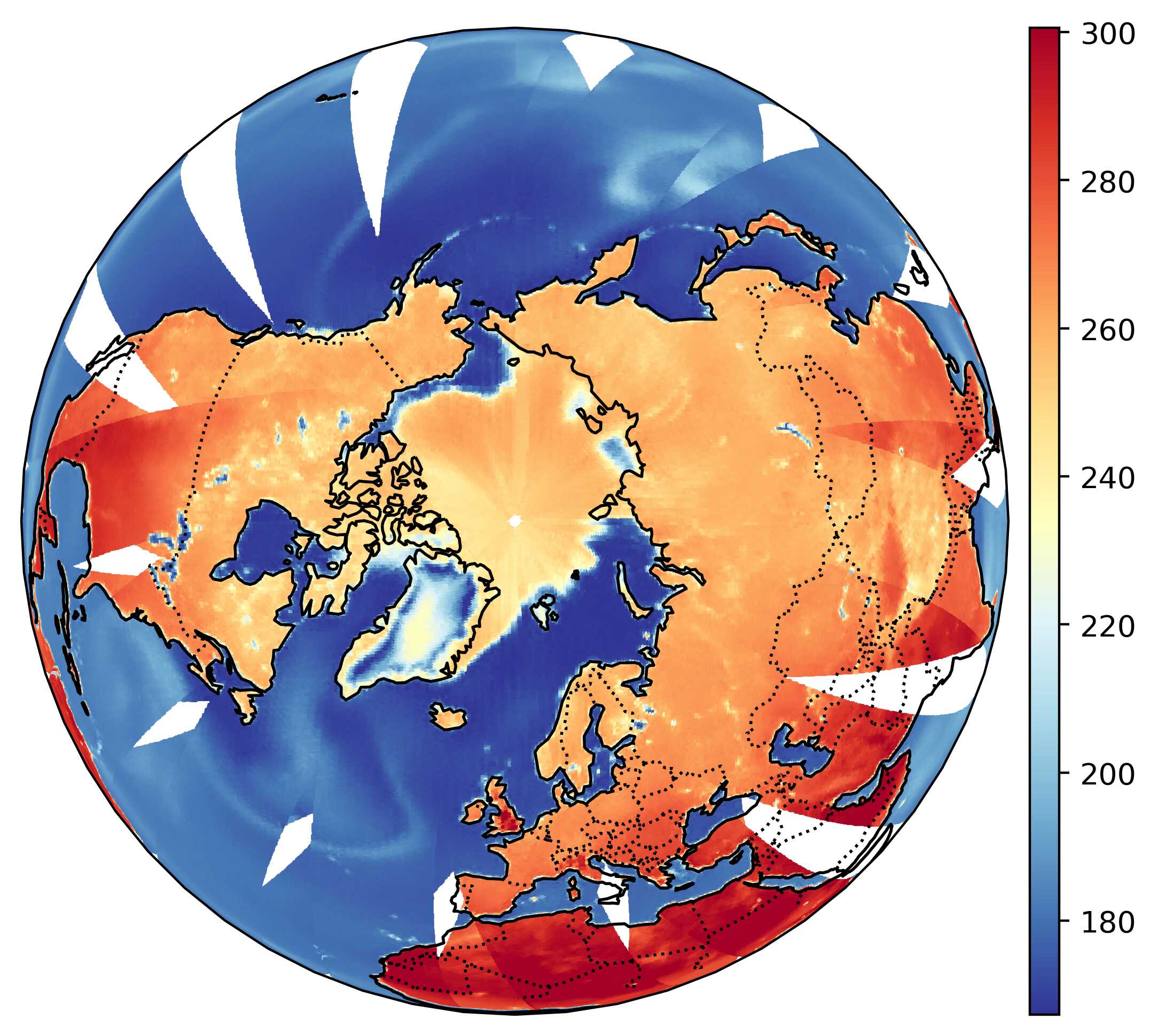} \includegraphics[height=.3\linewidth]{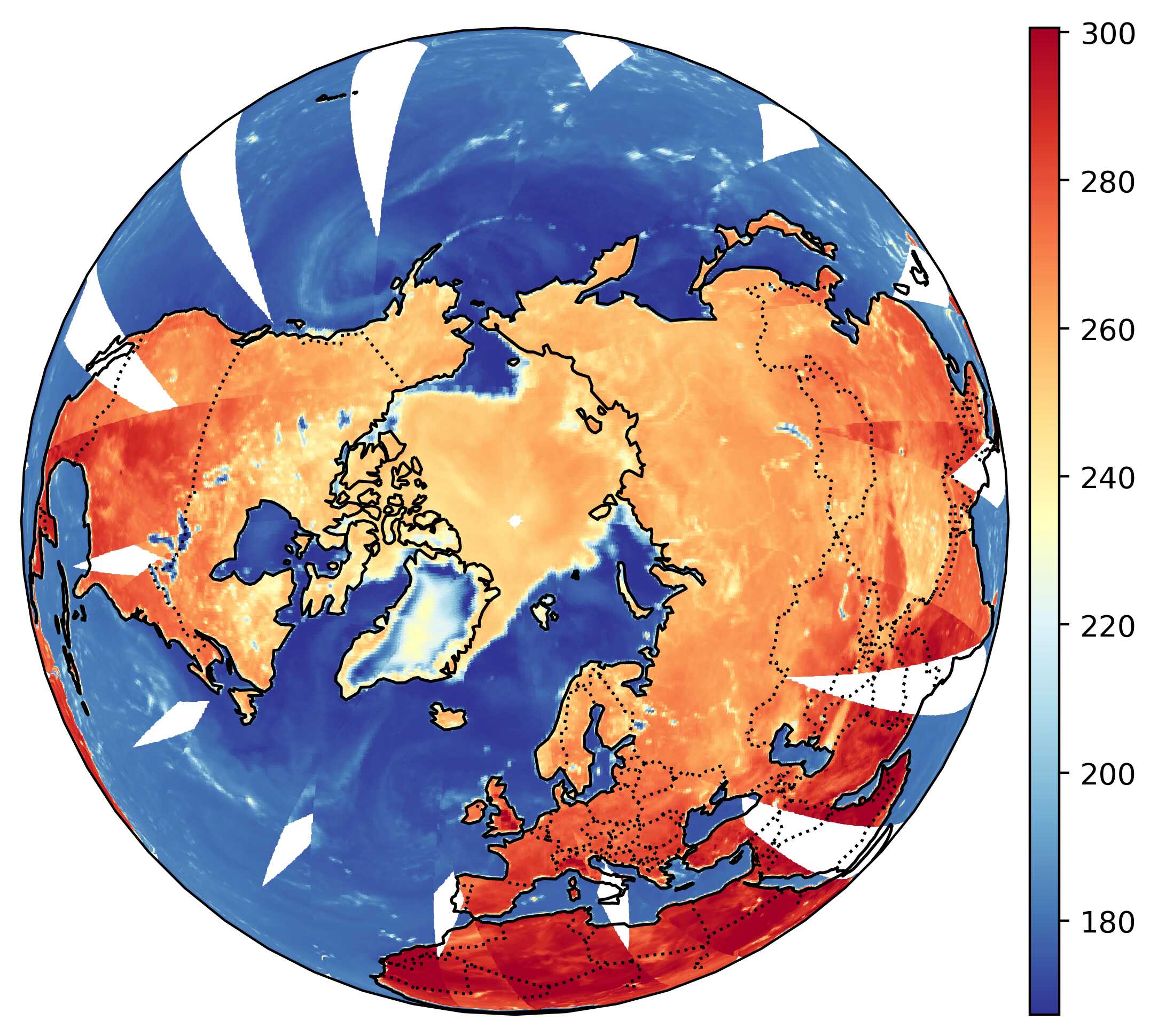}\includegraphics[height=.3\linewidth]{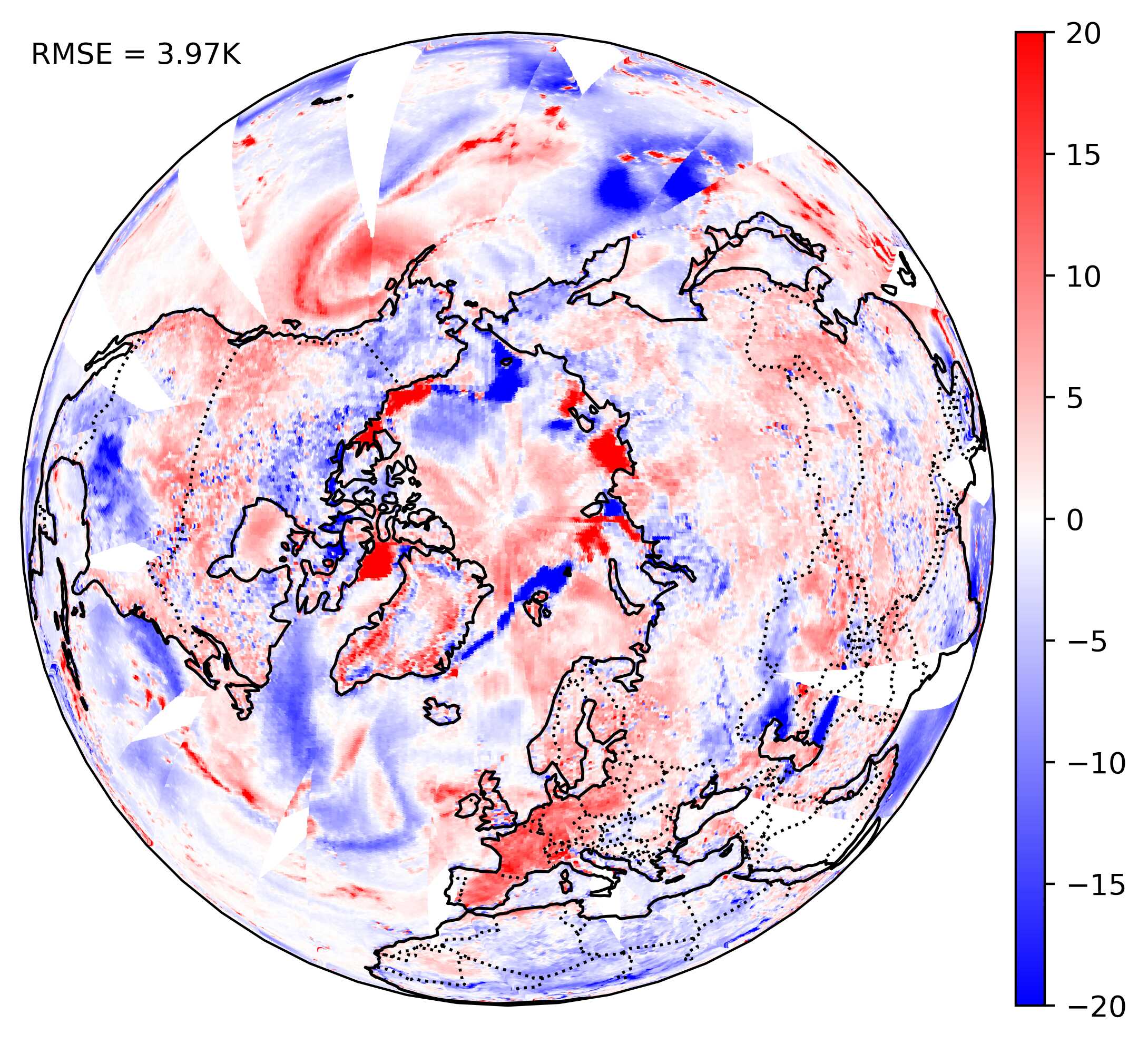}}
\caption{AMSR-2 channel 5 (10v) brightness temperatures (K): forecasted (left) observed (middle) and difference (observed minus forecast; right). The top left panel shows a 12-hour GraphDOP forecast (Oct 20, 2022, 21z - Oct 21, 2022, 09z), whereas the bottom left panel shows the sea ice signature forecasted at day 10 (Oct 29, 2022, 21z - Oct 30, 2022, 09z). The 10-day forecast shown here was initialised just before a rapid freezing event. The global forecast RMSE over the 12-hour window is shown on the right panel plot (top-left).}
\label{fig:sea_ice_amsr2}
\end{figure}

The top row of Figure \ref{fig:sea_ice_amsr2} shows the forecasted brightness temperatures (left) compared to target microwave brightness temperatures from AMSR-2 10GHz (V-pol) channel 5 (middle) over the first 12-hour window of a 10-day forecast, namely from Oct 20, 2022, 21z to Oct 21, 2022, 09z. AMSR-2 channel 5 brightness temperatures are sensitive to the ocean, sea ice and land. The 12-hour forecast in Figure \ref{fig:sea_ice_amsr2} is in very good agreement with the verifying observations. It shows a clear contrast between the land surface (e.g., Greenland) and sea ice, a strong indication that the model has correctly encoded the relationships between surface emissivity and measured brightness temperature into a coherent latent representation of sea ice dynamics.

The bottom row of Figure \ref{fig:sea_ice_amsr2} compares the forecast (left) to the target observations (middle) at a lead time of 10 days. A rapid freezing event occurred during this ten-day period. Remarkably, GraphDOP was able to forecast the signature of sea ice growth in the microwave brightness temperatures quite accurately. By combining the information in the initialising observations with its learned representation of the Earth system, the model was able to infer that the atmospheric conditions at the start of the forecast (not shown) were favourable for freezing to occur - namely, mostly clear skies that enhance radiative cooling at the surface, allowing heat to escape into space more effectively and leading to sea ice accumulation. We note that there is a slight difference in the ice boundaries around the Sea of Okhotsk (in the western Pacific Ocean), with GraphDOP predicting less sea ice accumulation than what was actually observed. The day-10 forecast also misses most of the small scale features in the brightness temperature signal over the ocean; we attribute this to the smoothing induced by the deterministic WMSE objective at long lead times.

\subsection{Hurricane Ian}
\label{ref:subsection-hurricane-ian}

\begin{figure}[h!]
\centering

\subfigure[Mean sea-level pressure (Pa)]{
    \begin{minipage}{\linewidth}
    \centering
    \includegraphics[trim=0 3 0 0, clip, height=.15\linewidth]{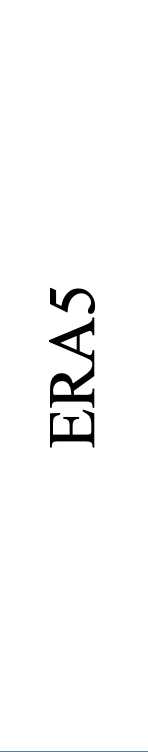}
    \includegraphics[height=.15\linewidth]{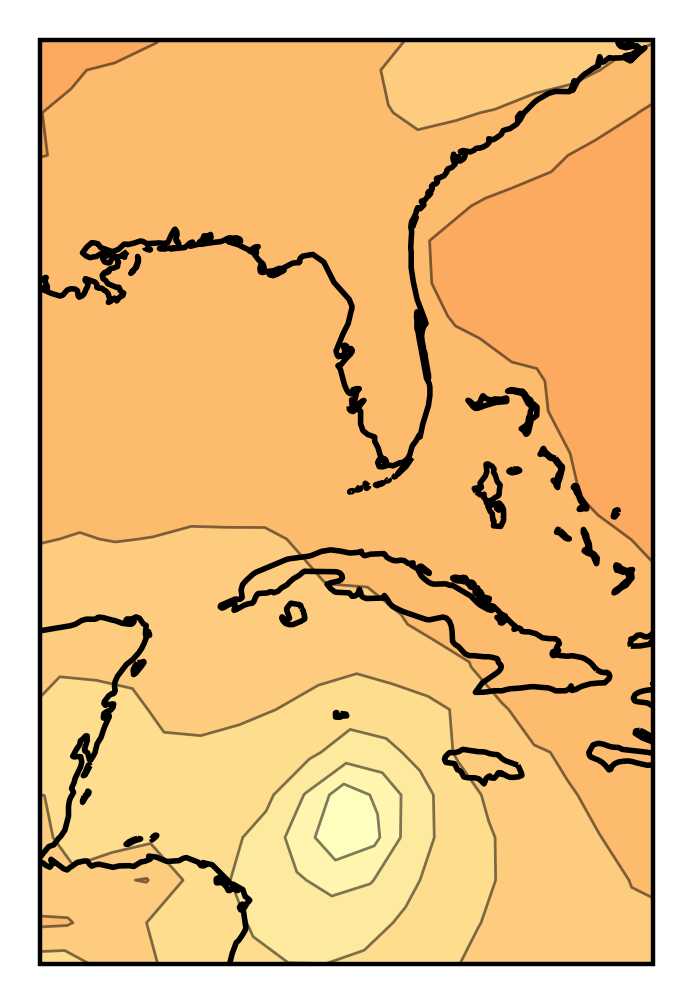}
    \includegraphics[height=.15\linewidth]{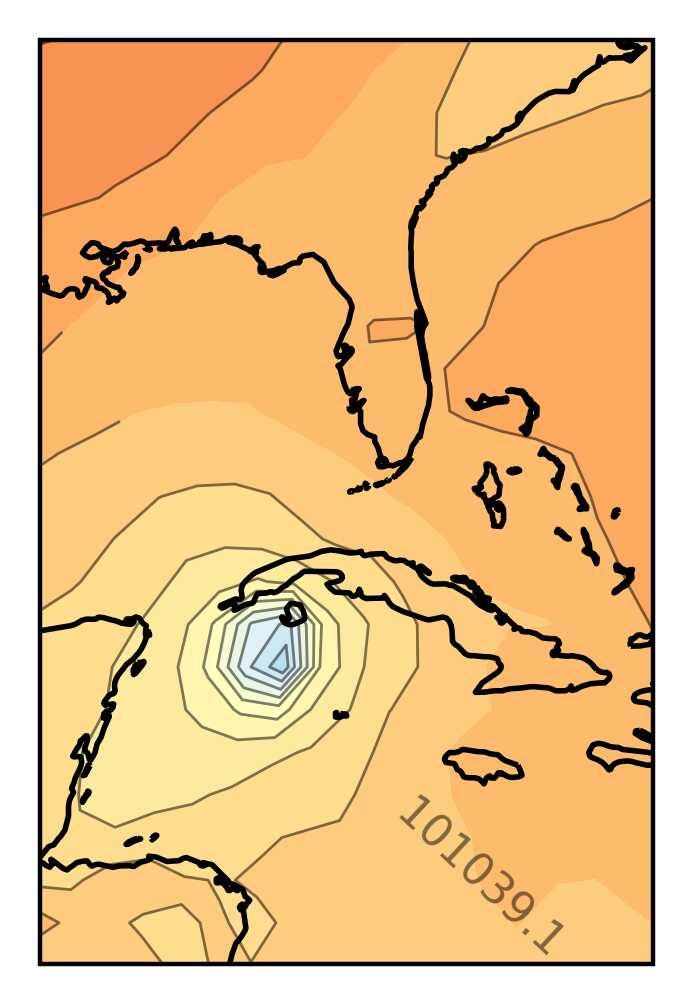}
    \includegraphics[height=.15\linewidth]{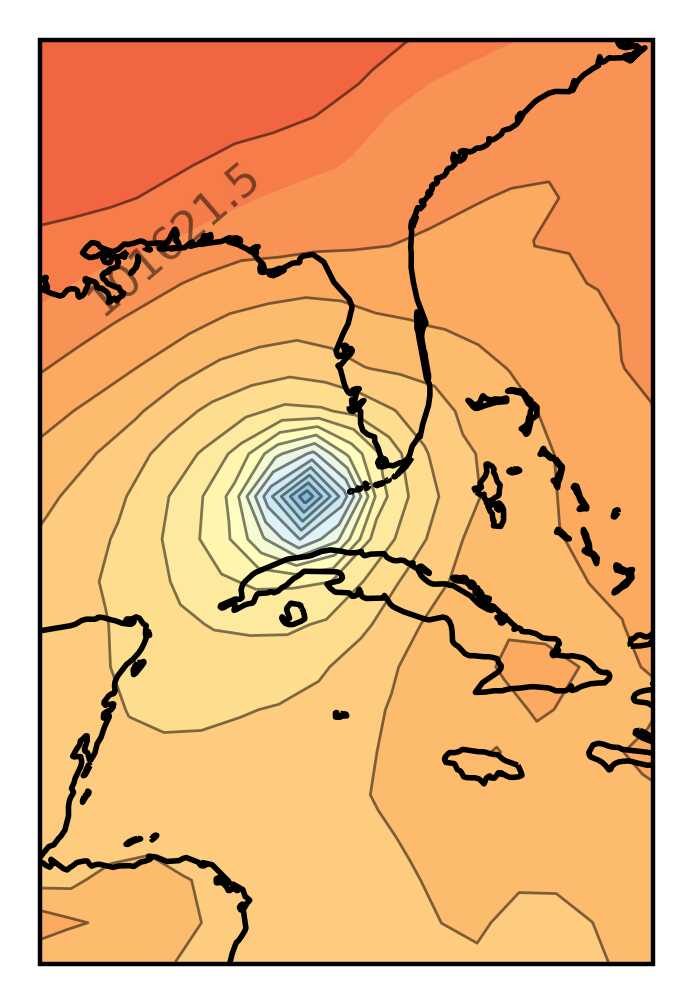}
    \includegraphics[height=.15\linewidth]{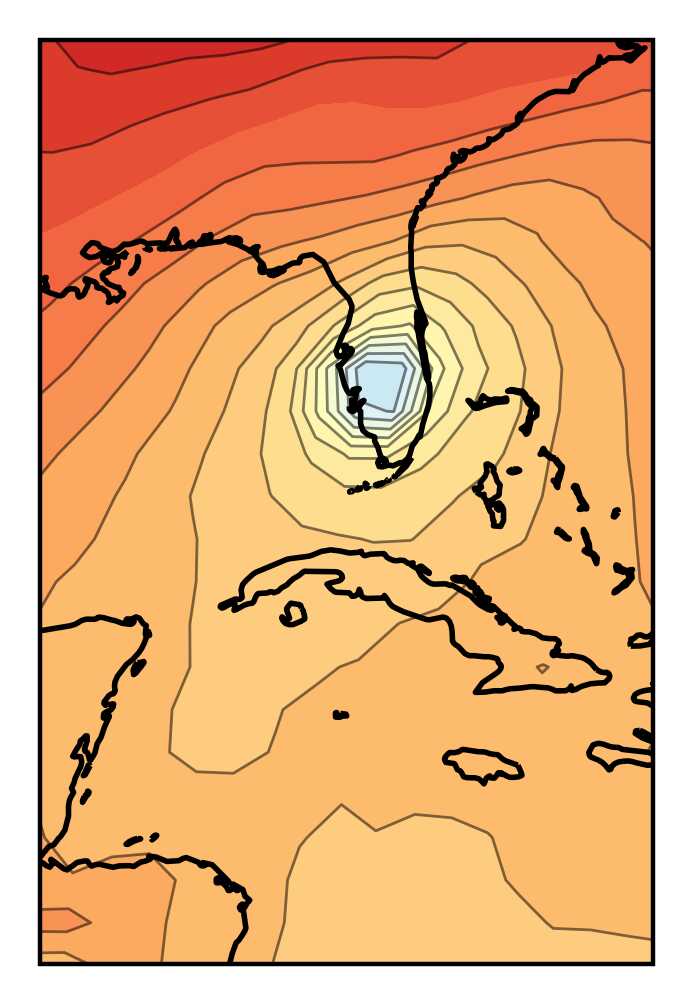}
    \includegraphics[height=.15\linewidth]{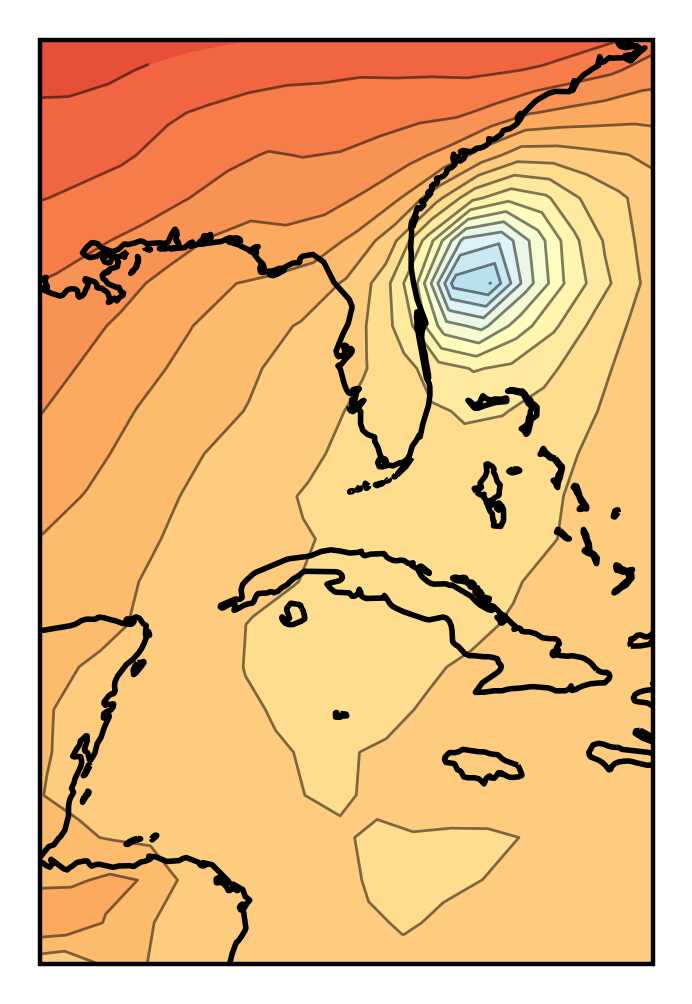}
    \includegraphics[height=.15\linewidth]{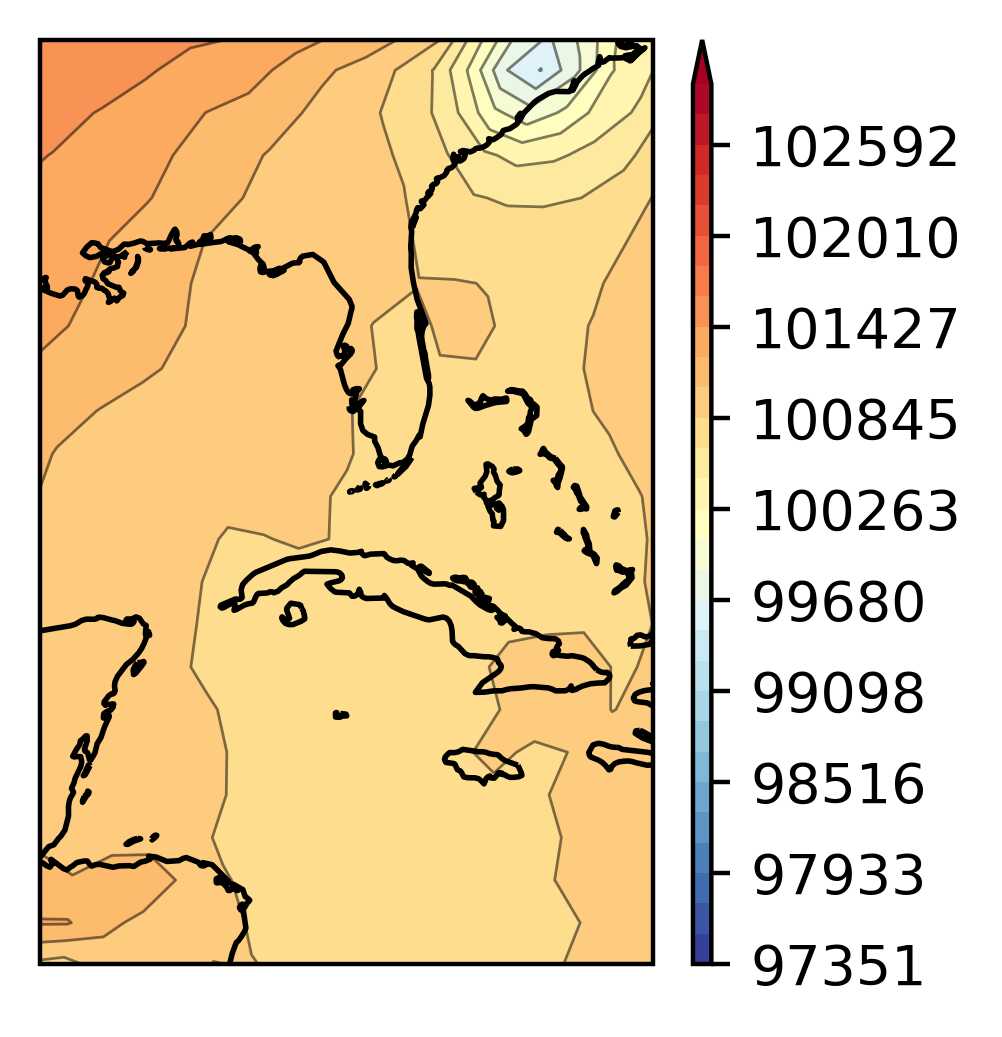}

    \vspace{0.5em}

    \includegraphics[trim=0 3 0 0, clip, height=.15\linewidth]{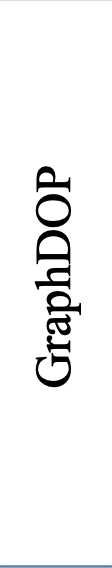}
    \includegraphics[height=.15\linewidth]{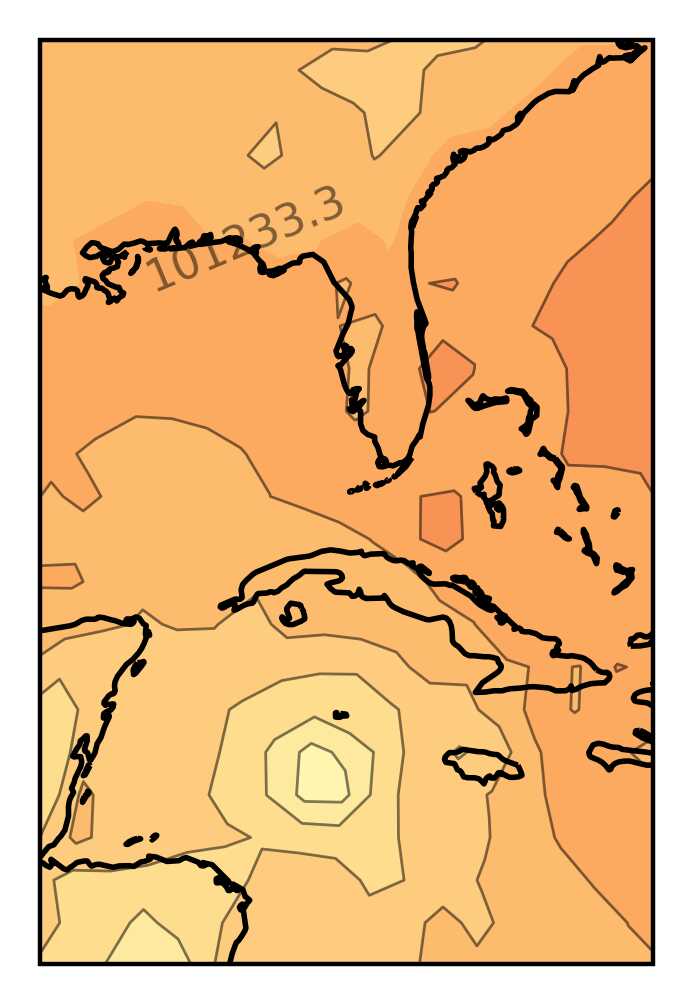}
    \includegraphics[height=.15\linewidth]{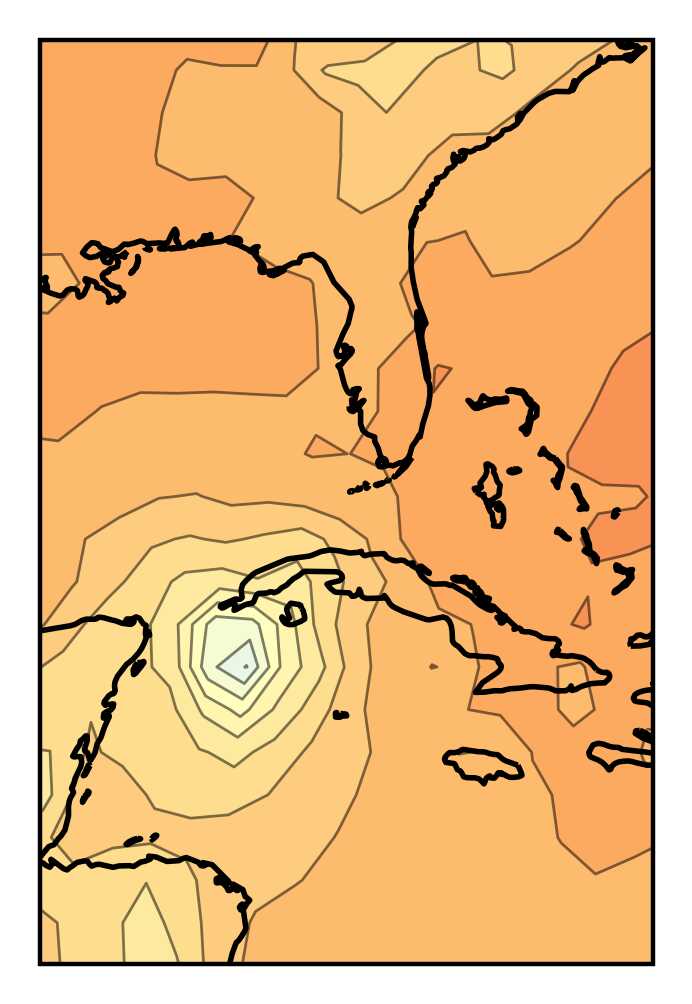}
    \includegraphics[height=.15\linewidth]{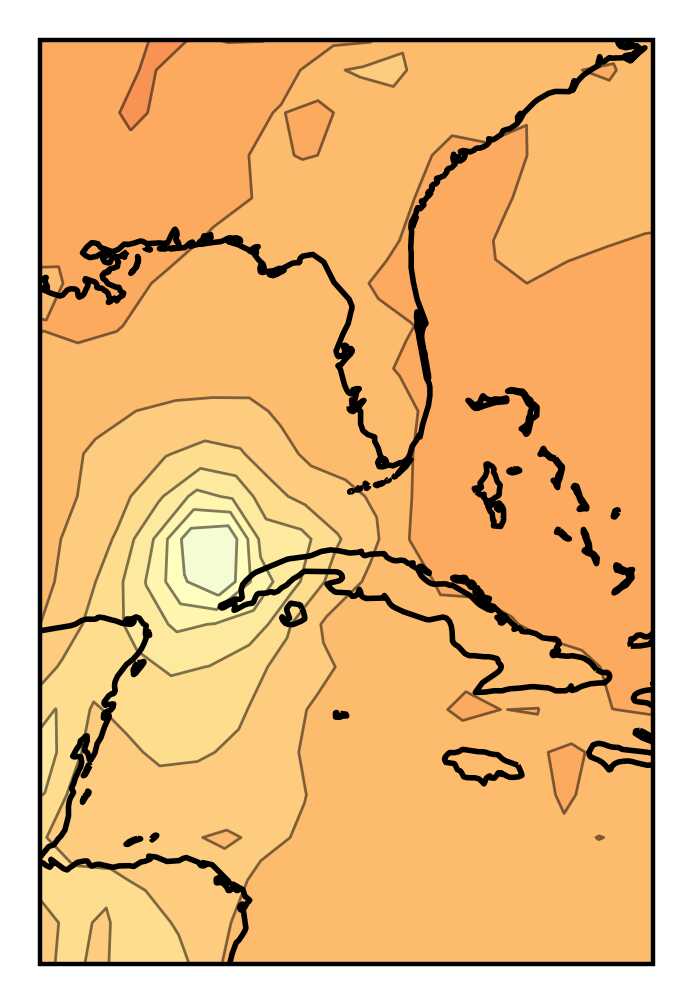}
    \includegraphics[height=.15\linewidth]{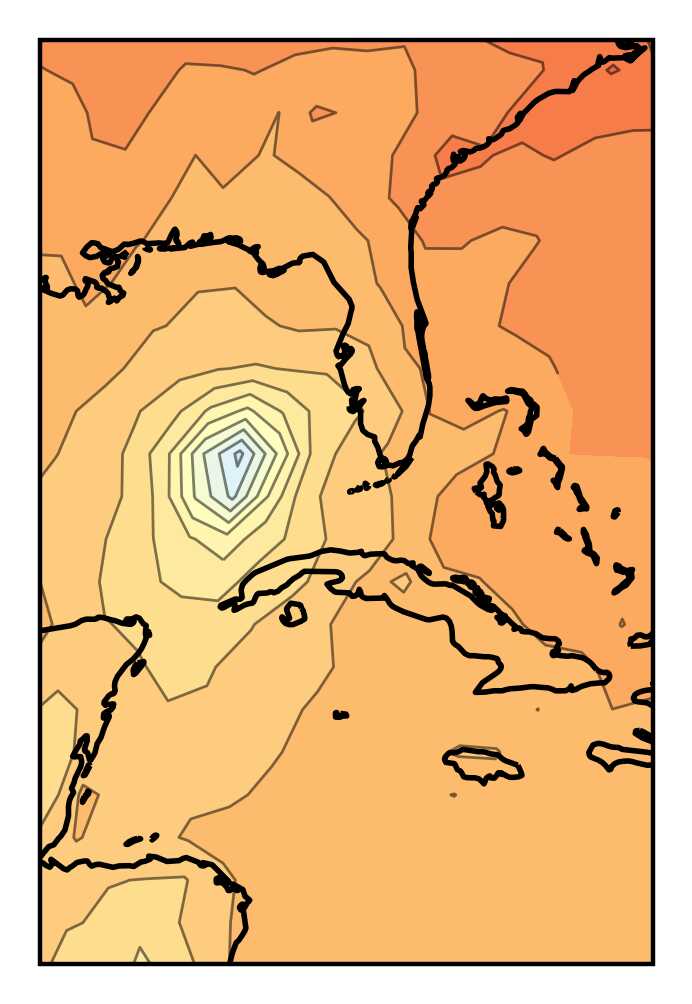}
    \includegraphics[height=.15\linewidth]{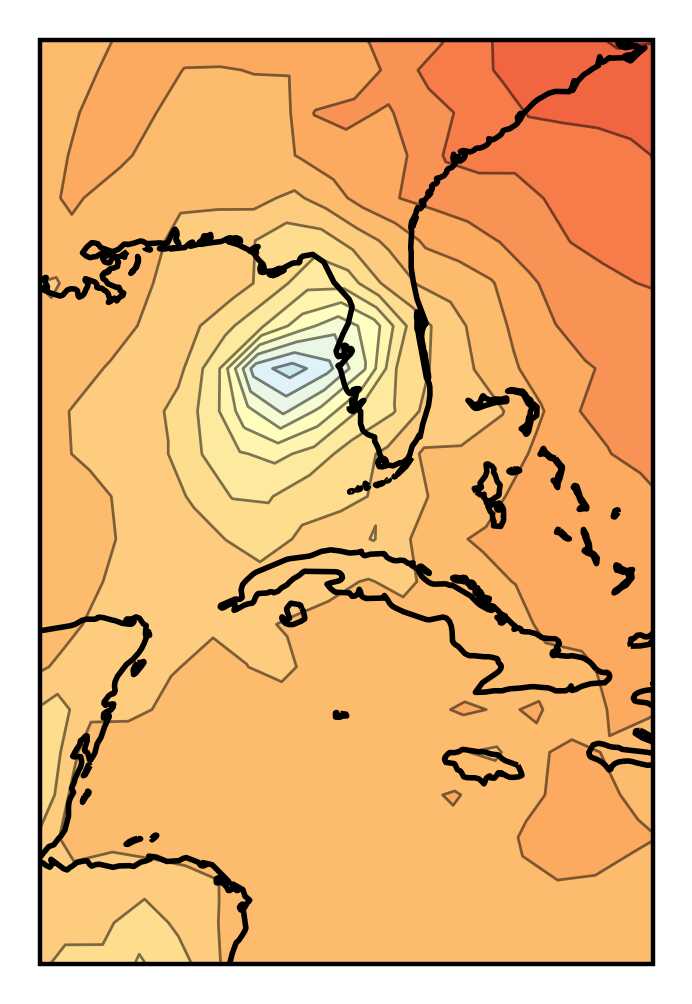}
    \includegraphics[height=.15\linewidth]{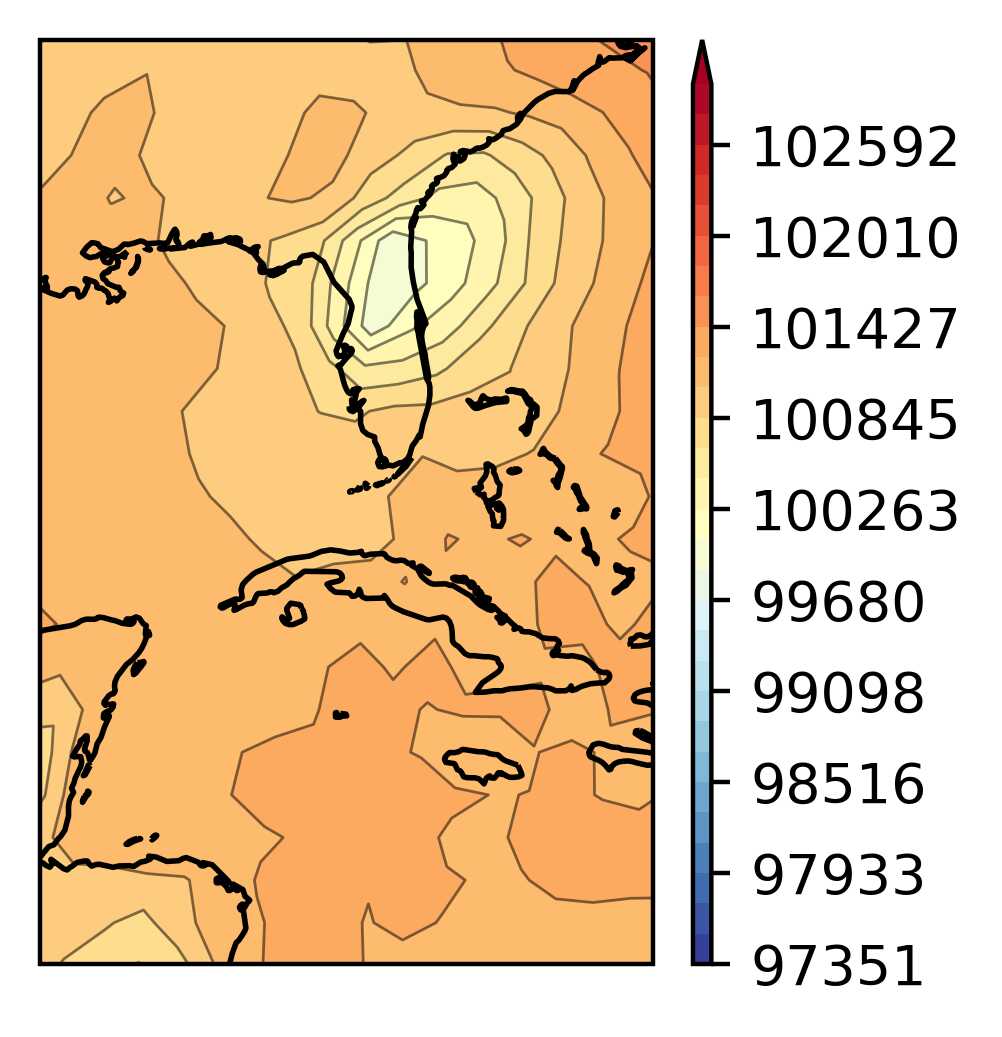}
    \end{minipage}
}

\vspace{1em}

\subfigure[Wind speed (m/s)]{
    \begin{minipage}{\linewidth}
    \centering
    \includegraphics[trim=0 3 0 0, clip, height=.15\linewidth]{images/ian/era.jpg}
    \includegraphics[height=.15\linewidth]{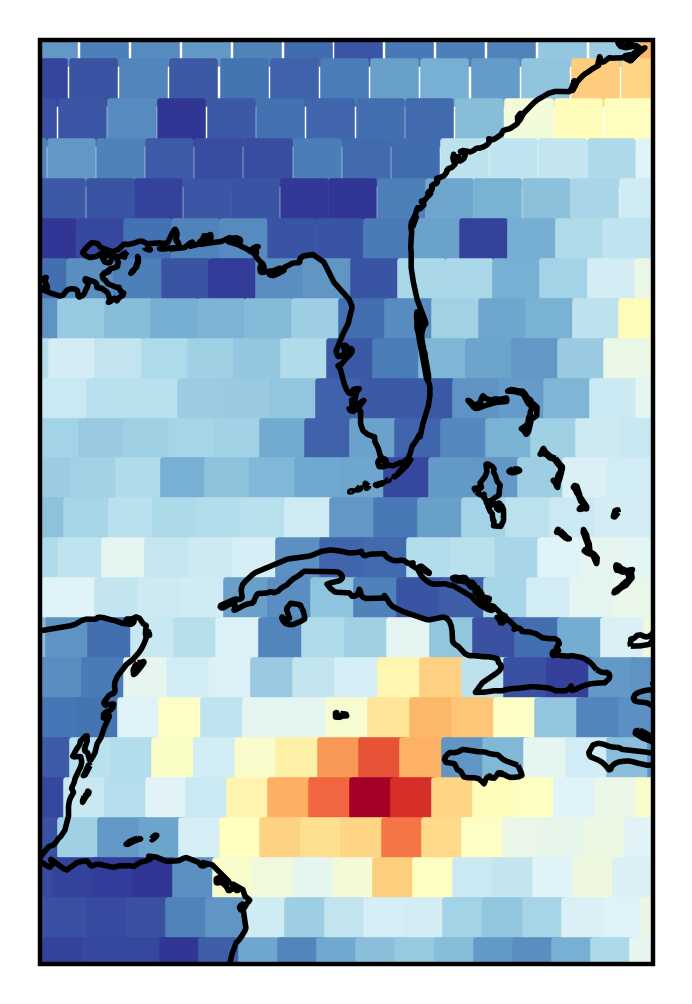}
    \includegraphics[height=.15\linewidth]{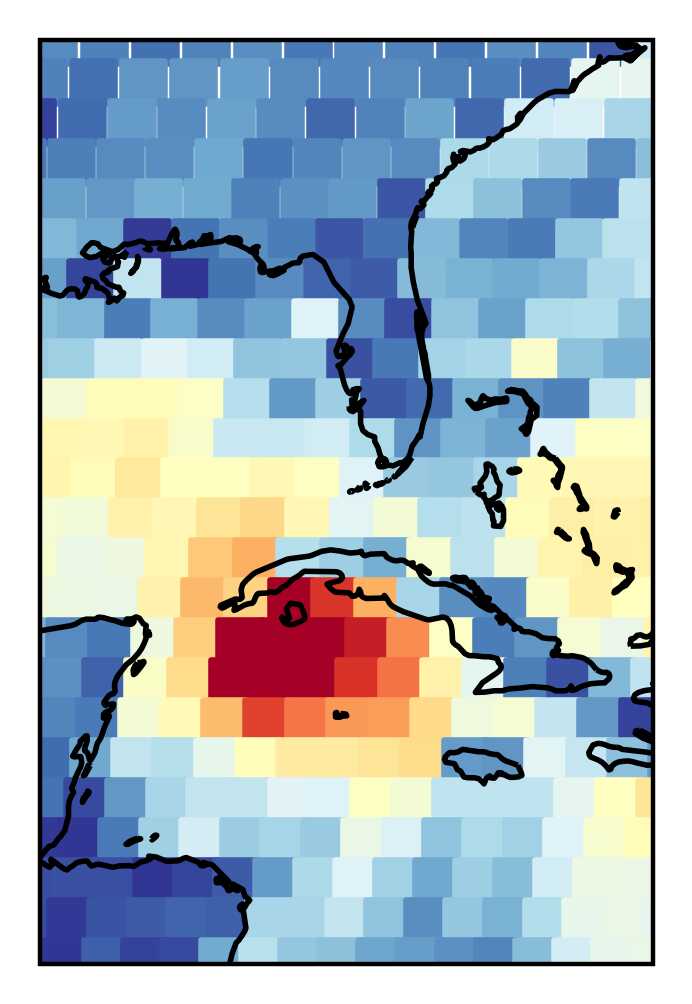}
    \includegraphics[height=.15\linewidth]{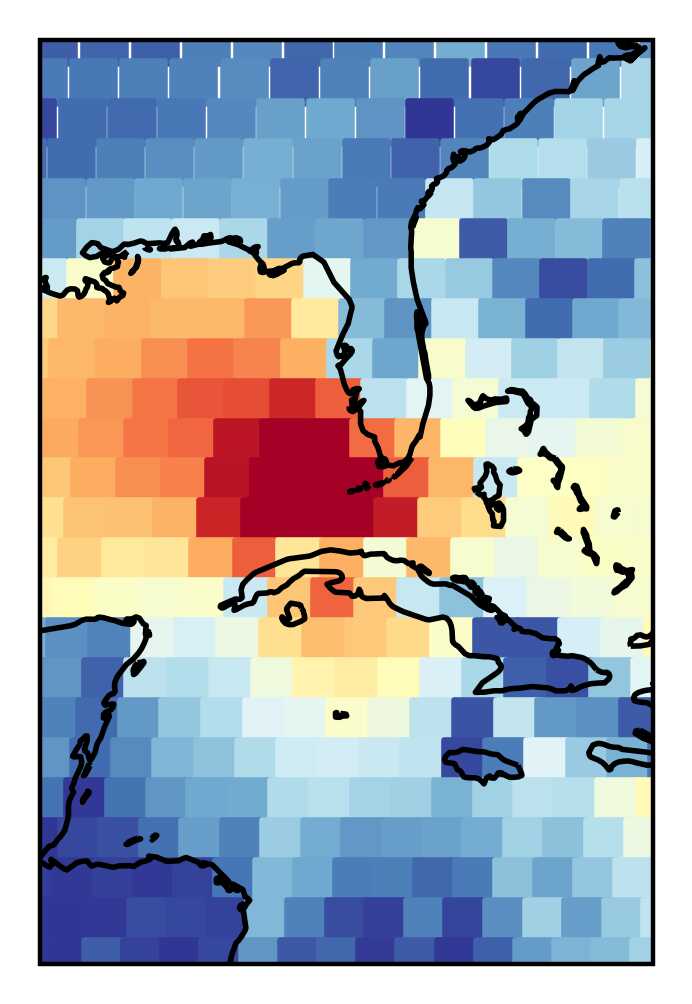}
    \includegraphics[height=.15\linewidth]{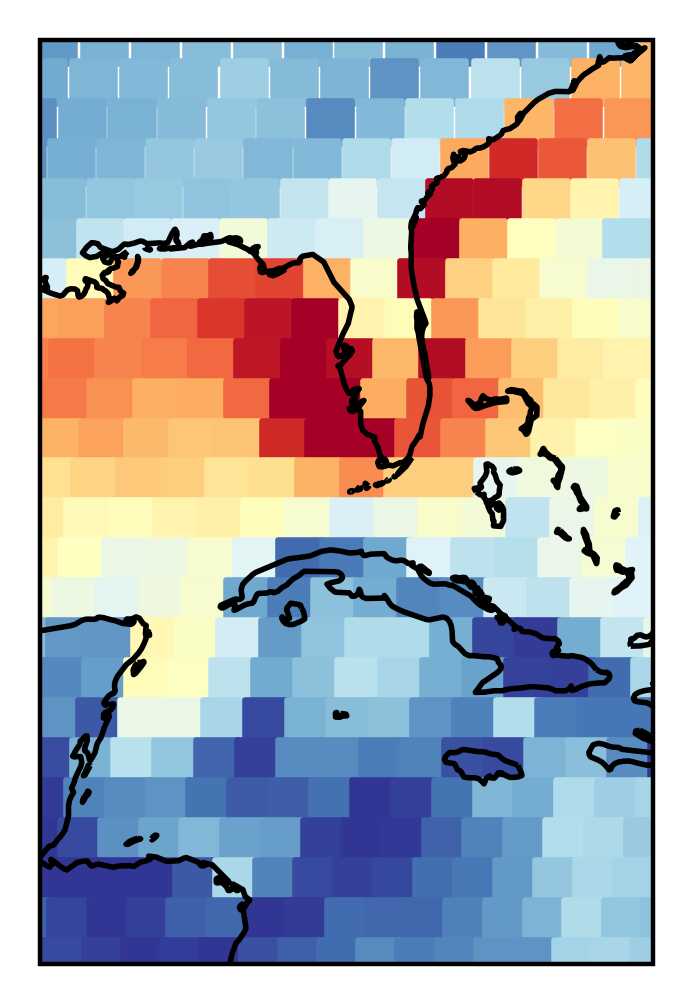}
    \includegraphics[height=.15\linewidth]{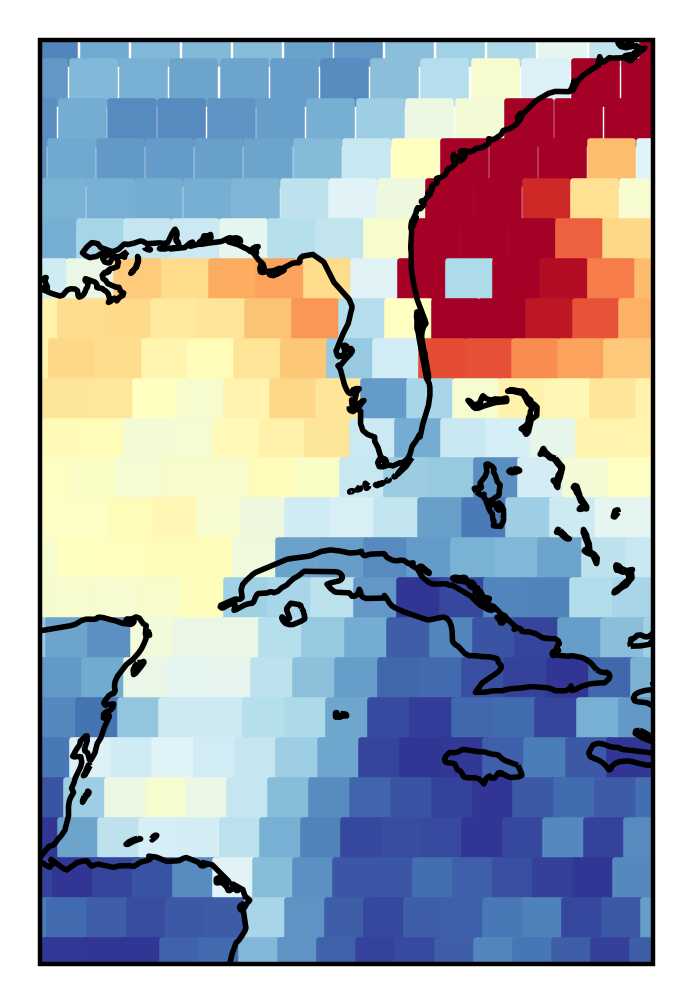}
    \includegraphics[height=.15\linewidth]{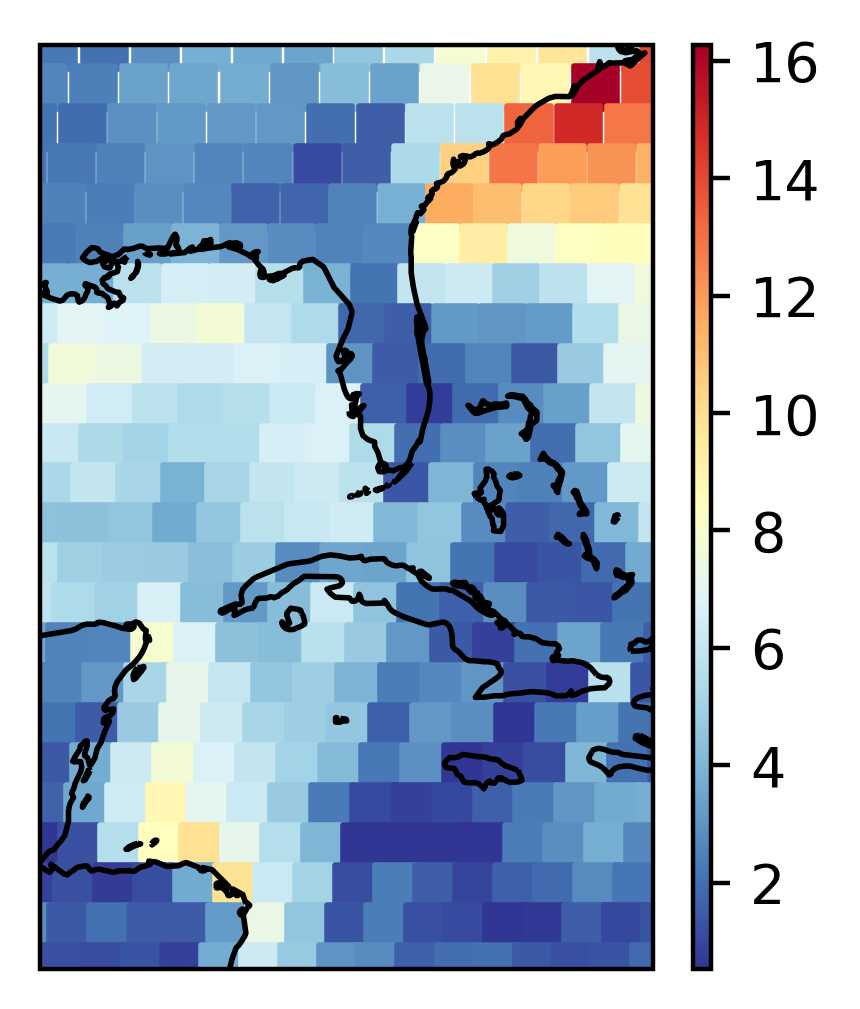}

    \vspace{0.5em}

    \includegraphics[trim=0 3 0 0, clip, height=.15\linewidth]{images/ian/dop.jpg}
    \includegraphics[height=.15\linewidth]{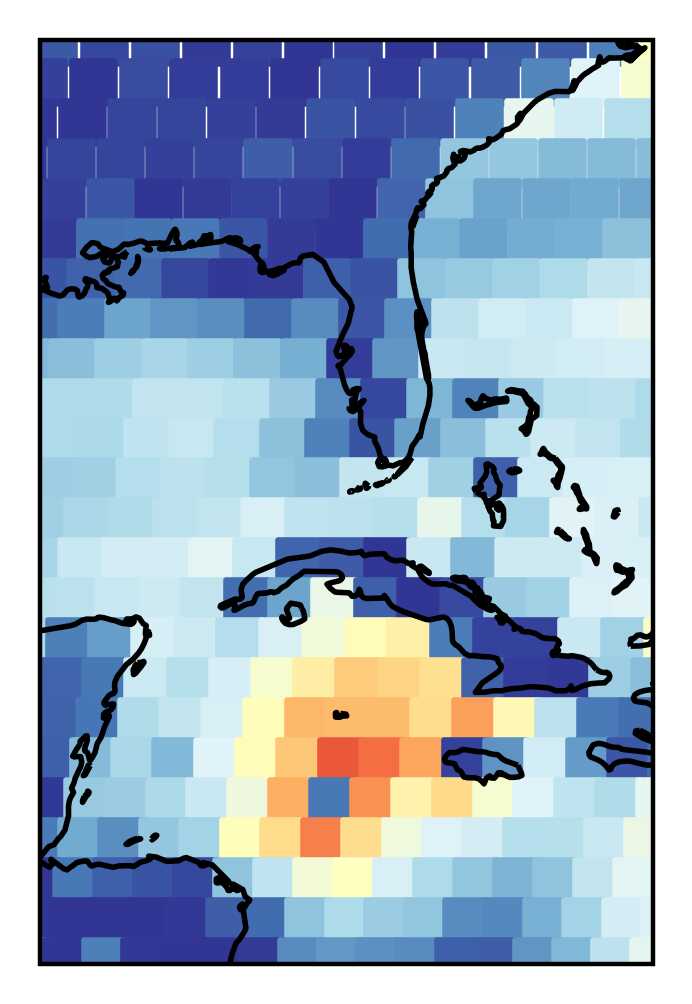}
    \includegraphics[height=.15\linewidth]{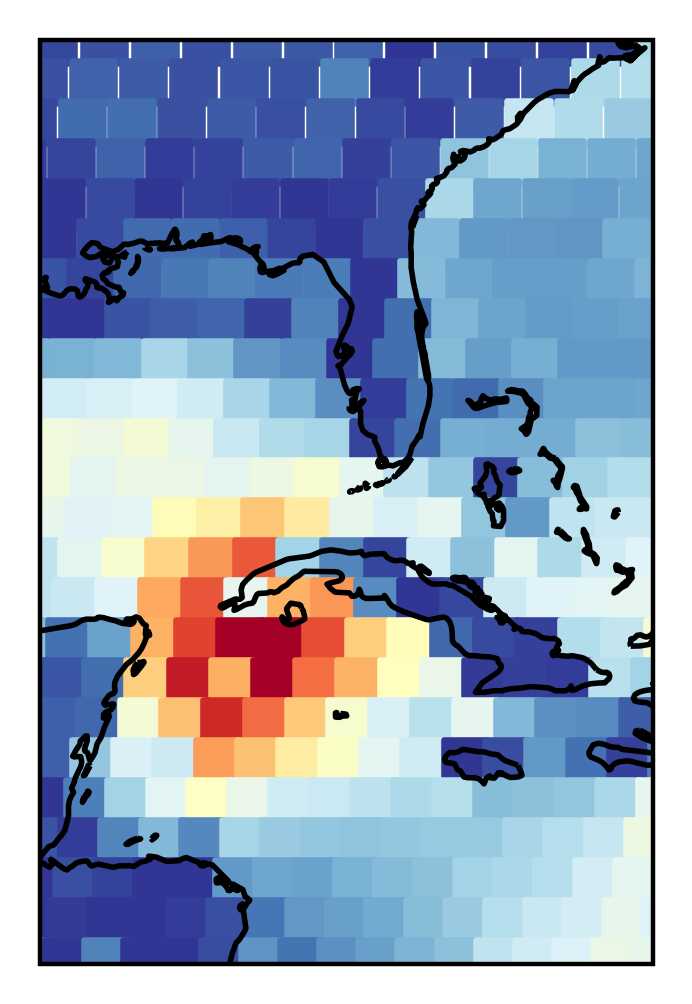}
    \includegraphics[height=.15\linewidth]{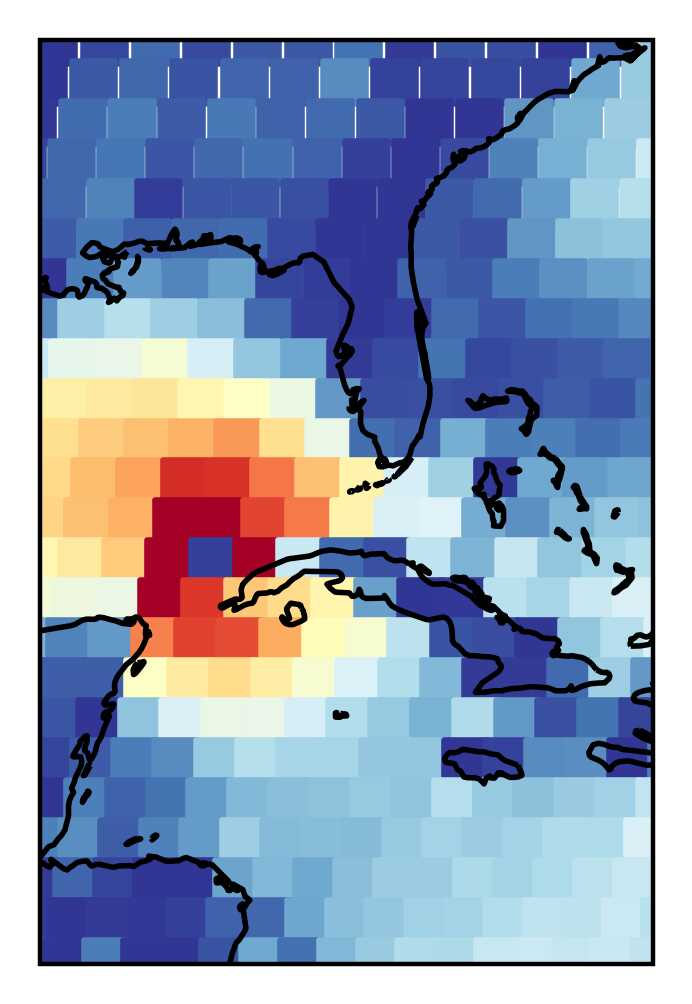}
    \includegraphics[height=.15\linewidth]{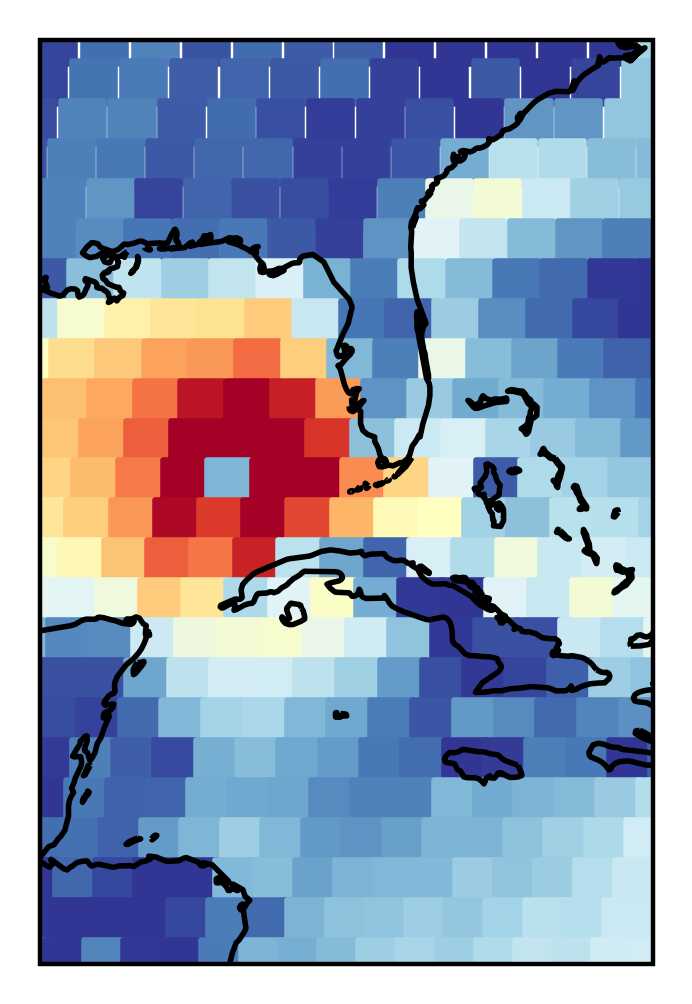}
    \includegraphics[height=.15\linewidth]{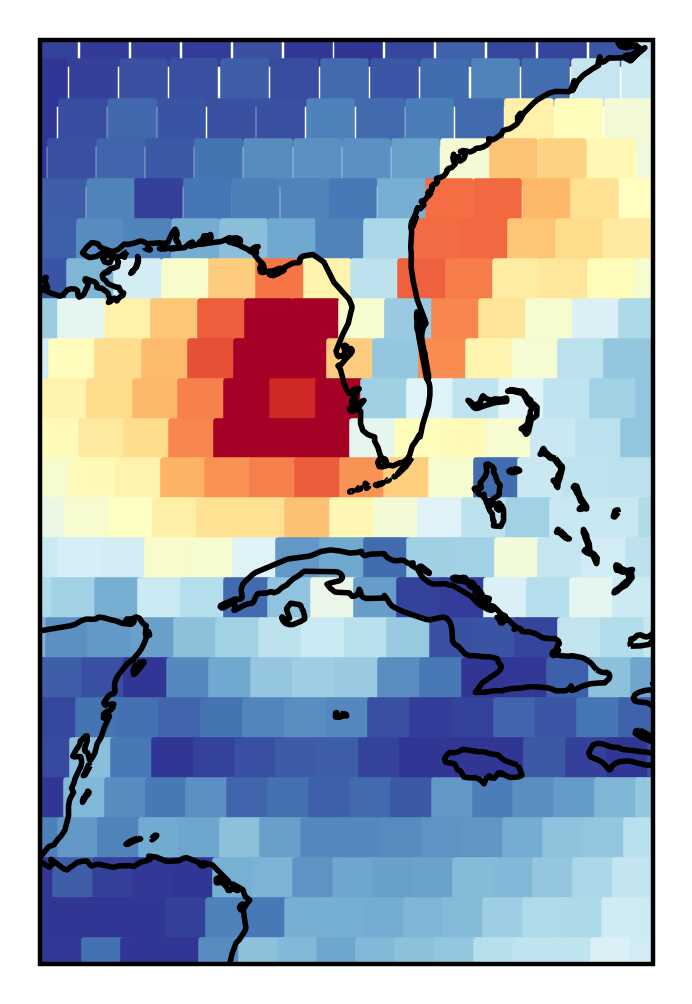}
    \includegraphics[height=.15\linewidth]{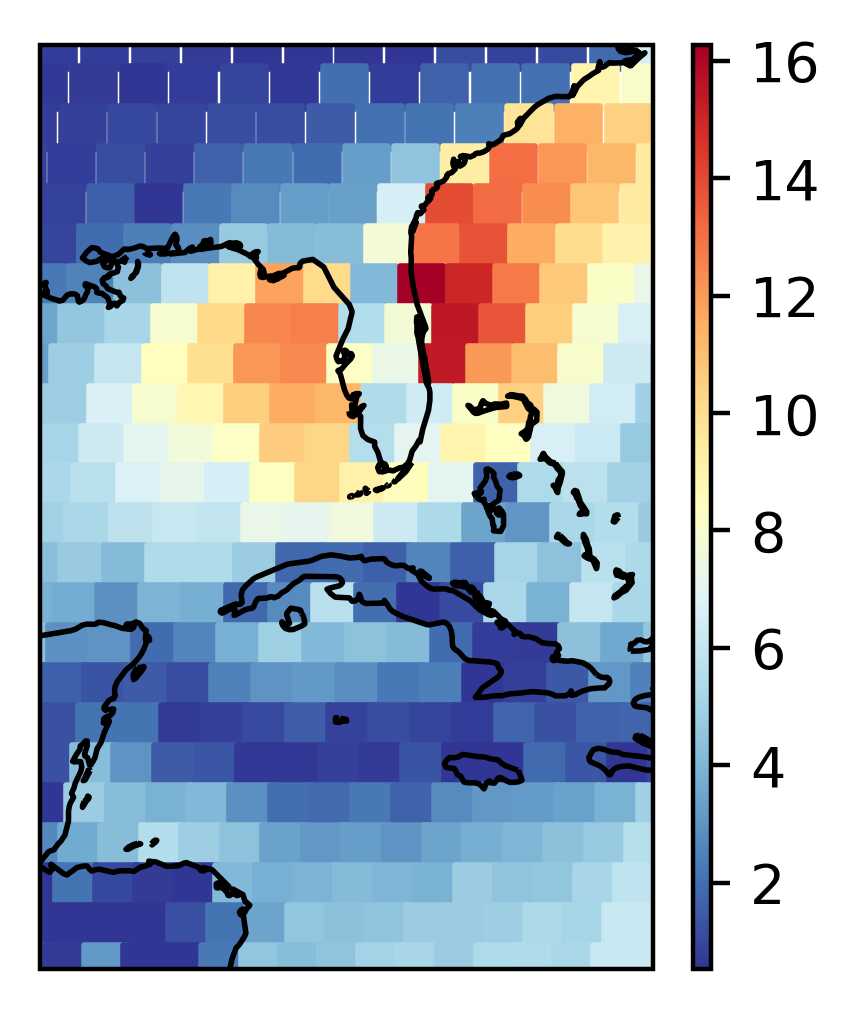}
    \end{minipage}
}

\vspace{1em}

\subfigure[Significant wave height (m)]{
    \begin{minipage}{\linewidth}
    \centering
    \includegraphics[trim=0 3 0 0, clip, height=.15\linewidth]{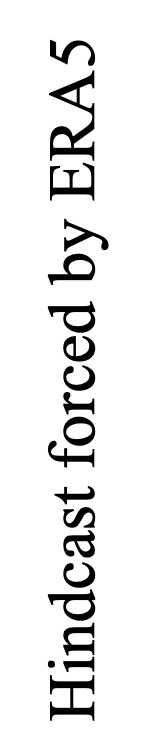}
    \includegraphics[height=.15\linewidth]{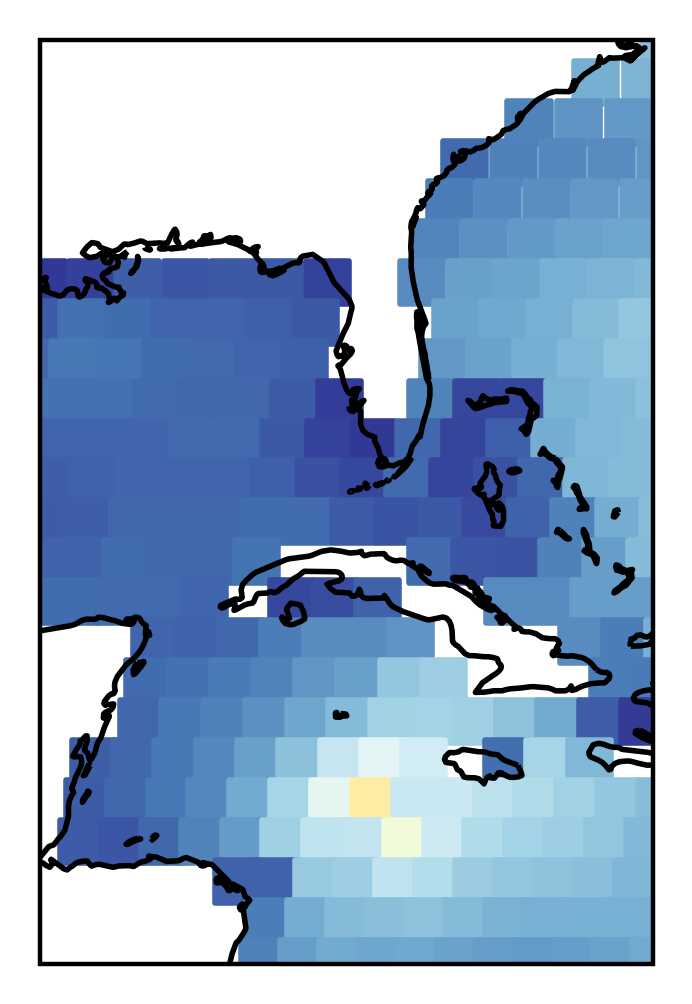}
    \includegraphics[height=.15\linewidth]{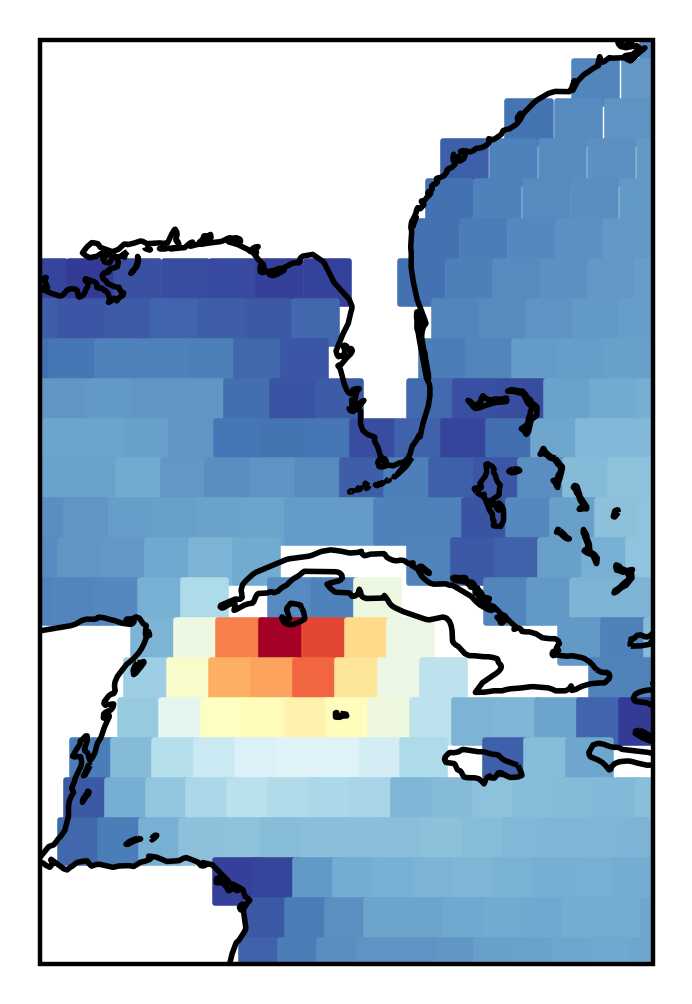}
    \includegraphics[height=.15\linewidth]{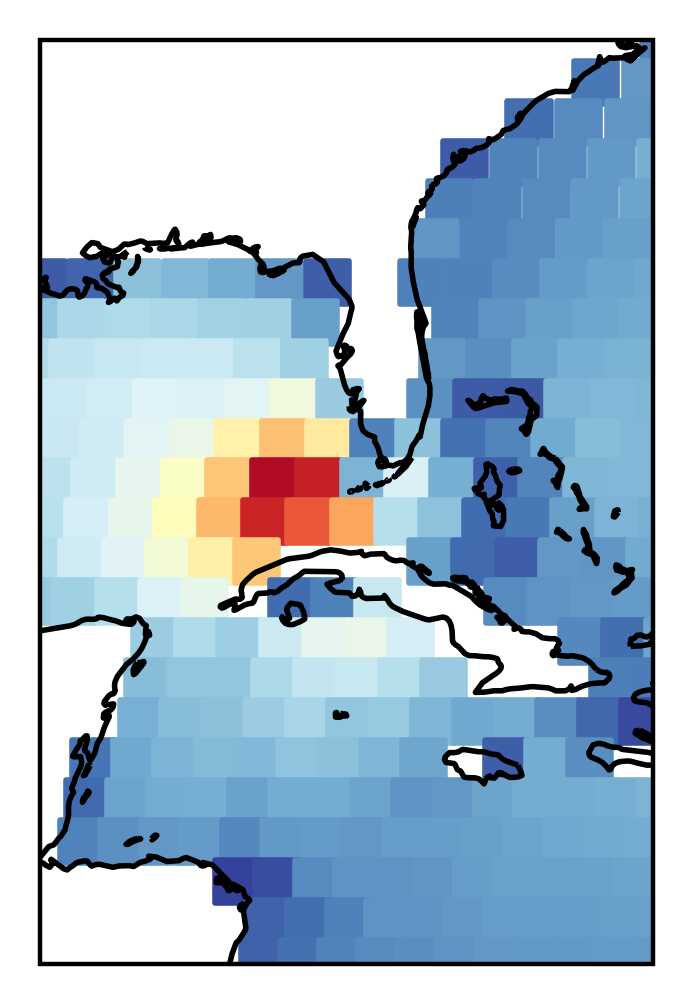}
    \includegraphics[height=.15\linewidth]{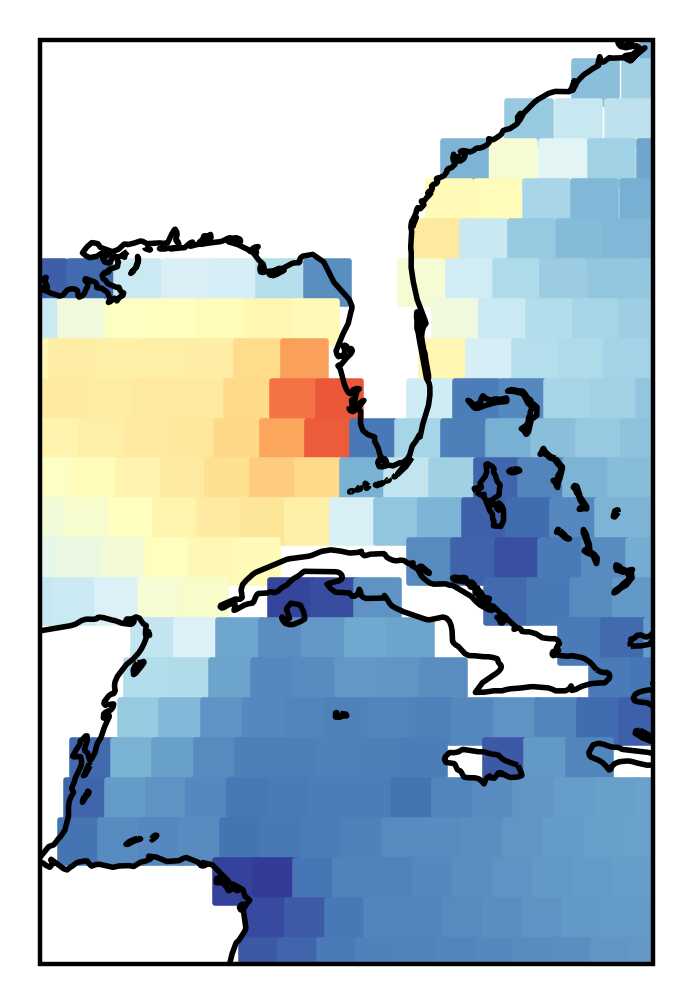}
    \includegraphics[height=.15\linewidth]{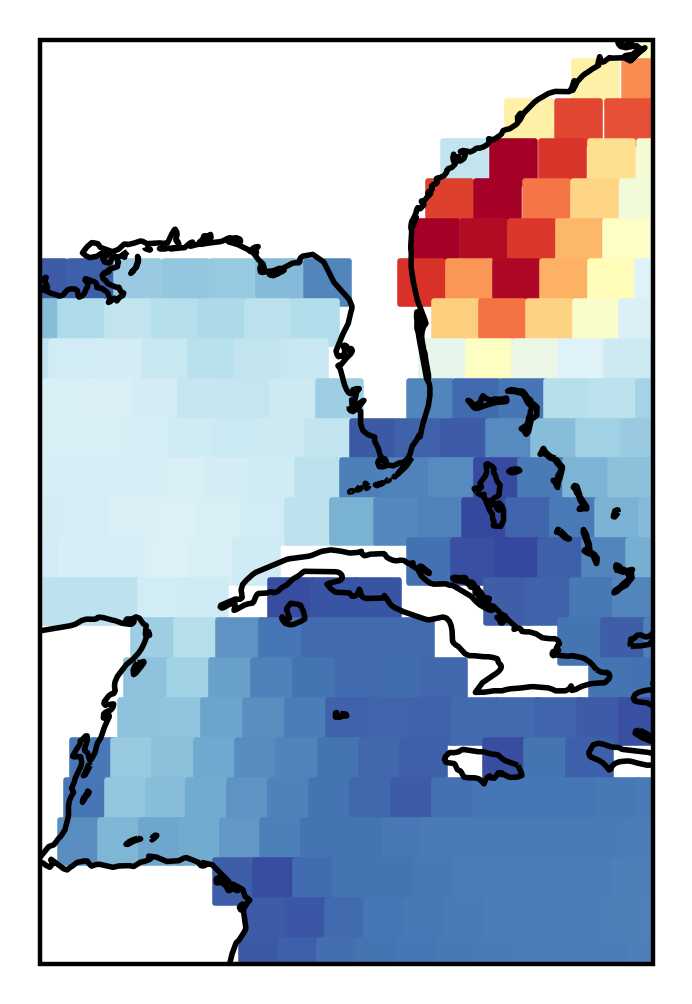}
    \includegraphics[height=.15\linewidth]{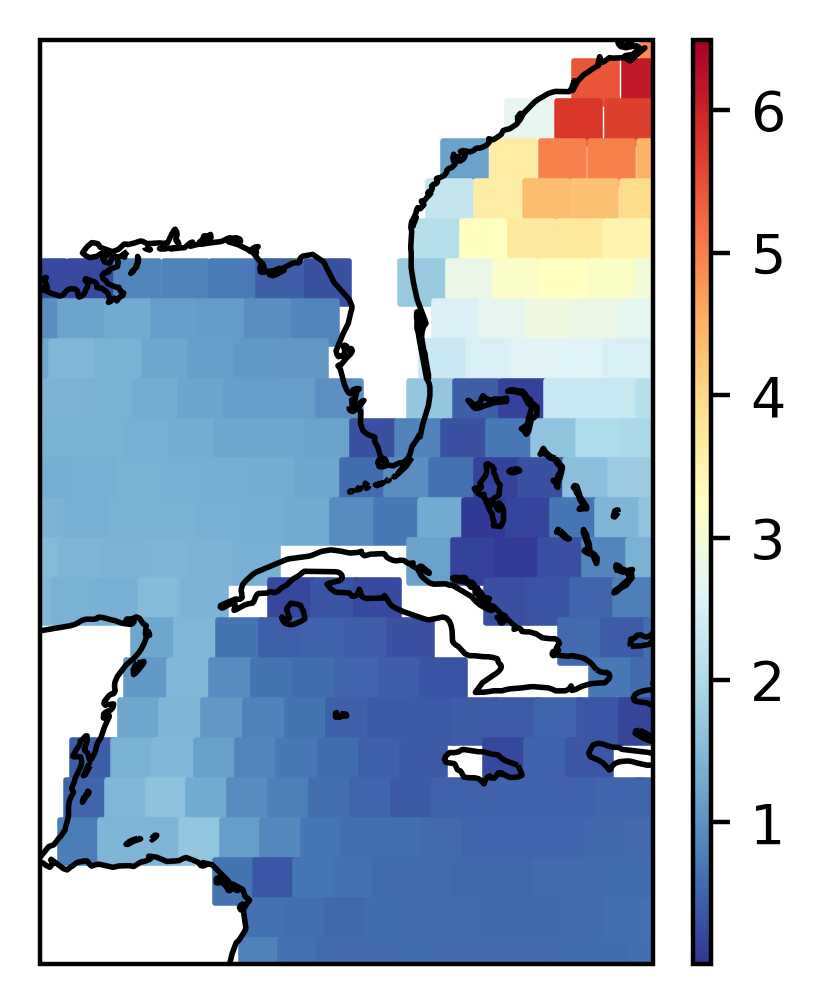}

    \vspace{0.5em}

    \includegraphics[trim=0 3 0 0, clip, height=.15\linewidth]{images/ian/dop.jpg}
    \includegraphics[height=.15\linewidth]{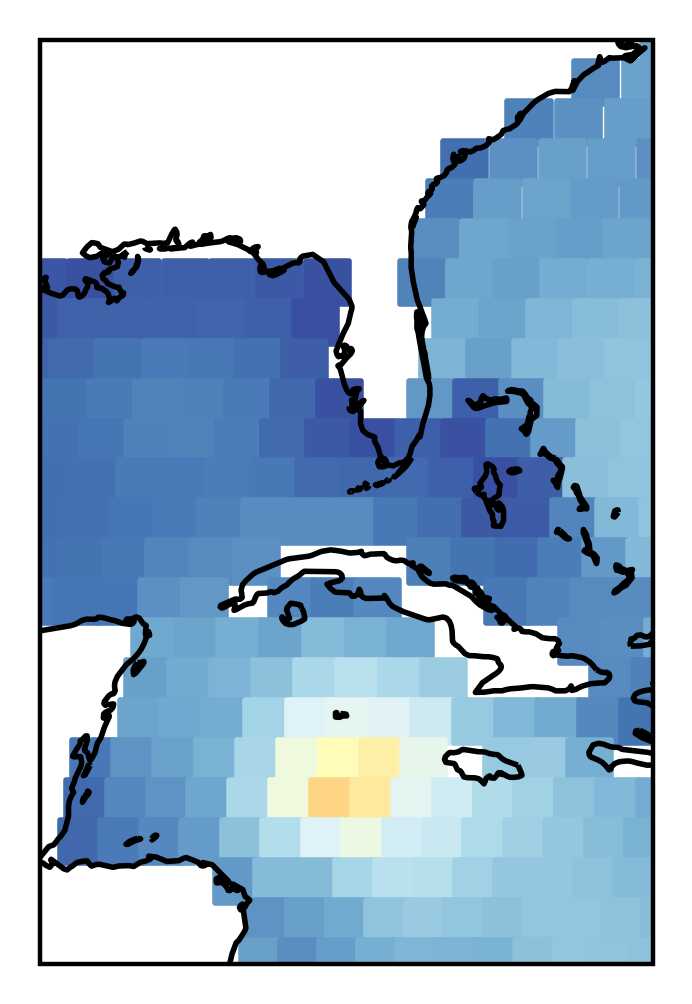}
    \includegraphics[height=.15\linewidth]{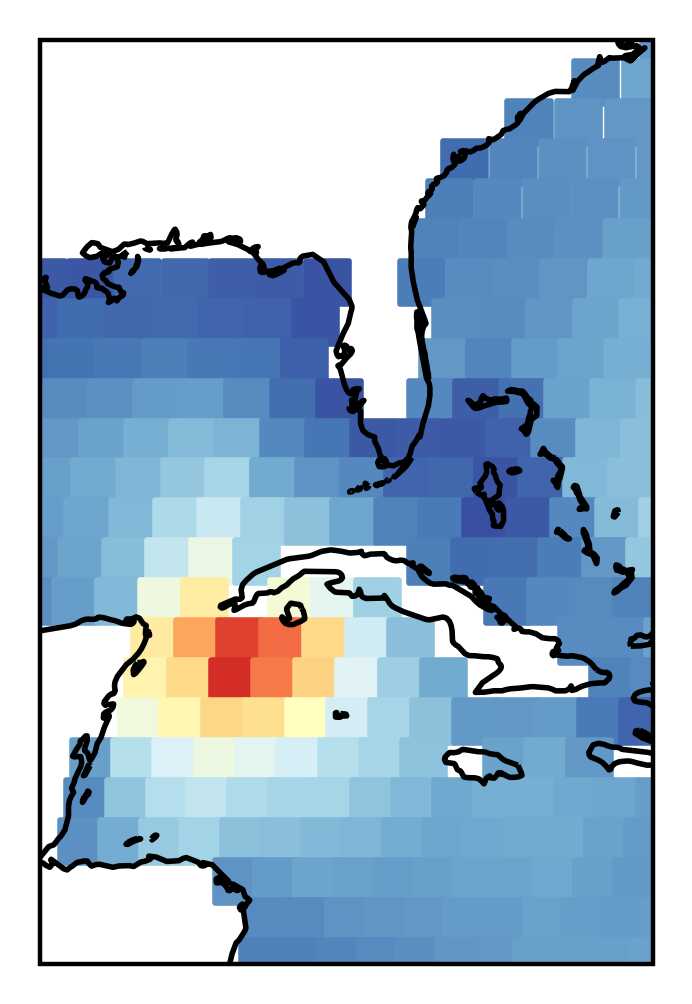}
    \includegraphics[height=.15\linewidth]{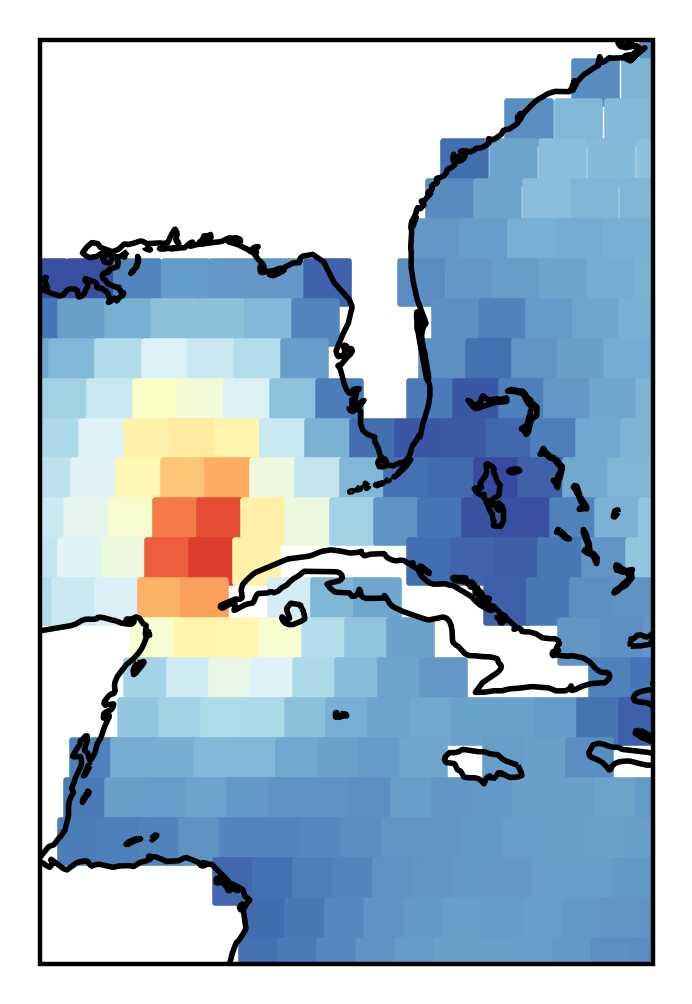}
    \includegraphics[height=.15\linewidth]{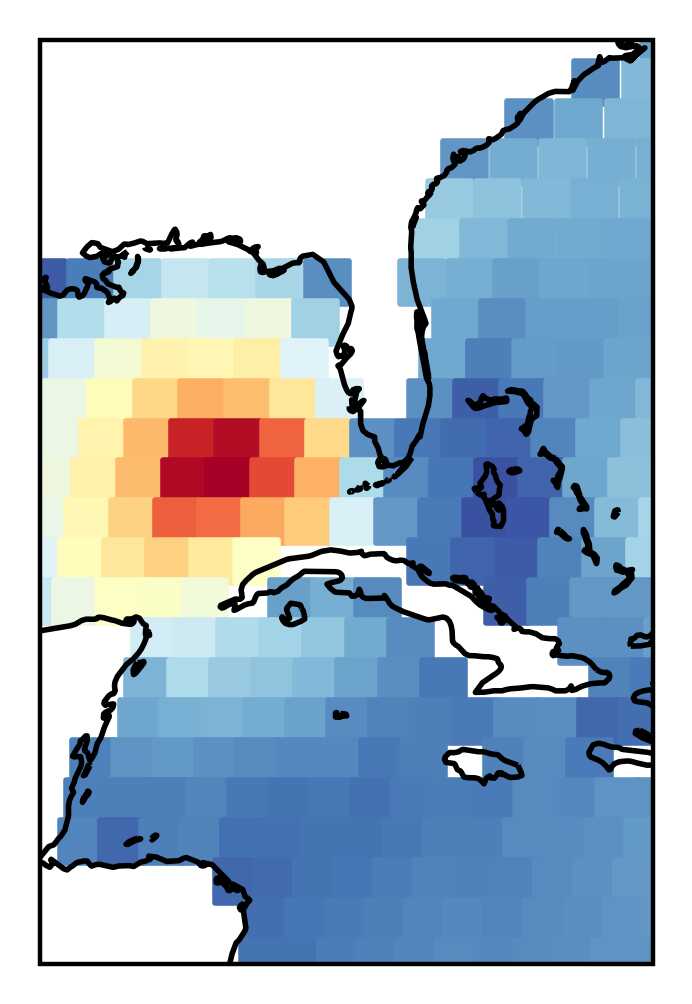}
    \includegraphics[height=.15\linewidth]{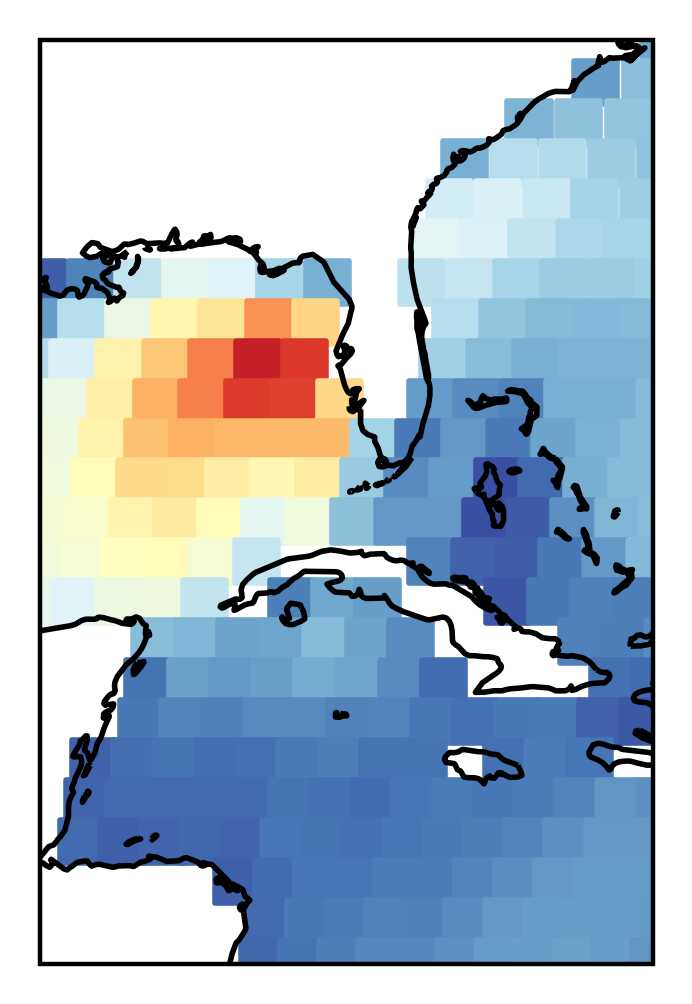}
    \includegraphics[height=.15\linewidth]{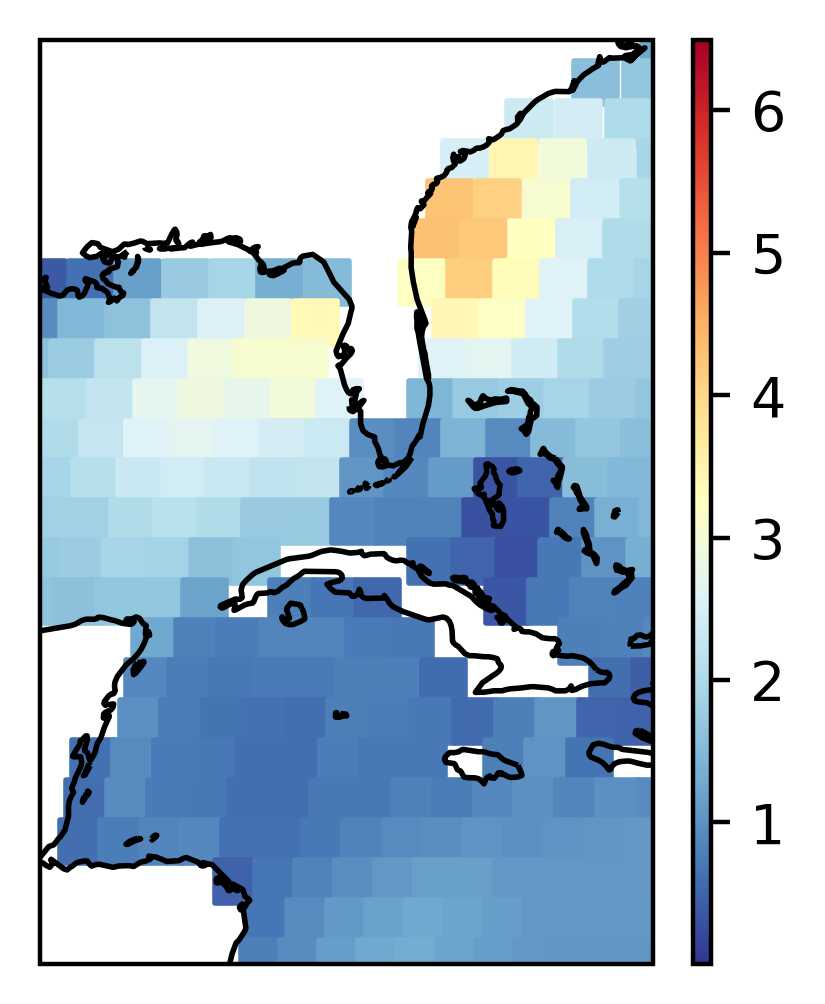}
    \end{minipage}
}

\caption{ERA5 reanalysis (top) and GraphDOP gridded forecasts (bottom) mean sea-level pressure (a), wind speed (b) and significant wave height (c) at 00z over the course of six days (Sept 26, 2022 to Oct 1, 2022).}
\label{fig:ian}

\end{figure}

This section examines the forecasts produced by GraphDOP for Hurricane Ian, a category-5 (major) hurricane that occurred in September 2022 and was one of the most devastating tropical cyclones to make landfall in the US state of Florida since Hurricane Michael in 2018 \citep{osti_10492534}. A GraphDOP forecast was initialised from 12-hours of observations on September 24, 2022, 09z - 21z. Figure \ref{fig:ian} shows the mean sea-level pressure (a), wind speed (b) and significant wave height (c) from the ERA5 reanalysis (top rows) and from the GraphDOP o96 gridded forecasts (bottom rows), at 00z over 6 days, from September 26, 2022 to October 1, 2022, in 24-hour steps.

From Figure \ref{fig:ian}(a), we can see that GraphDOP performs well in predicting the general trajectory and evolution of the hurricane's low-pressure centre, keeping in relatively close agreement with the ERA5 reanalysis for the first three days. The model also captures the deepening of the central pressure, a key indicator of cyclone intensification.

However, the model significantly underestimates the storm's forward speed. As a result, while the general forecast trajectory remains correct, Ian's forecasted progress is slower than that captured in the ERA5 reanalysis. This delay can also be seen when looking at Figure \ref{fig:ian}(b) and (c). In the GraphDOP forecast, Ian persists longer over western Florida. This extended fetch (i.e., the uninterrupted distance that wind travels over water) allows the hurricane to transfer more energy into the ocean and create larger waves. The resultant wave activity persists for a couple of days after the winds have subsided, which is physically sensible. The ERA5 reanalysis for days five and six shows stronger sustained winds over the Atlantic Ocean, to the east of Florida, whereas GraphDOP forecasts winds of a somewhat lower magnitude and for a shorter period. Consequently, GraphDOP does not forecast substantial wave activity east of Florida.

Figure \ref{fig:ian}(b) also shows that the GraphDOP network picks up on more subtle features such as the hurricane's eye (a circular zone of fair weather at the storm’s centre). This shows up as a blue pixel in the middle of the strong winds showing in red. The eye is particularly visible in the sixth day of the ERA5 reanalysis, but also almost every day for the GraphDOP forecast, correctly placed at the centre of the strong winds. Presumably the dropsondes available in the training dataset for other tropical cyclones would have targetted the eyes and the eye walls quite densely, allowing the network to pick up on the existence of such features. 

The errors in propagation speed notwithstanding, it is particularly encouraging to see that the wave, wind speed and mean sea level pressure forecasts calculated by GraphDOP for hurricane Ian appear to be largely consistent and physically sensible.

\section{Discussion and outlook}
\label{ref:section-discussion-outlook}

The results presented in the previous section have demonstrated that GraphDOP is able to combine information from diverse observation types into a coherent internal representation of the Earth System state. The relationships that the network learns between different observed variables have been shown to generalise to areas where no observations exist - for example, forecasts of upper-level winds compare well with ERA5 even in areas where there is no radiosonde or aircraft coverage. Short-range forecasts for various geophysical variables compare closely with an independent ERA5 reference that was not used in the training process. While the GraphDOP medium-range forecast skill is not yet close to matching state-of-the-art NWP performance, the fact that the forecasts are skilful out to day 5 provides a promising indication that it might be possible for neural networks to learn Earth System dynamics from level-1 and level-2 Earth System observations alone.

We believe multiple pathways exist for improvement. First, refinements to the training approach, particularly in the weighting of observations, warrant exploration. Many satellite sounding channels used in this study peak in the stratosphere and above, resulting in relatively little weight given to crucial tropospheric channels. The loss function could be modified to increase the weight of channels sensitive to tropospheric temperature. A careful assignment of observation errors is crucial in traditional 4D-Var \citep{weston2014obserrors}, and established error covariance models used in physics-based systems may serve as guidance for improving GraphDOP. For cloud-sensitive channels, error models could be implemented to give more weight to clear-sky brightness temperatures, preventing the loss from being dominated by cloud signals.

Architectural refinements may lead to further improvements. A key limitation in these experiments stems from the use of a 12-hour input observation window. Experience with physics-based models indicates that more information about the recent evolution of the atmosphere is required to achieve good skill scores at the medium range. While GraphDOP can support longer windows (cf. Figure \ref{fig:dop-model-schematic}), in this study we chose to use an input window aligned with ECMWF's 12-hour assimilation window. Traditional data assimilation systems carry forward information from earlier windows via a background state. Addressing this through longer input windows or introducing an observation-based prior through a recurrent architecture could substantially improve performance. Sensitivity tests have revealed that using a shorter decoder output interval in conjunction with autoregressive time-stepping in the latent space (cf. Figure \ref{fig:dop-model-schematic}) can lead to sharper predictions. For example, Figure \ref{fig:SEVIRI_72h} in the Appendix shows predicted brightness temperatures from a network trained using a 3-hour decoder output interval. This suggests that significant improvements are possible through architectural modifications.

Increasing the network size offers another pathway to improved performance. Experimentation has demonstrated that forecast skill improves with increased parameter count (not shown). Further increases in size would likely enable more sophisticated representations of atmospheric dynamics.

The training dataset will also be expanded both temporally and in terms of instrument diversity. While most reanalysis data-driven models train on 40 years of data, this study used only 18 years. Additionally, incorporating more instruments to increase data diversity could enable the model to learn richer representations of atmospheric state and dynamics.

Finally, the impact of data quality merits further investigation. Despite attempts to remove low-quality observations in this study, many remained in the dataset. For instance, several degraded AMSU-A channels were included from some satellites in the training dataset. The impact of more rigorous quality control on model performance remains to be determined.

One obvious limitation of using observations as targets in the training is that predictions can only be made for physical variables for which we have direct observations. For example, although the results in this paper showed accurate predictions of sea ice in radiance space, it is not possible to make predictions for more abstract quantities, such as sea ice concentration, without having direct observations of that variable.

Ongoing work focuses on evaluating AI-DOP predictions from a coupled perspective. The aim here is to assess the extent to which the current graph and transformer-based AI-DOP models are able to leverage interface observations and correctly capture the interactions between different components of the Earth System. Another research direction currently being investigated is probabilistic forecasting, through diffusion-based training \citep{karras2022elucidating,alexe2024ecmwfnl} and proper score optimization \citep{lang2024ensscore}. It is expected that probabilistic training will produce considerably sharper meteorological features in the GraphDOP forecasts.

Furthermore, the AI-DOP models can provide insight into the information content of observations and their contribution to medium-range forecast skill using, e.g., adjoint sensitivities that can be calculated relatively easily in a data-driven, fully differentiable system. Finally, we are already exploring the construction of hybrid end-to-end systems, where encoder / decoder elements of GraphDOP are employed to augment the AIFS with direct observation information to further improve skill (e.g. of surface parameter forecasts). 

While the results presented herein give good cause for optimism, it remains to be determined whether an end-to-end data-driven system trained and initialized exclusively from observations can compete at the medium range with a state-of-the-art physics-based system such as the IFS. AI-DOP remains a very active and exciting research direction at ECMWF.

\subsubsection*{Acknowledgments}
We thank Sarah Keeley for discussions and advice on sea ice, Joshua Kousal for feedback on the Hurricane Ian use case, Hans Hersbach for guidance with the ERA5 climatology, Alan Geer for advice on the observation-space evaluation methodology, Christian Lessig, Peter D\"uben, Patricia de Rosnay, James Steer, Florence Rabier, Andrew Brown and many other colleagues at ECMWF for engaging discussions about AI-DOP and feedback on the present manuscript. We acknowledge the Partnership for Advanced Computing in Europe (PRACE) for awarding us access to Leonardo, CINECA, Italy, and the EuroHPC Joint Undertaking (JU) for awarding us project access to the EuroHPC supercomputer MareNostrum5, hosted by the Barcelona Supercomputing Centre, through a EuroHPC JU Special Access call. 

\section*{Appendix}

Table \ref{table:dop-observations} lists the parameters (e.g., satellite channels or physical variables measured) for each observing instrument used in GraphDOP. Instruments are grouped by category.

\begin{figure}[!ht]
\centering
\includegraphics[width = \linewidth]{images/gridded_forecasts/legend.jpg}
\subfigure[Sea surface temperature]{\includegraphics[trim=0 0 55 0, clip, height=.16\linewidth]{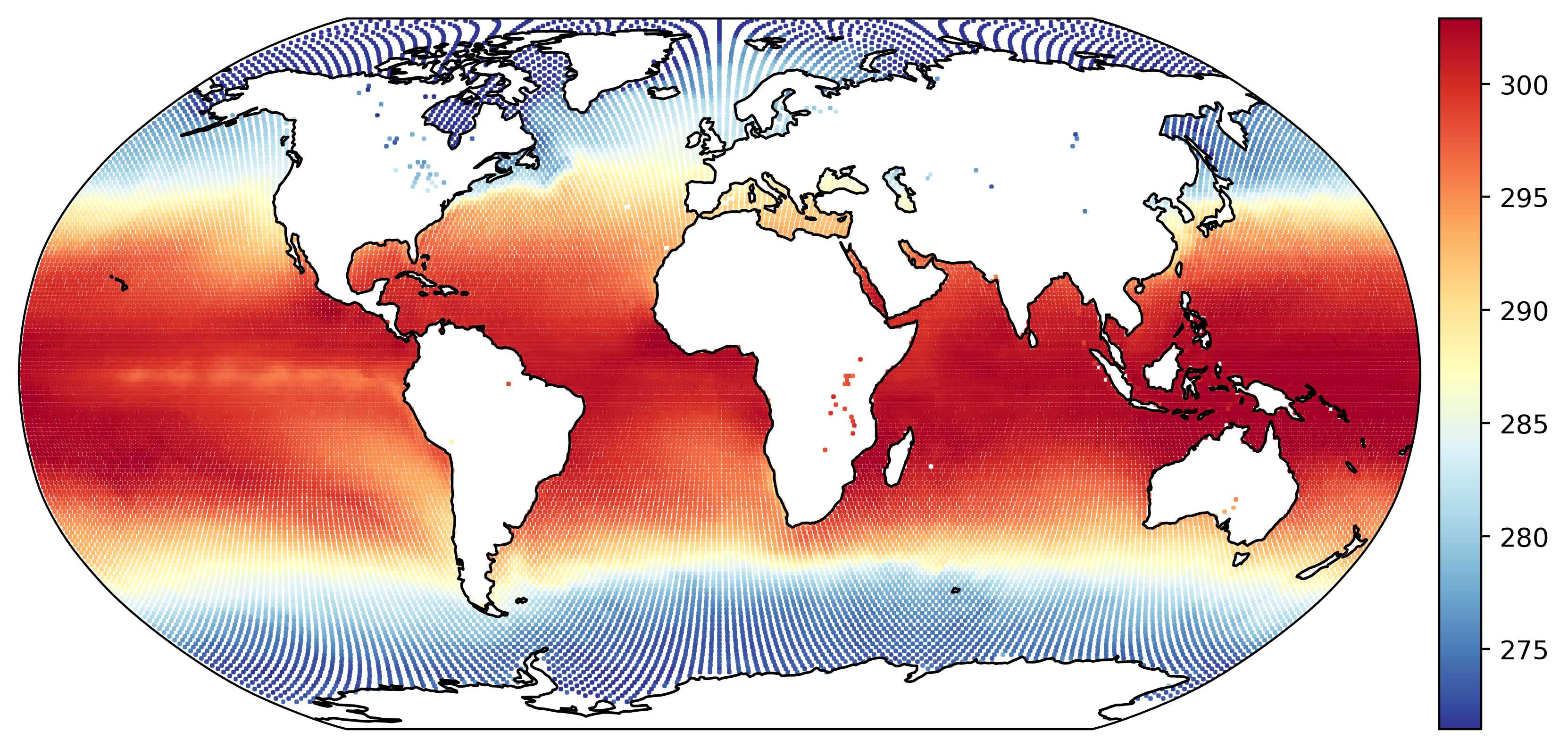} \includegraphics[height=.16\linewidth]{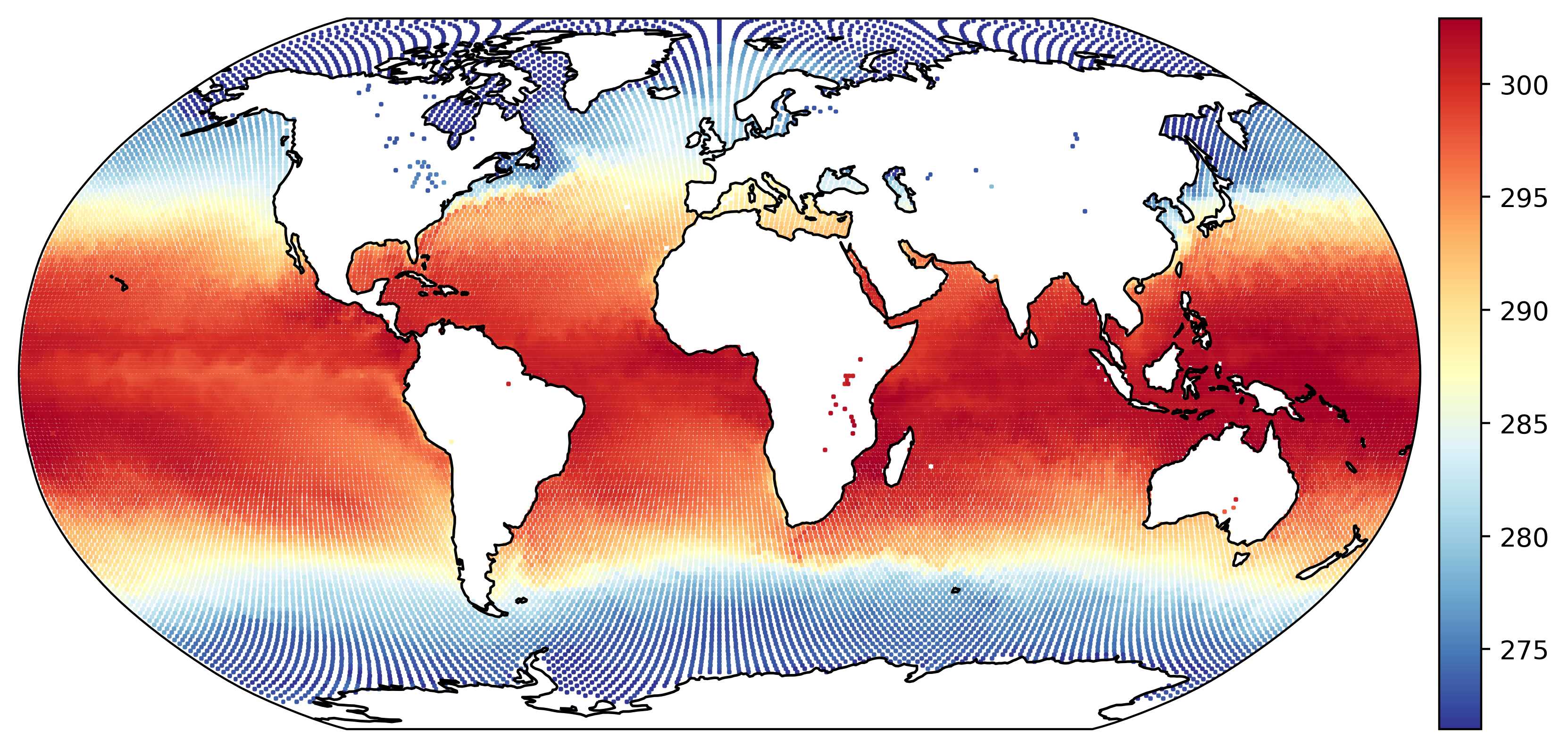}\includegraphics[height=.16\linewidth]{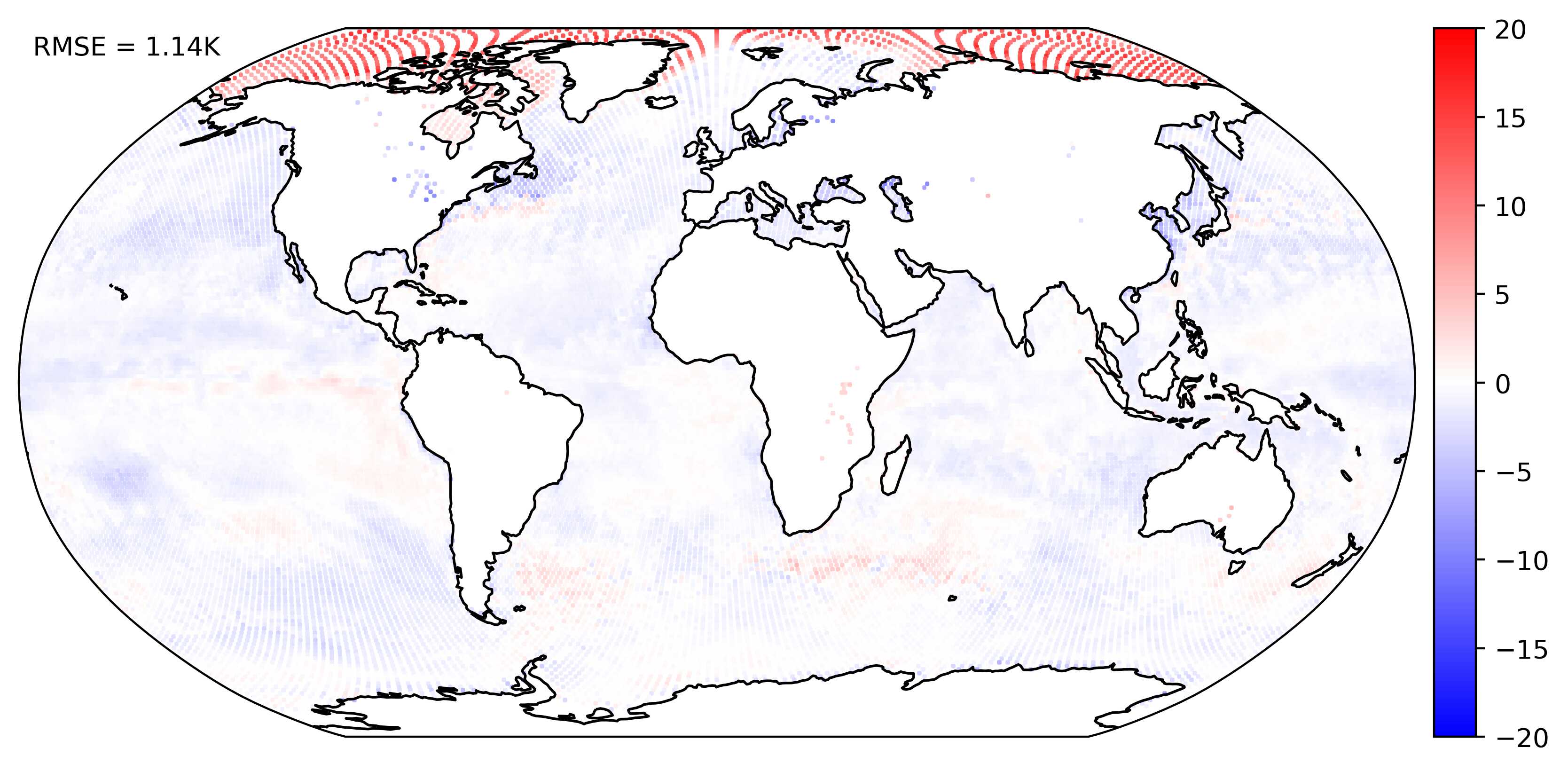}} \\
\subfigure[2-meter temperature]{\includegraphics[trim=0 0 55 0, clip, height=.16\linewidth]{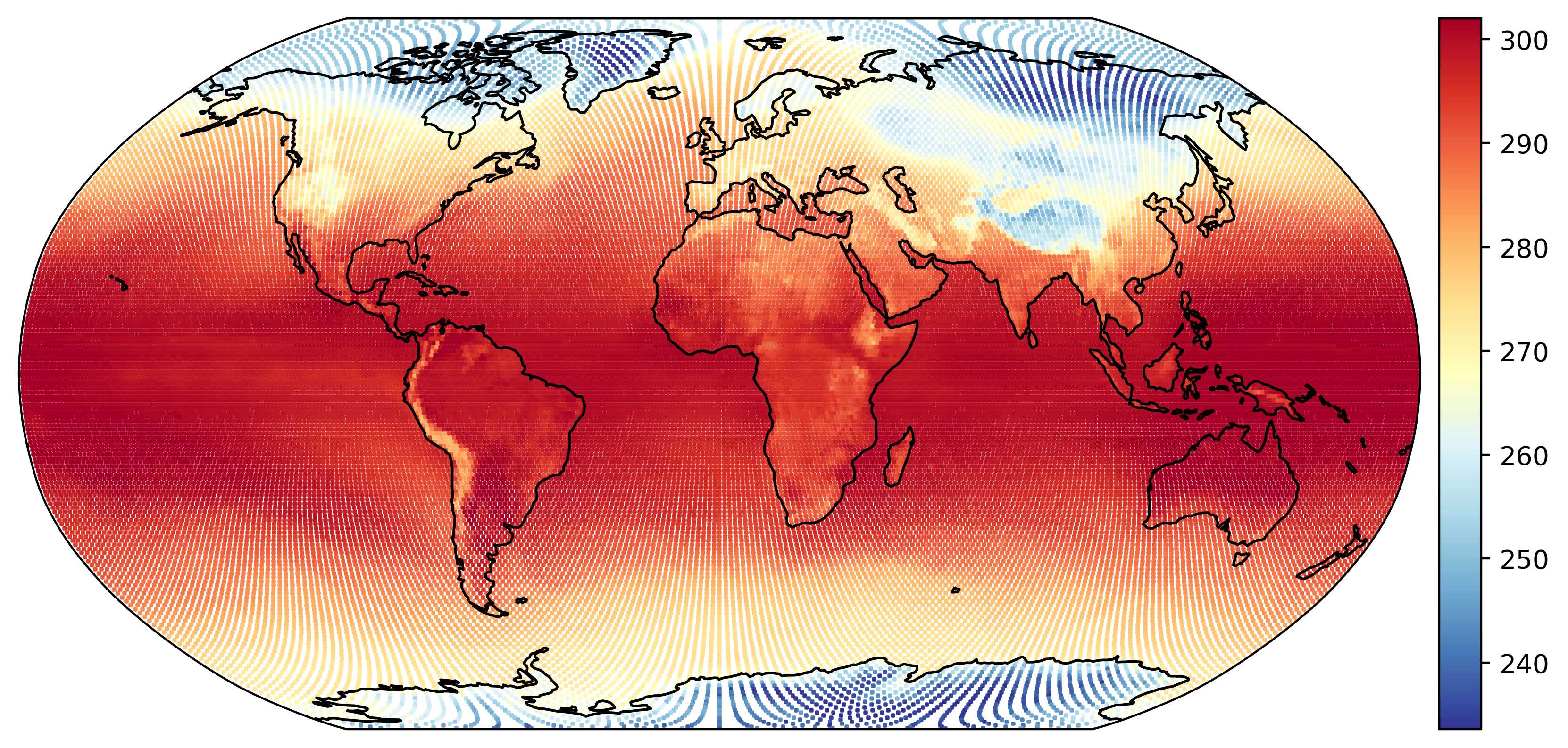} \includegraphics[height=.16\linewidth]{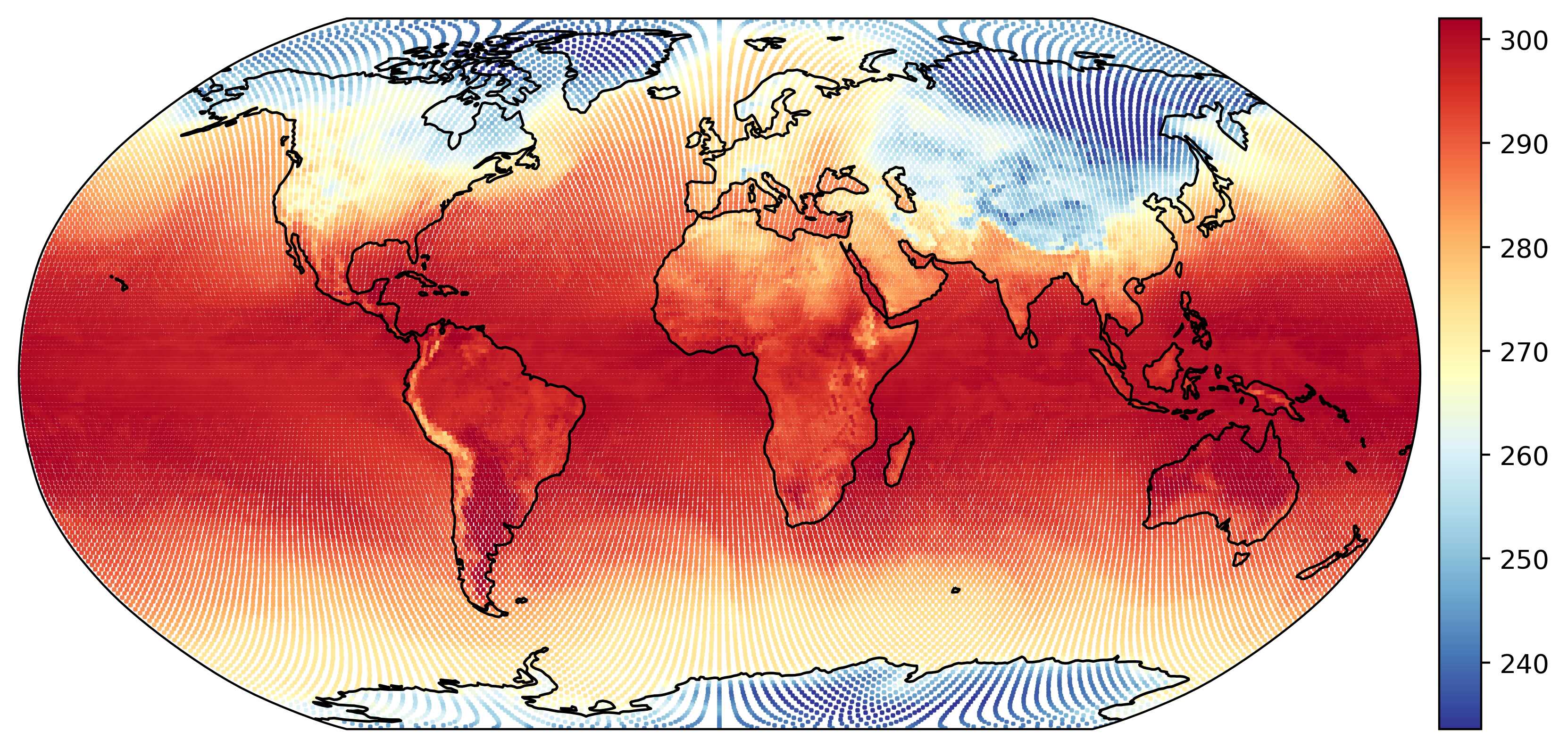}\includegraphics[height=.16\linewidth]{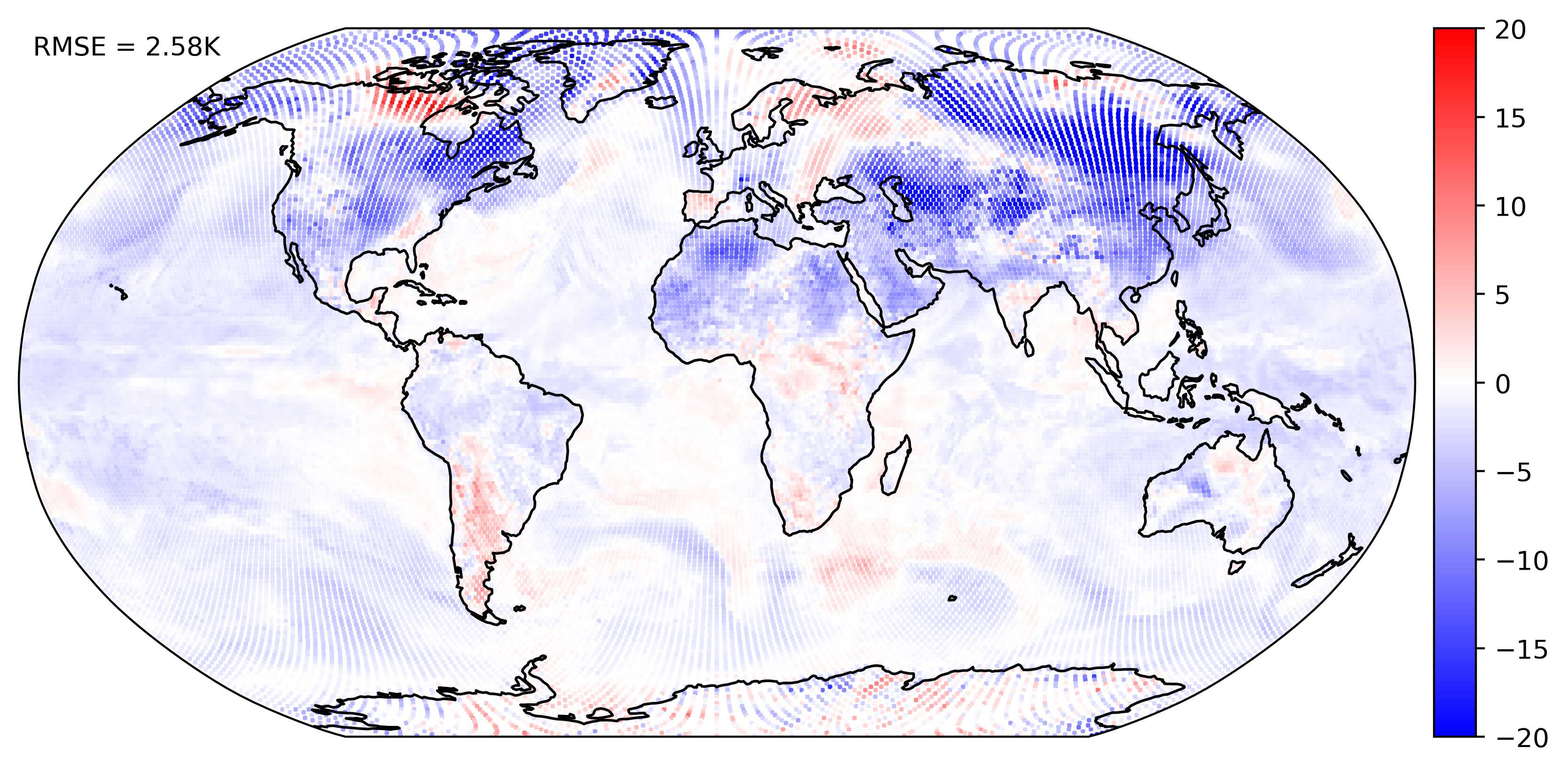}}\\
\subfigure[10-meter wind speed]{\includegraphics[trim=0 0 55 0, clip, height=.16\linewidth]{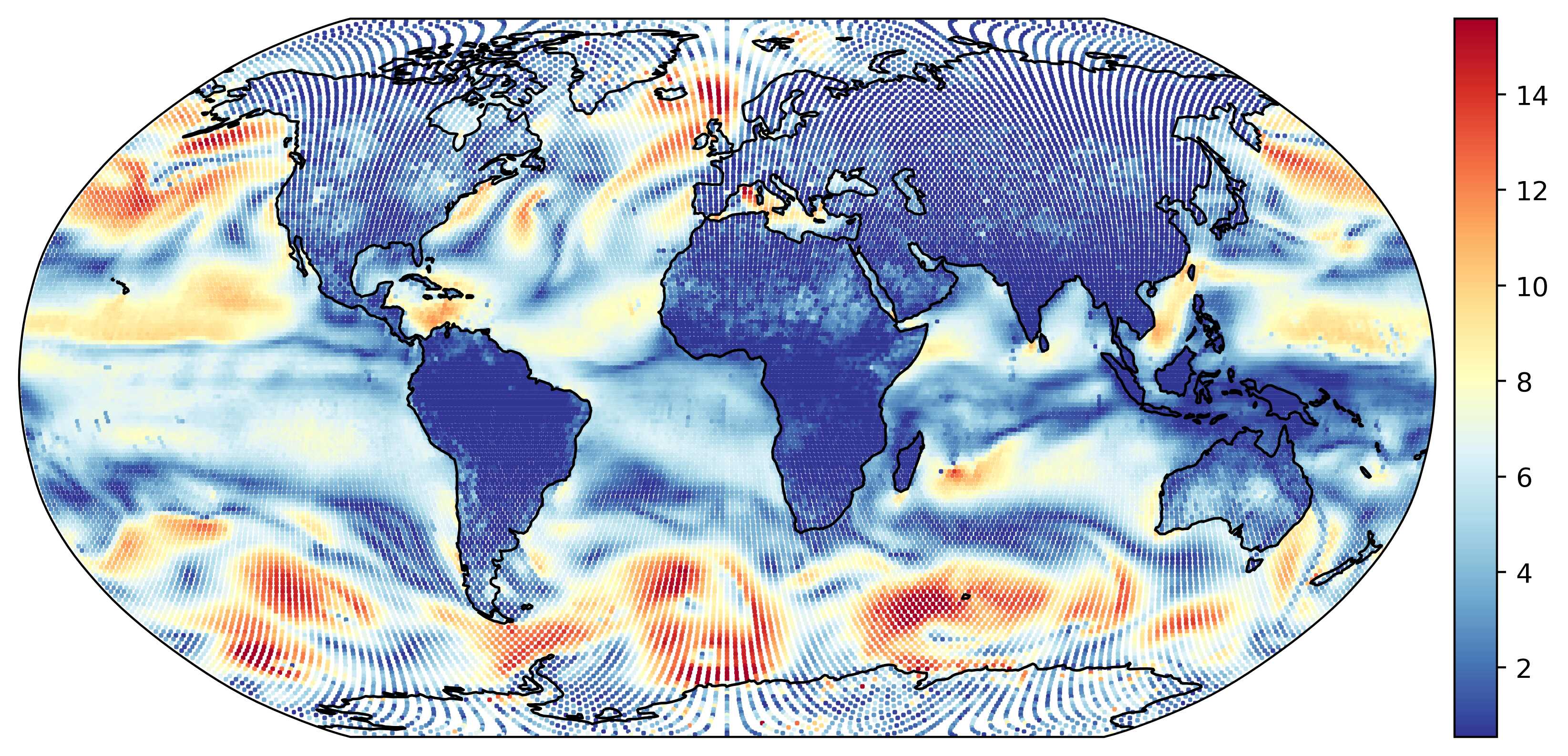}
\includegraphics[height=.16\linewidth]{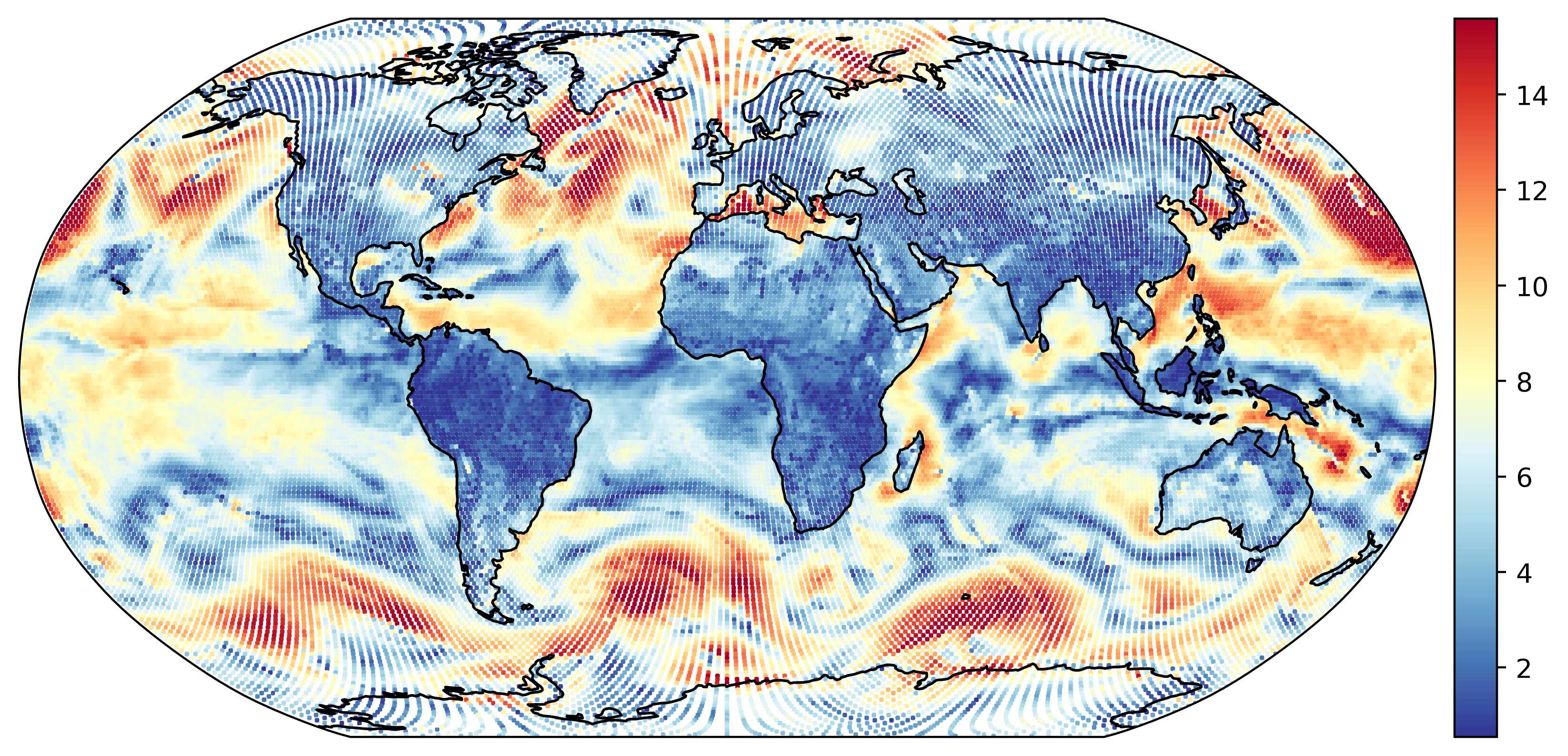}\includegraphics[height=.16\linewidth]{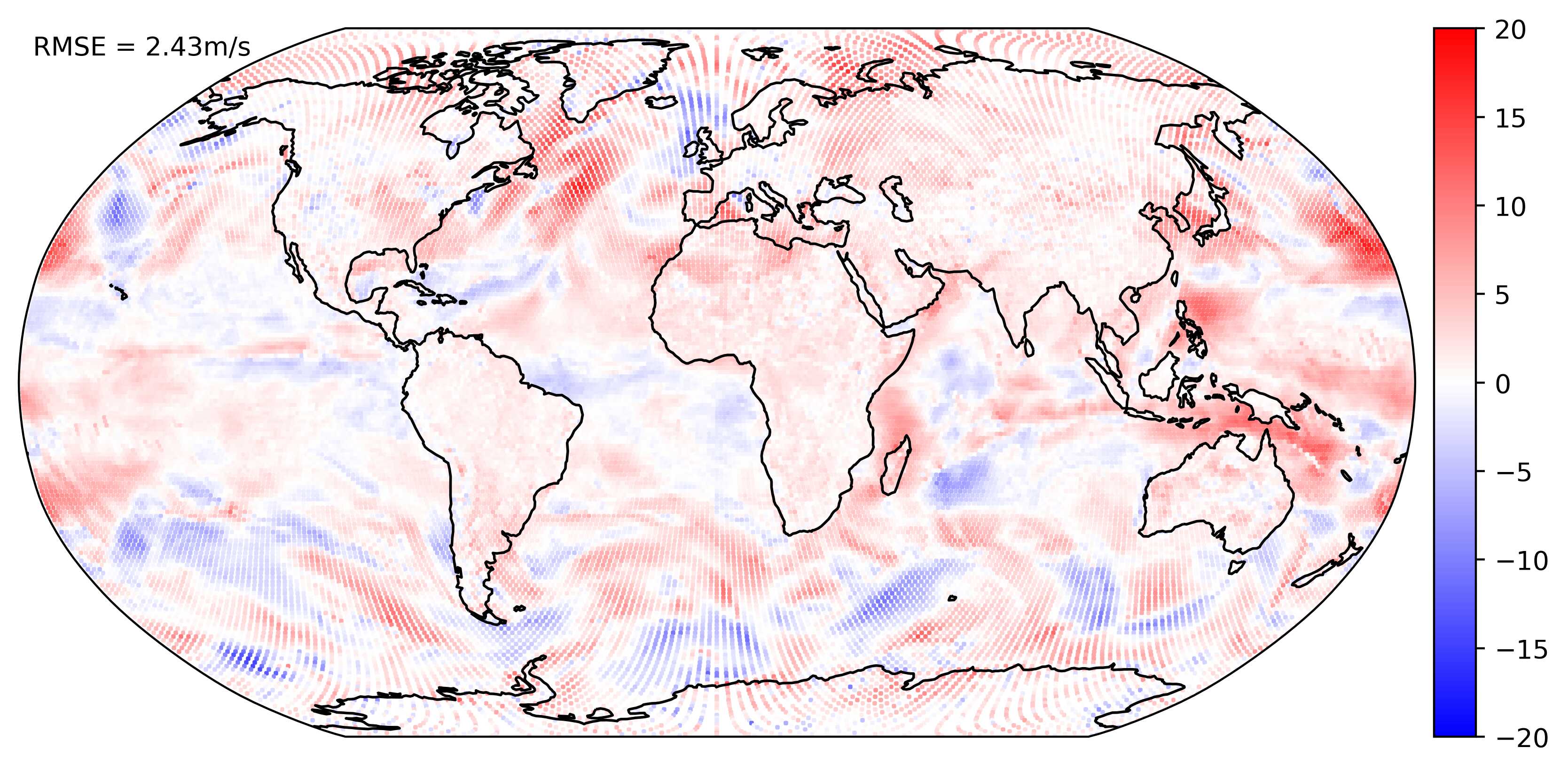}}\\
 \subfigure[Wind speed at 200hPa]{\includegraphics[trim=0 0 55 0, clip, height=.16\linewidth]{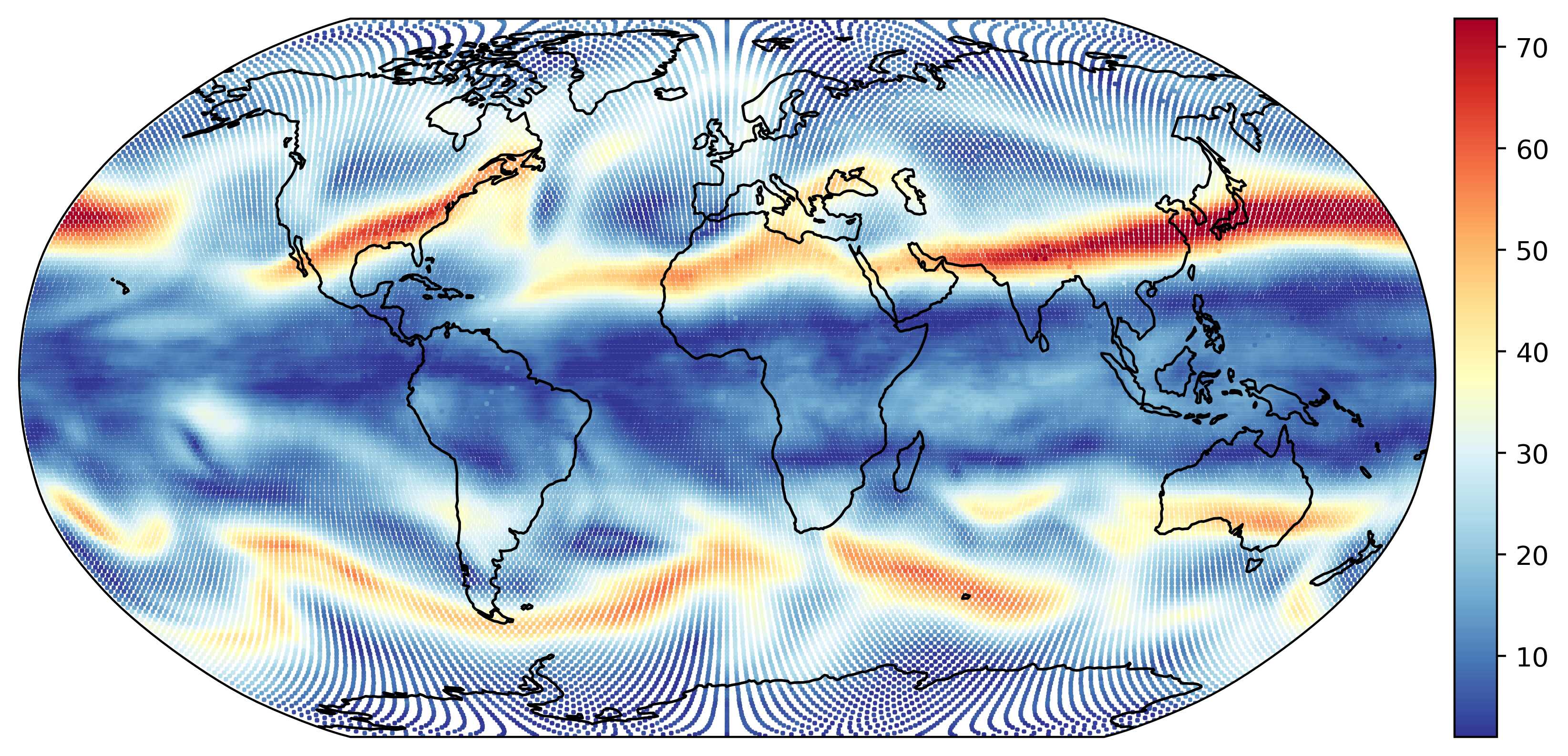}
 \includegraphics[height=.16\linewidth]{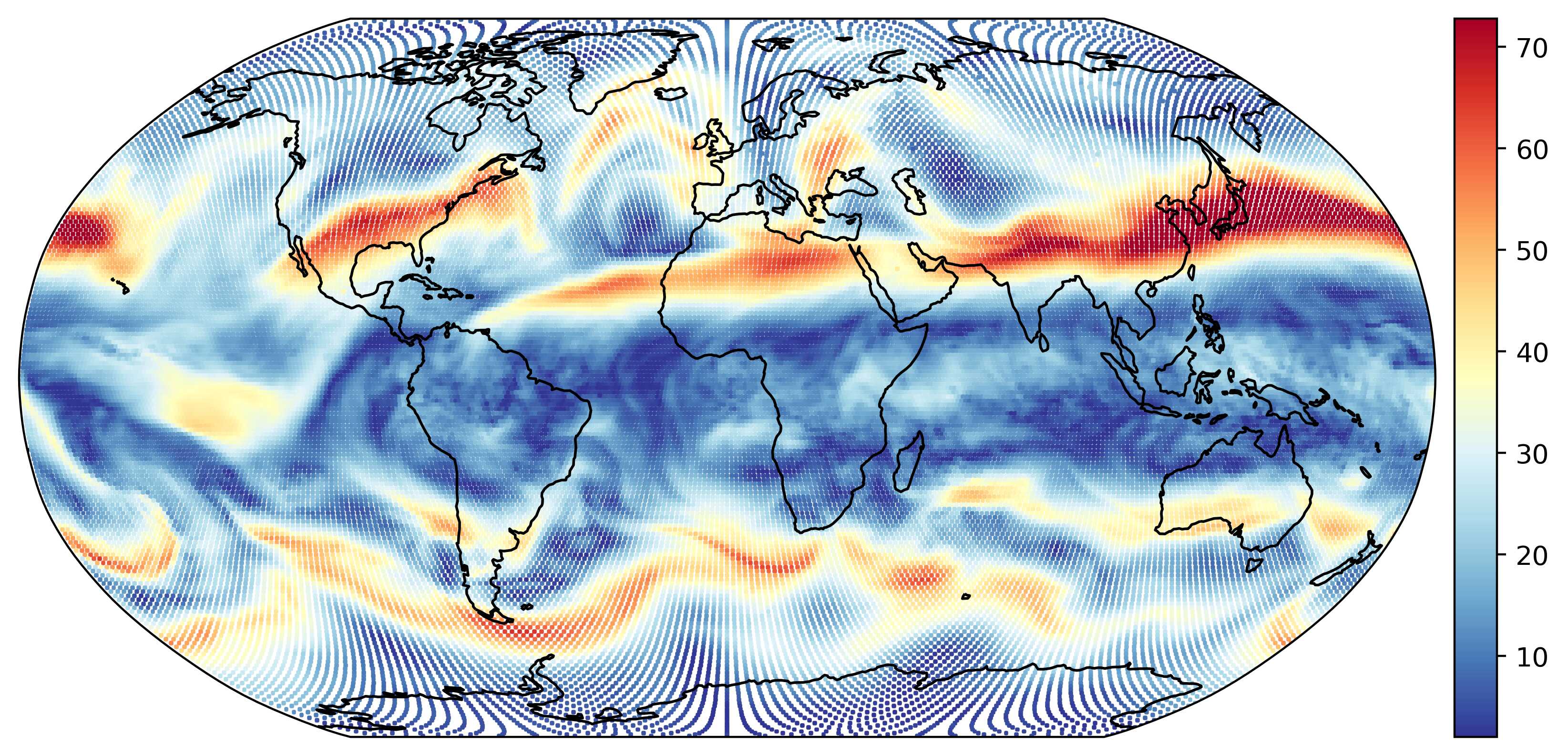}\includegraphics[height=.16\linewidth]{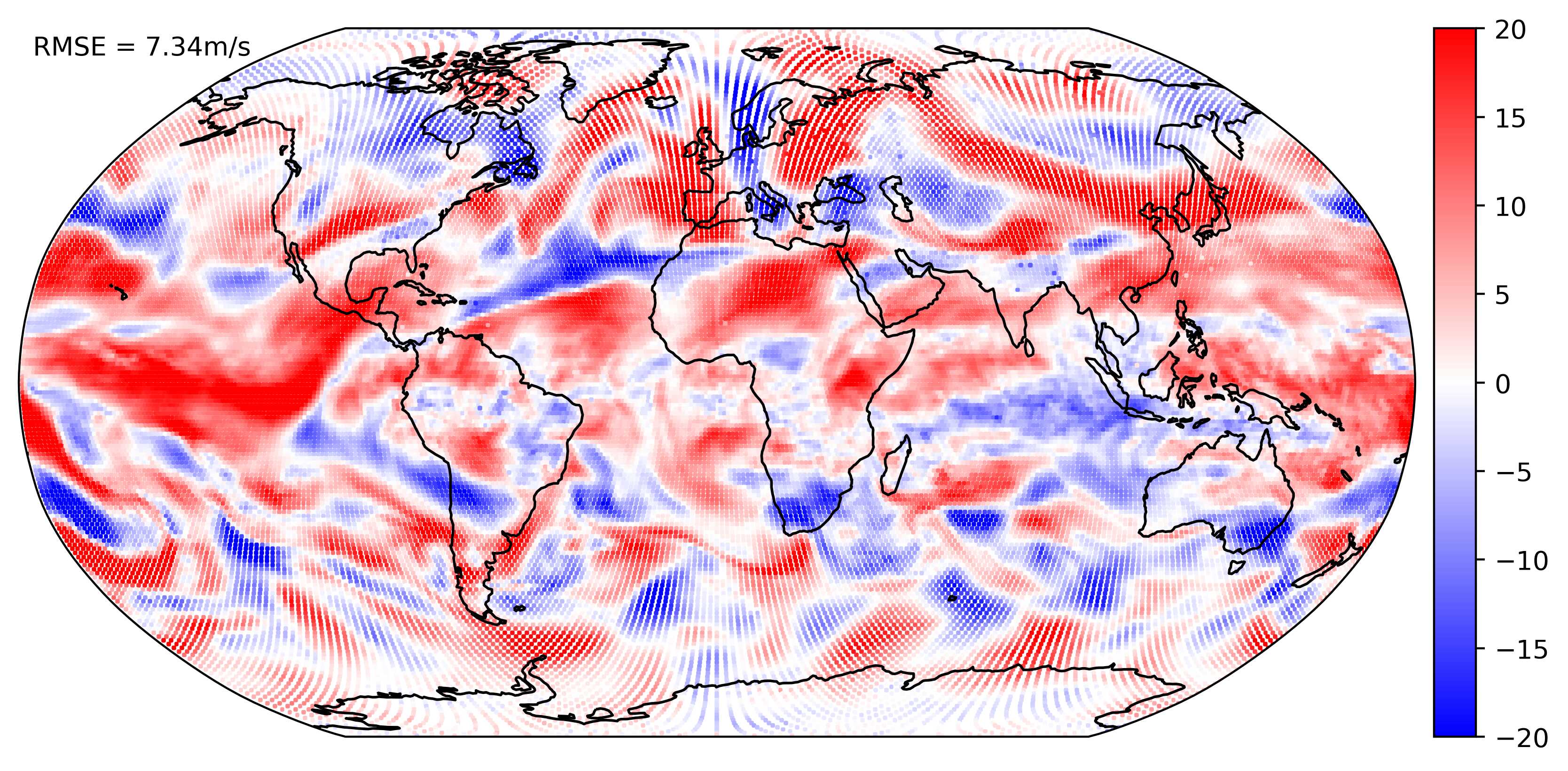}}\\
 \subfigure[Temperature at 850hPa]{\includegraphics[trim=0 0 55 0, clip, height=.16\linewidth]{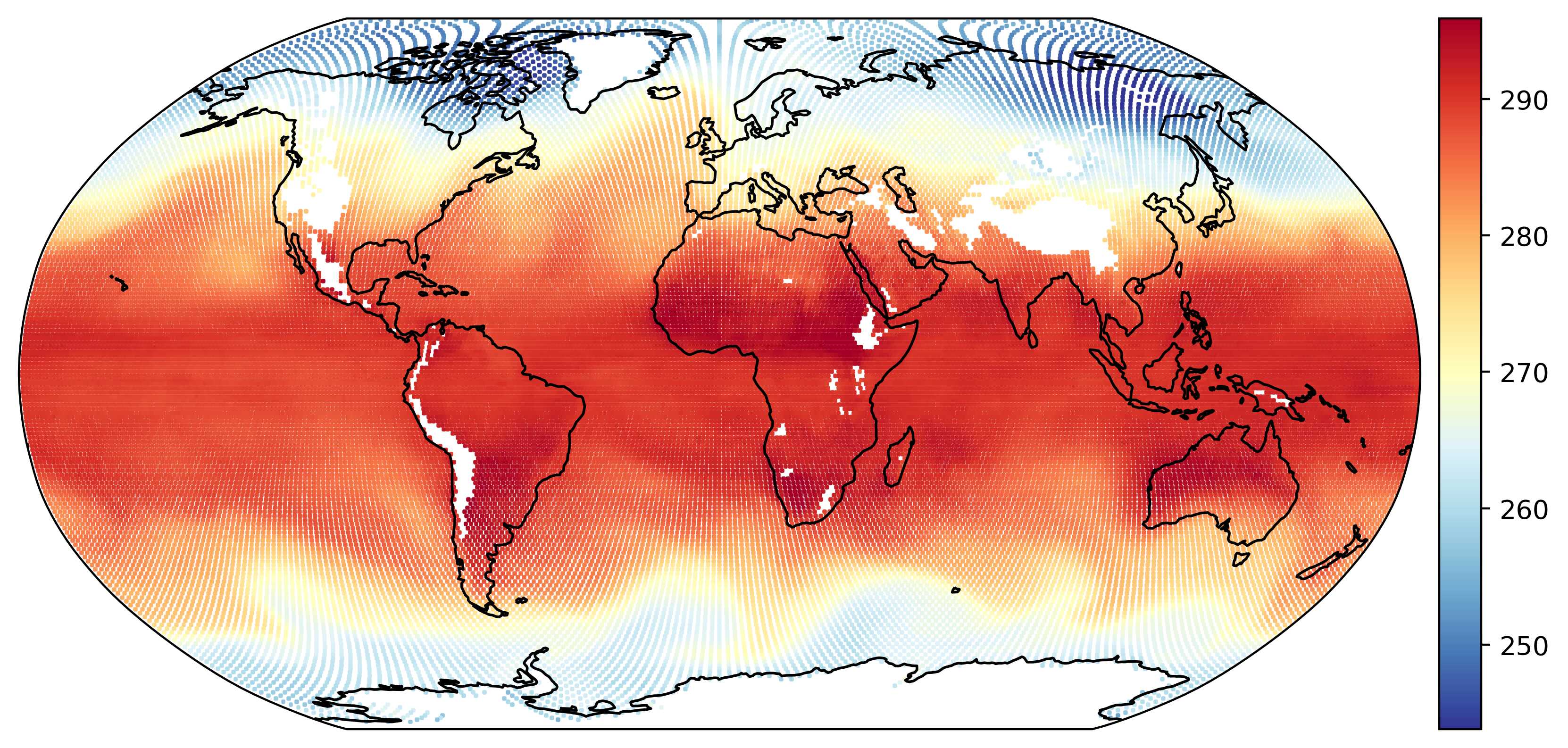} \includegraphics[height=.16\linewidth]{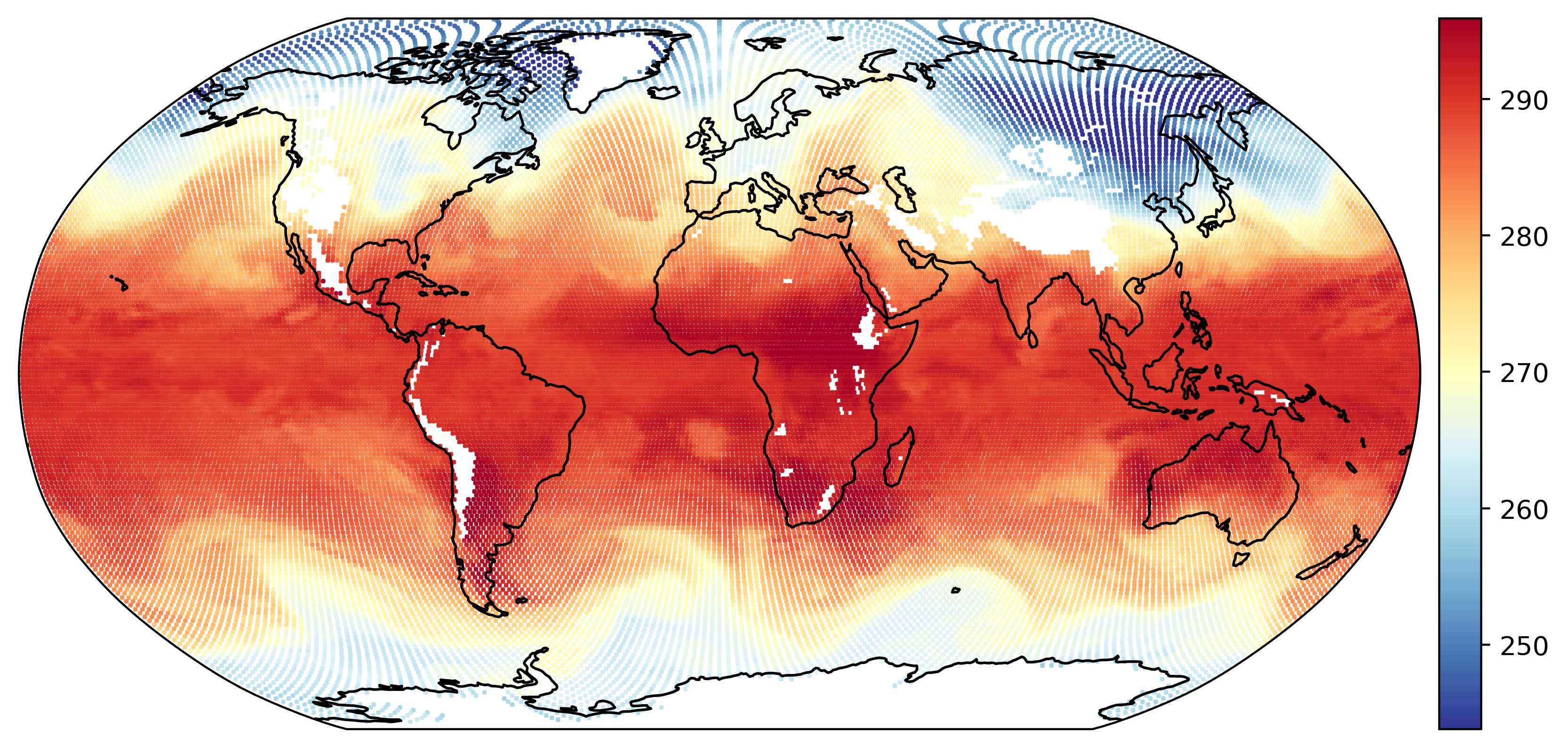}\includegraphics[height=.16\linewidth]{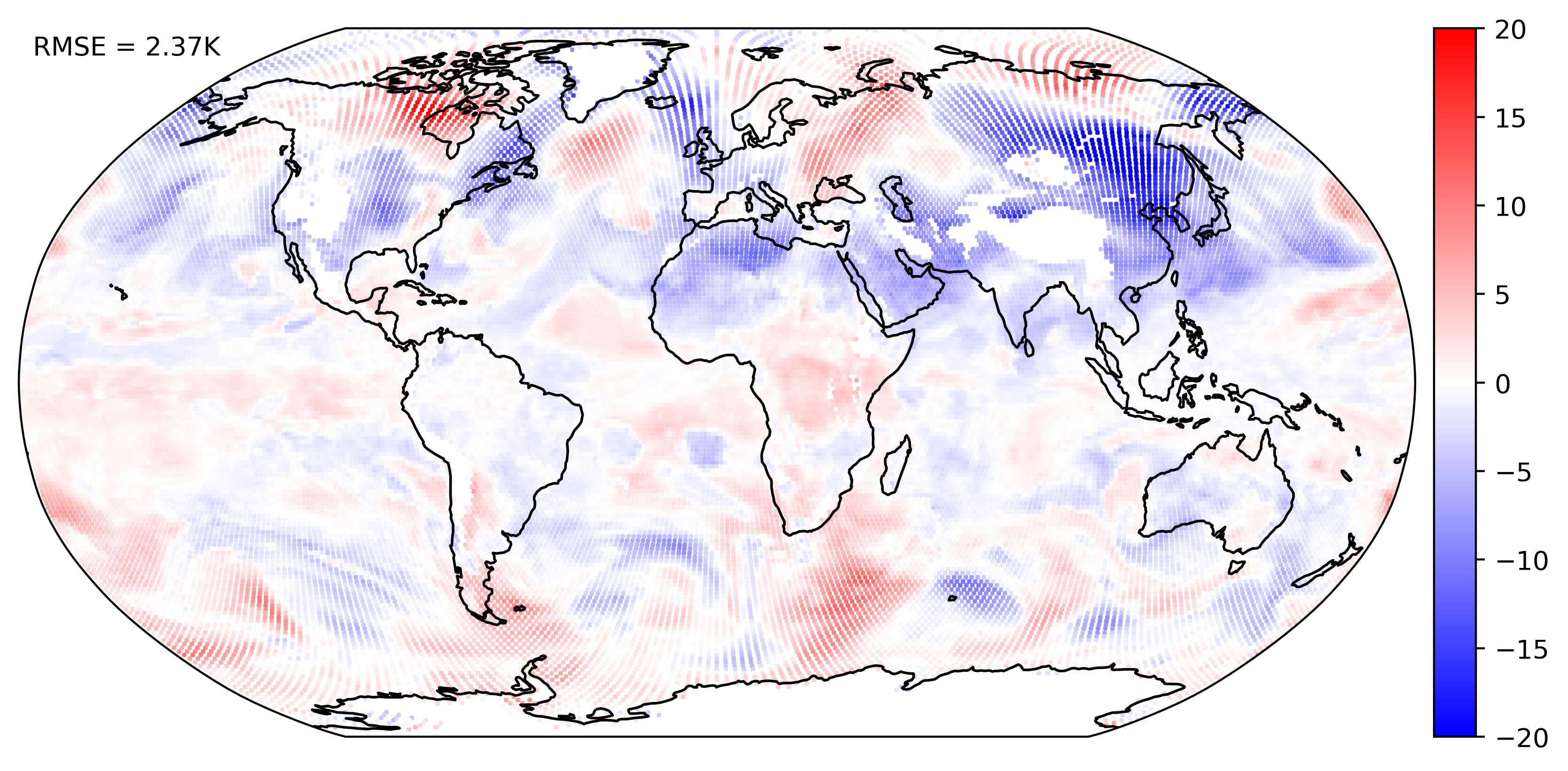}}
 \caption{Gridded five-day forecasts (Jan 20, 2023, 12z; right) compared to the ERA5 reanalysis field (middle).}
 \label{fig:gridded_examples_day5}
 \end{figure}

Figure \ref{fig:gridded_examples_day5} shows day-five gridded forecasts of sea surface temperature (SST), 2-meter temperature, 10-meter wind speed, wind speed at 200~hPa and temperature at 850~hPa. 

\begin{figure}[!ht]
\centering
\includegraphics[width = .9\linewidth]{images/seaice/legend.jpg}
\subfigure[t+10h (Jan 2, 2023, 09:45z) 45 minutes after the end of the input window]{\includegraphics[trim=0 0 45 0, clip, height=.3\linewidth]{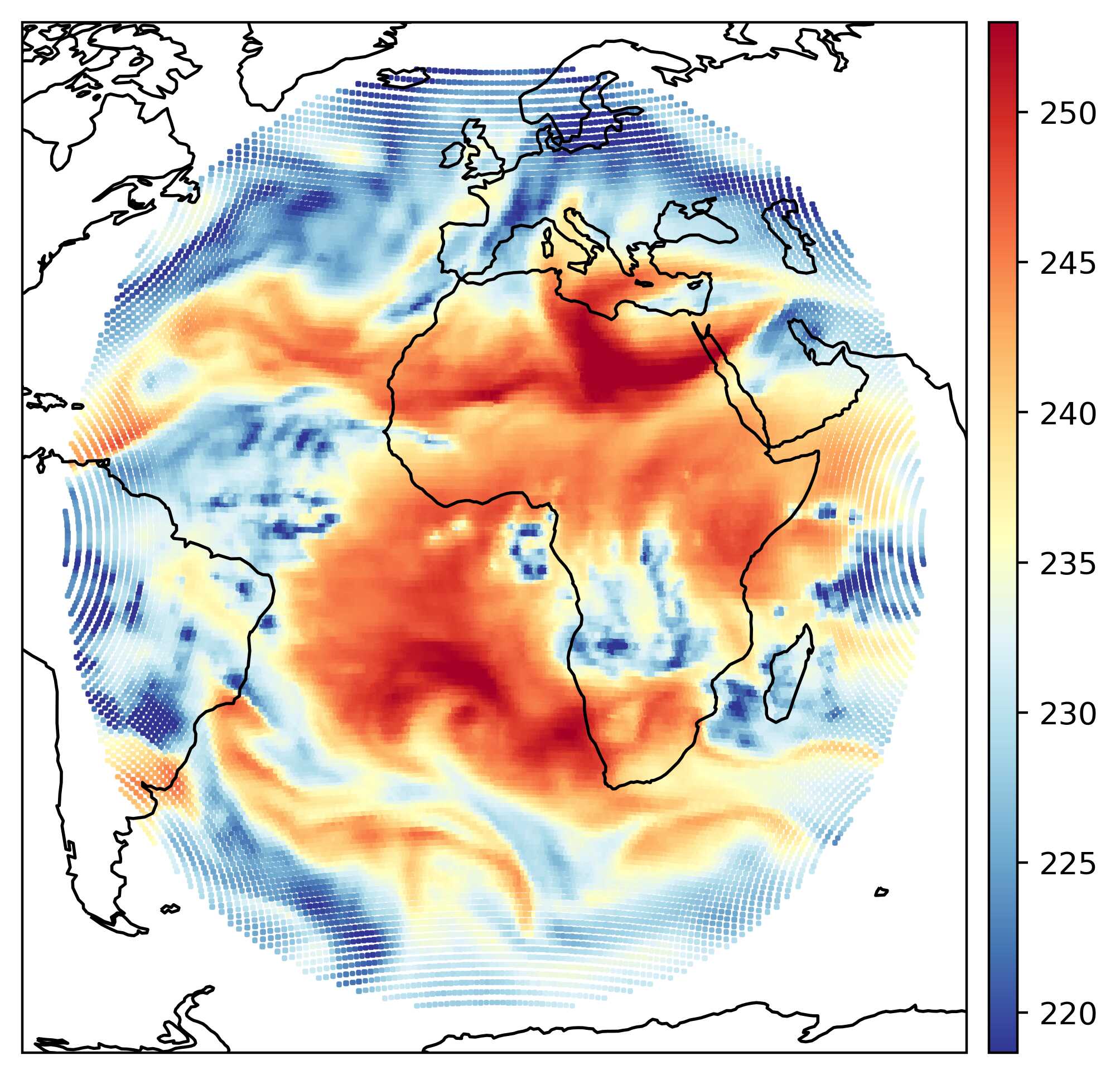}\includegraphics[height=.3\linewidth]{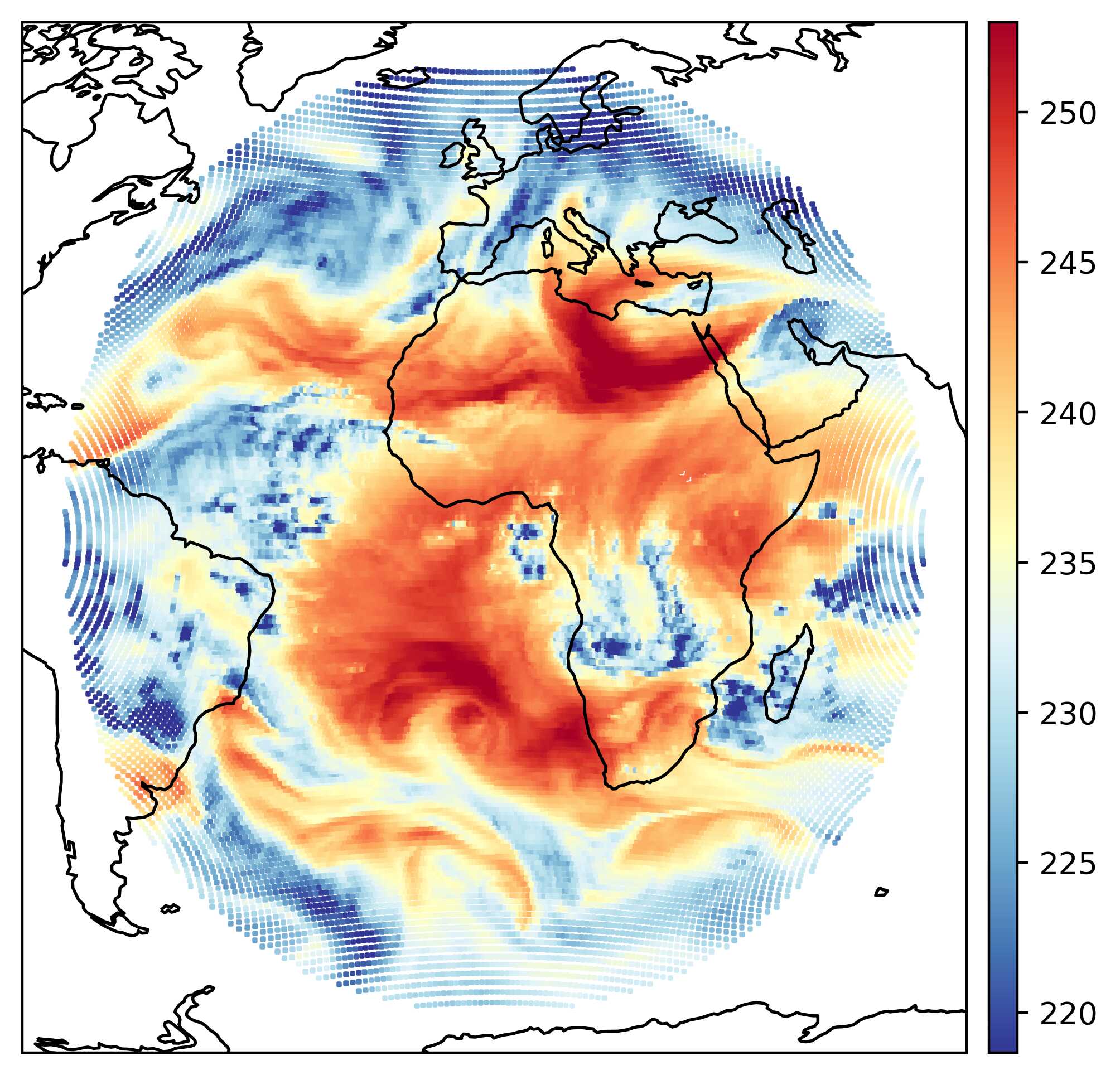}\includegraphics[height=.305\linewidth]{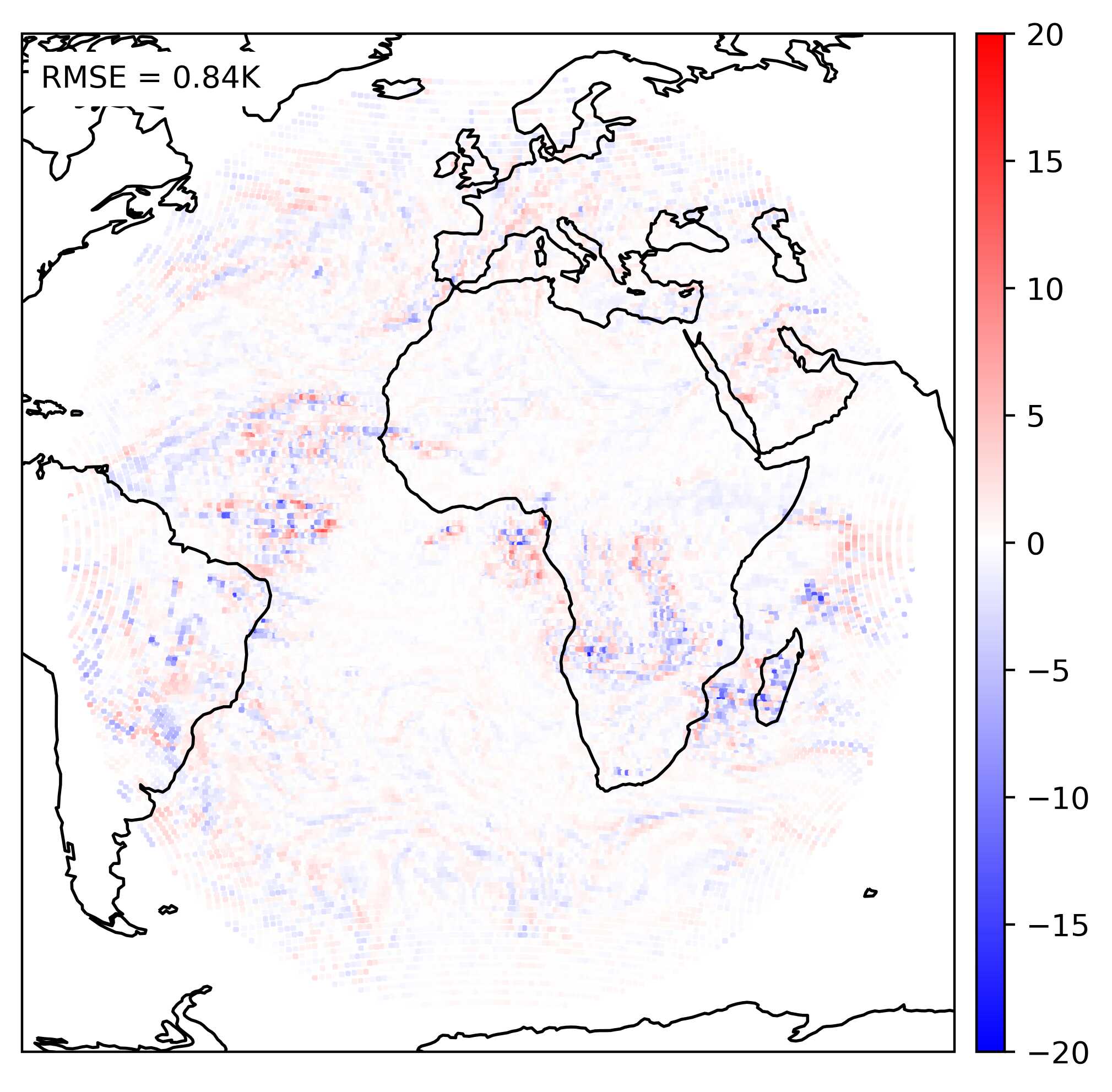}} \\
\subfigure[t+33h (Jan 3, 2023, 08:45z)]{\includegraphics[trim=0 0 45 0, clip, height=.3\linewidth]{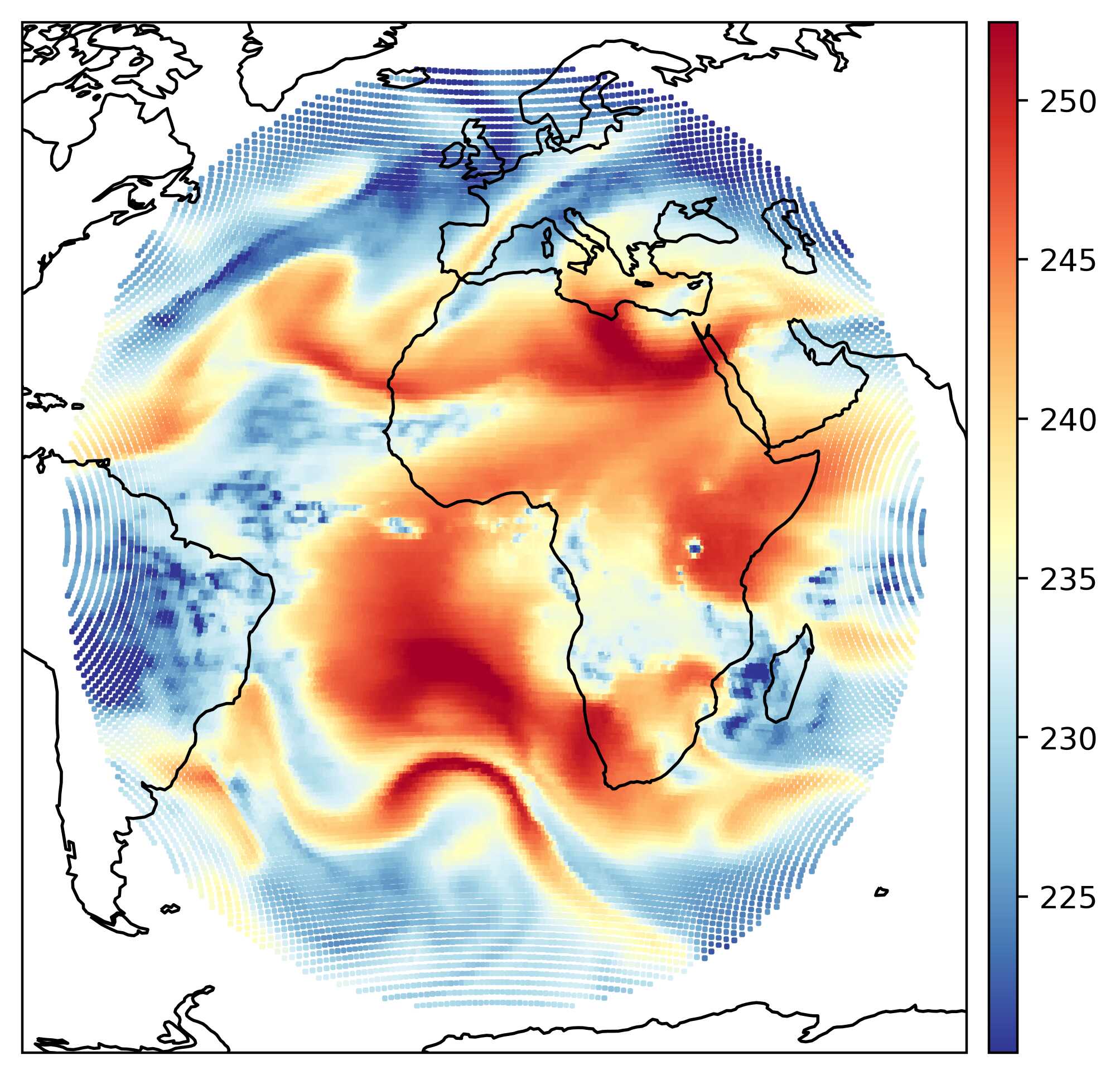}\includegraphics[height=.3\linewidth]{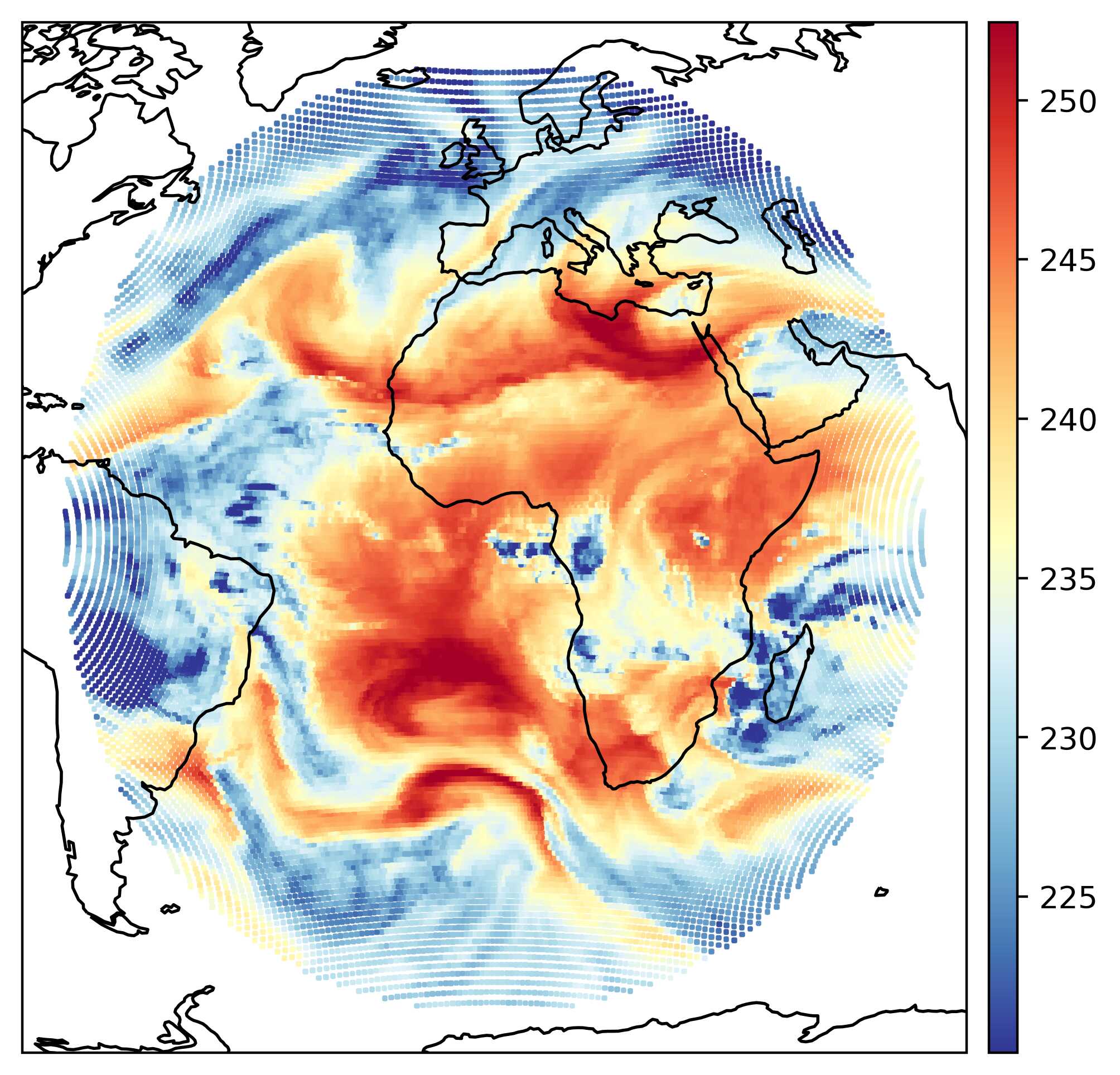}\includegraphics[height=.305\linewidth]{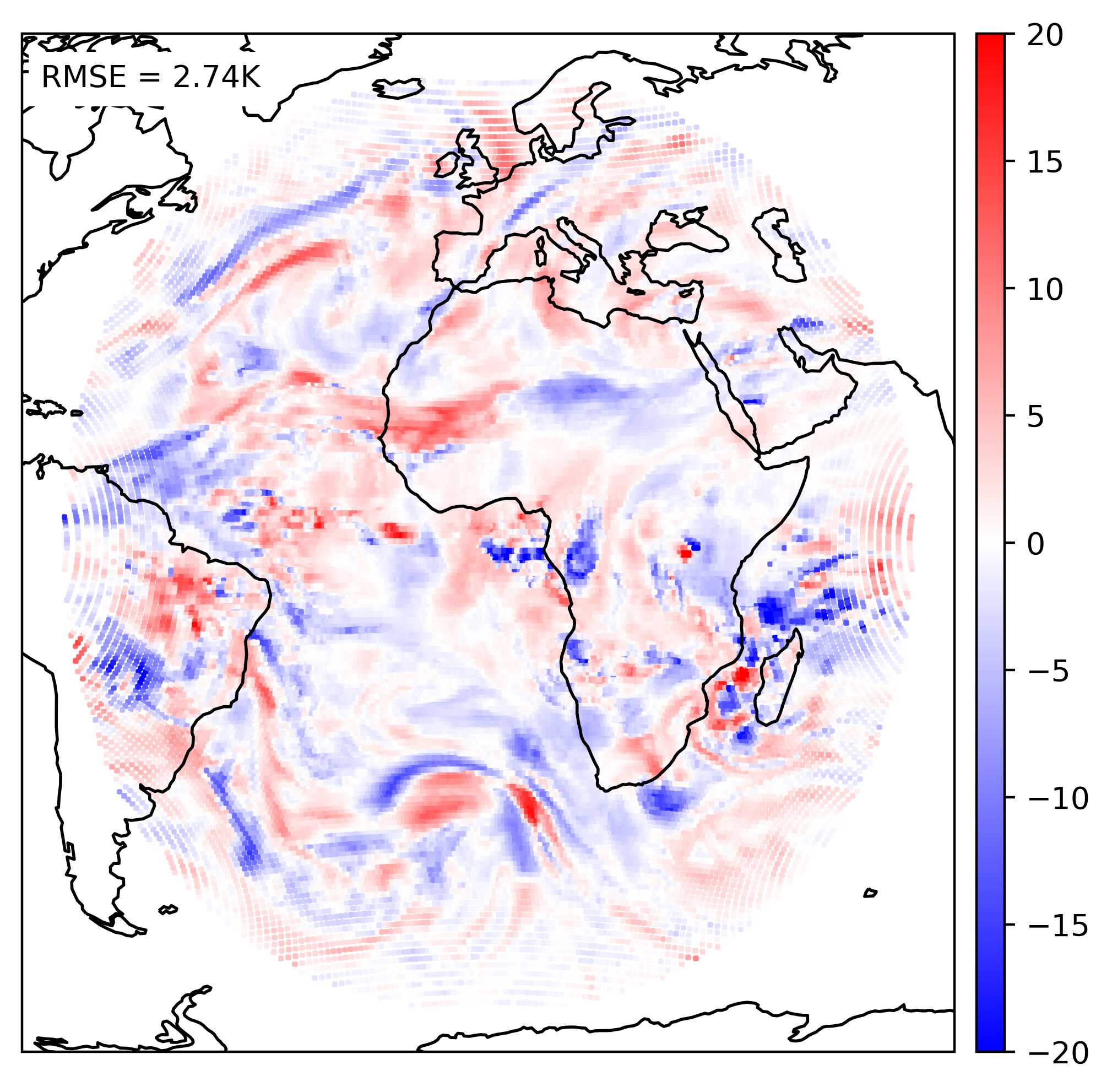}} \\
\subfigure[t+63h (Jan 4, 2023, 14:45z)]{\includegraphics[trim=0 0 45 0, clip, height=.3\linewidth]{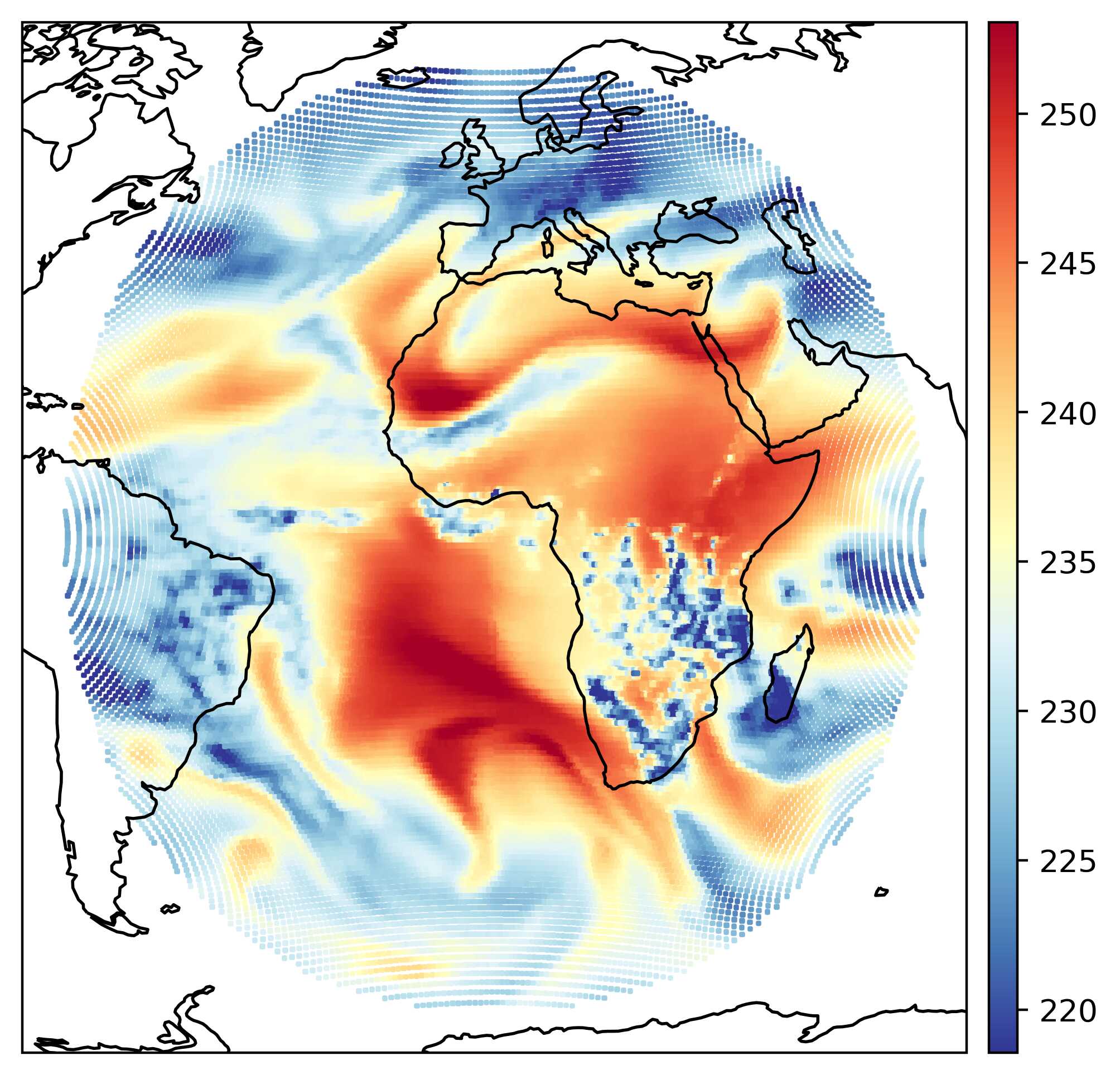}\includegraphics[height=.3\linewidth]{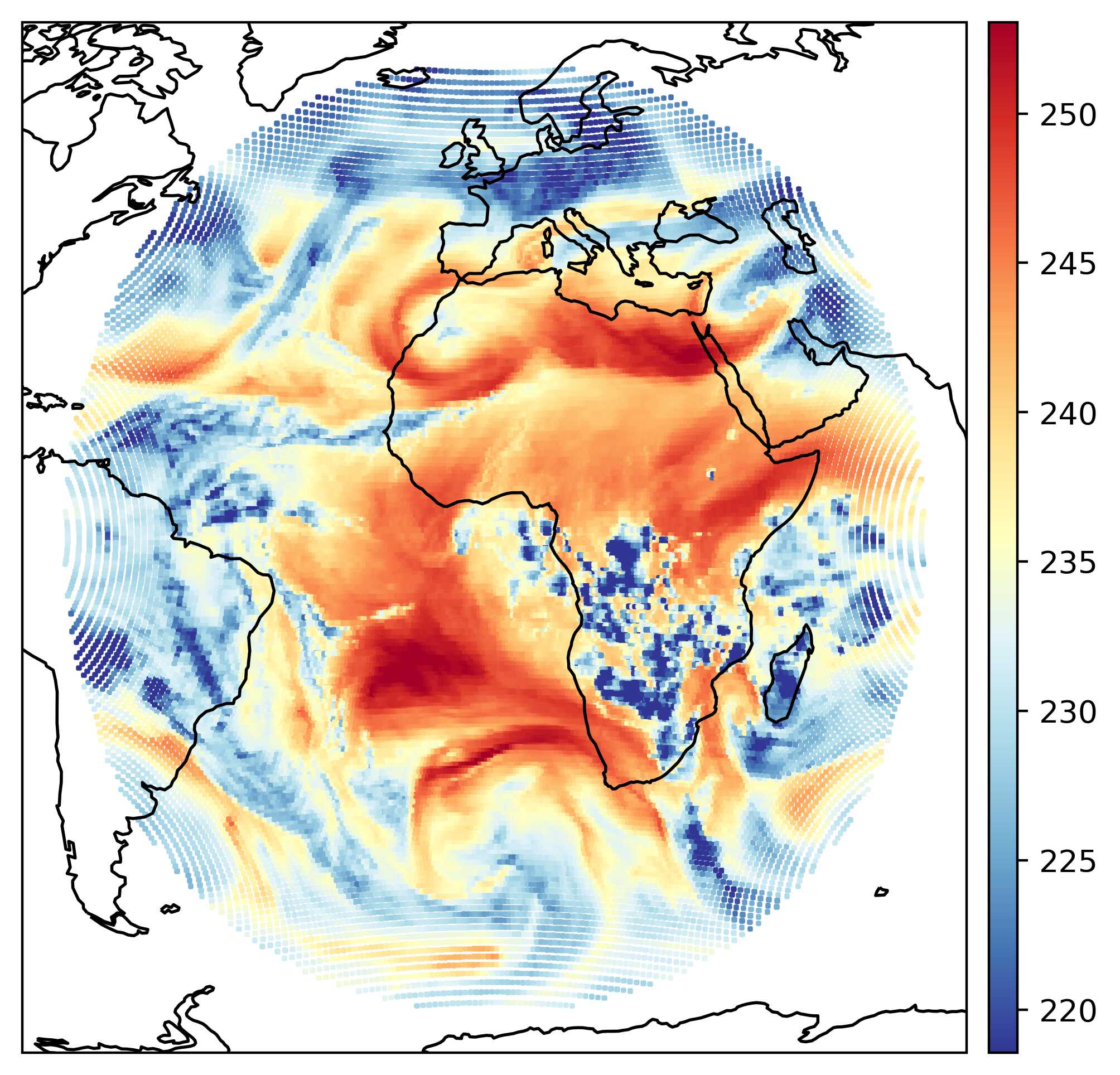}\includegraphics[height=.305\linewidth]{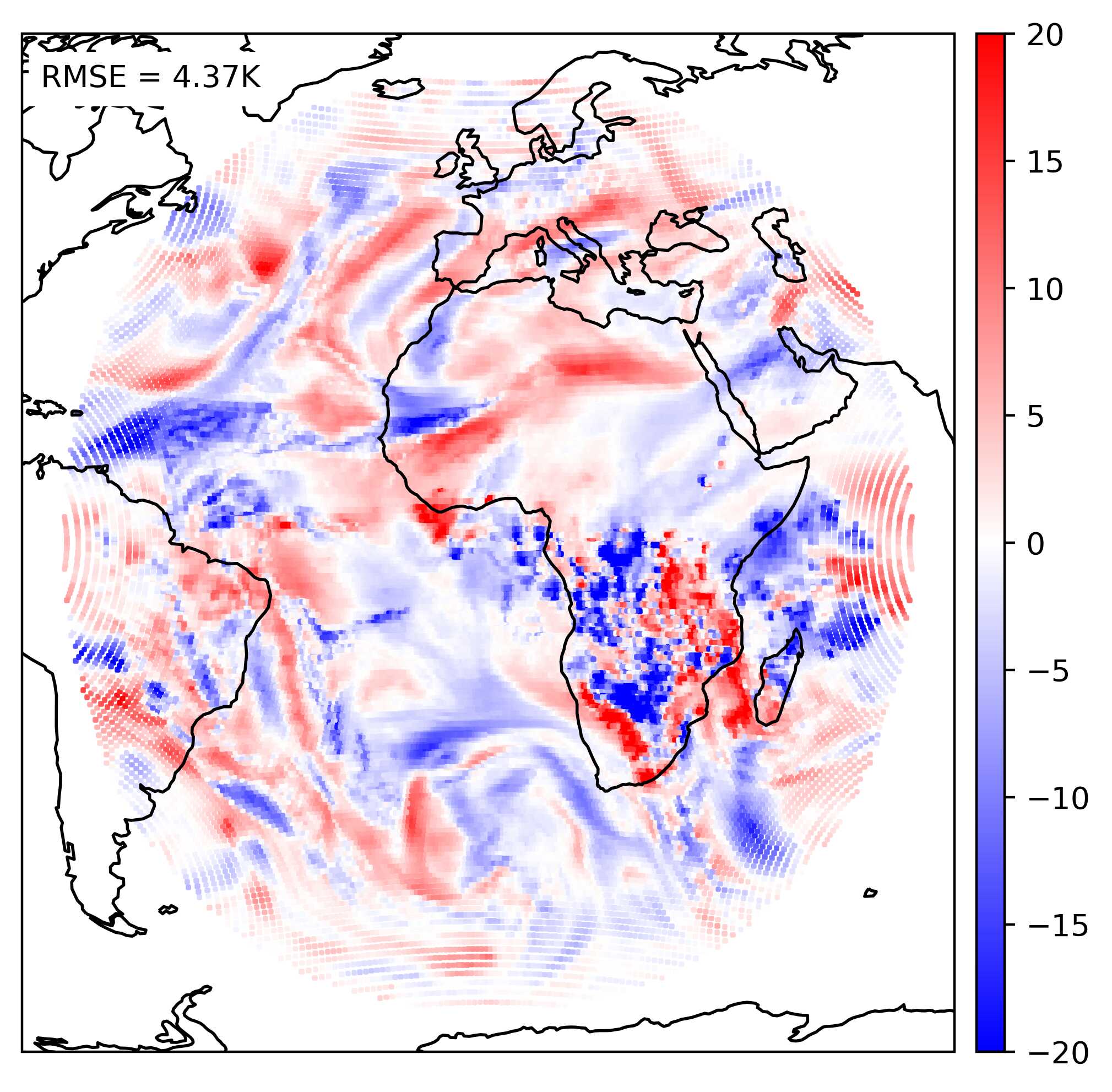}} \\
\caption{SEVIRI water‐vapour 6.2 channel 5 brightness temperatures (K): forecasted (left), observed (middle), and difference (observed minus forecast; right). We show forecast for lead times of 10, 33, and 63 hours. The forecast RMSE for the sample is printed in the top left corner.}
\label{fig:SEVIRI_72h}
\end{figure}

Figure \ref{fig:SEVIRI_72h} shows SEVIRI water‐vapour 6.2-$\mu$m channel 5 brightness temperatures forecasts from a slightly different model architecture where the processor is taking 3-hour steps through a 12-hour target output window (latent space rollout; cf. figure \ref{fig:dop-model-schematic}).

\begin{table}[!htbp]
\centering
\begin{tabular}{|l|l|c|l|}
\hline
\textbf{Category} & \textbf{Instrument} & \textbf{Period} & \textbf{Parameters} \\
\hline
\multirow{6}{*}{Microwave Sounders} 
& NPP ATMS & 2012-2023 & channels 1-22 \\
& NOAA 20 ATMS & 2018-2023 & channels 1-22 \\
& NOAA 15 AMSU-A & 1999-2023 & channels 1-15 \\
& NOAA 18 AMSU-A & 2005-2023 & channels 1-15 \\
& METOP-A AMSU-A & 2006-2021 & channels 1-15 \\
& METOP-B AMSU-A & 2012-2023 & channels 1-15 \\
\hline
\multirow{2}{*}{Microwave Imagers}
& DMSP 17 SSMIS & 2009-2023 & channels 1-24 \\
& GCOM-W AMSR-2 & 2012-2023 & channels 1-14 \\
\hline
\multirow{5}{*}{Infrared Sounders}
& METOP-A IASI & 2007-2023 & 17 channels \\
& METOP-B IASI & 2013-2023 & 17 channels \\
& METOP-C IASI & 2019-2023 & 17 channels \\
& NPP CrIS & 2012-2023 & 15 channels \\
& AQUA AIRS & 2002-2023 & 15 channels \\
\hline
\multirow{3}{*}{Visible}
& METOP-A AVHRR & 2007-2023 & visible channel \\
& METOP-B AVHRR & 2013-2023 & visible channel \\
& METOP-C AVHRR & 2019-2023 & visible channel \\
\hline
\multirow{4}{*}{Geostationary Infrared}
& Meteosat 8 SEVIRI & 2004-2007 & channels 4-11 \\
& Meteosat 8 IODC SEVIRI & 2017-2022 & channels 4-11 \\
& Meteosat 9 SEVIRI & 2007-2012 & channels 4-11 \\
& Meteosat 10 SEVIRI & 2012-2018 & channels 4-11 \\
& Meteosat 11 SEVIRI & 2018-2023 & channels 4-11 \\
\hline
\multirow{3}{*}{Radio-occultation}
& METOP-A GPSRO & 2008-2021 & bending angle \\
& METOP-B GPSRO & 2012-2023 & bending angle \\
& METOP-C GPSRO & 2018-2023 & bending angle \\
\hline
\multirow{3}{*}{Scatterometer}
& METOP-A ASCAT & 2006-2021 & backscatter coefficient sigma0 (3 beams) \\
& METOP-B ASCAT & 2013-2023 & sigma0 (3 beams) \\
& METOP-C ASCAT & 2020-2023 & sigma0 (3 beams) \\
\hline
\multirow{1}{*}{Radar altimeter}
 & SARAL RALT & 2014-2023 & significant wave height, surface wind speed \\
\hline
\multirow{11}{*}{Conventional - surface}
& Automatic Land SYNOP & 1979-2023 & ps, t2m, rh2m, u10, v10 \\
& Manual Land SYNOP & 1979-2023 & ps, t2m, rh2m, u10, v10 \\
& BUFR Land SYNOP & 2014-2023 & ps, t2m, rh2m, u10, v10 \\
& SHIP & 1979-2023 & ps, t2m, rh2m, u10, v10 \\
& BUFR SHIP SYNOP & 2014-2023 & ps, t2m, rh2m, u10, v10 \\
& Abbreviated SHIP & 1979-2023 & ps, t2m, rh2m, u10, v10 \\
& METAR & 2004-2023 & ps, t2m, rh2m, u10, v10 \\
& Automatic METAR & 2013-2023 & ps, t2m, rh2m, u10, v10 \\
& In-situ snow reports & 2014-2023 & snow depth \\
& DRIBU & 1979-2023 & ps,sst \\
& BUFR Drifting Buoys & 2016-2023 & ps, sst \\
\hline
\multirow{4}{*}{Conventional - sonde}

& TEMP SHIP & 1979-2023 & z, t, td, u, v and selected pressure levels \\
& BUFR SHIP TEMP & 2014-2023 & z, t, td, u, v and selected pressure levels \\
& Land TEMP & 1979-2023 & z, t, td, u, v and selected pressure levels \\
& Dropsondes & 1980-2023 & z, t, td, u, v and selected pressure levels \\
\hline
\multirow{4}{*}{Conventional - aircraft}
& AIREP & 1979-2023 & t, u, v \\
& AMDAR & 1992-2023 & t, u, v \\
& ACARS & 1979-1996 & t, u, v \\
& WIGOS AMDAR & 2014-2023 & t, u, v \\
\hline
Surface radar & NEXRAD radar & 2009-2023 & 1h accumulation \\
\hline
\end{tabular}
\caption{Input and output observations used in the current version of GraphDOP.}
\label{table:dop-observations}
\end{table}

\bibliographystyle{unsrtnat}
\bibliography{dop}

\begin{thebibliography}{63}
\providecommand{\natexlab}[1]{#1}
\providecommand{\url}[1]{\texttt{#1}}
\expandafter\ifx\csname urlstyle\endcsname\relax
  \providecommand{\doi}[1]{doi: #1}\else
  \providecommand{\doi}{doi: \begingroup \urlstyle{rm}\Url}\fi

\bibitem[Pathak et~al.(2022)Pathak, Subramanian, Harrington, Raja, Chattopadhyay, Mardani, Kurth, Hall, Li, Azizzadenesheli, and Hassanzadeh]{pathak2022fourcastnet}
J.~Pathak, S.~Subramanian, P.~Harrington, S.~Raja, A.~Chattopadhyay, M.~Mardani, T.~Kurth, D.~Hall, Z.~Li, K.~Azizzadenesheli, and P.~Hassanzadeh.
\newblock {F}our{C}ast{N}et: A global data-driven high-resolution weather model using adaptive fourier neural operators.
\newblock \emph{arXiv preprint arXiv:2202.11214}, Feb 22 2022.

\bibitem[Lam et~al.(2023)Lam, Sanchez-Gonzalez, Willson, Wirnsberger, Fortunato, Alet, Ravuri, Ewalds, Eaton-Rosen, Hu, Merose, Hoyer, Holland, Vinyals, Stott, Pritzel, Mohamed, and Battaglia]{lam2022graphcast}
Remi Lam, Alvaro Sanchez-Gonzalez, Matthew Willson, Peter Wirnsberger, Meire Fortunato, Ferran Alet, Suman Ravuri, Timo Ewalds, Zach Eaton-Rosen, Weihua Hu, Alexander Merose, Stephan Hoyer, George Holland, Oriol Vinyals, Jacklynn Stott, Alexander Pritzel, Shakir Mohamed, and Peter Battaglia.
\newblock Learning skillful medium-range global weather forecasting.
\newblock \emph{Science}, 382\penalty0 (6677):\penalty0 1416–1421, December 2023.
\newblock ISSN 1095-9203.
\newblock \doi{10.1126/science.adi2336}.
\newblock URL \url{http://dx.doi.org/10.1126/science.adi2336}.

\bibitem[Bi et~al.(2023)Bi, Xie, Zhang, et~al.]{bi2023accurate}
K.~Bi, L.~Xie, H.~Zhang, et~al.
\newblock Accurate medium-range global weather forecasting with 3{D} neural networks.
\newblock \emph{Nature}, 619:\penalty0 533--538, 2023.
\newblock \doi{10.1038/s41586-023-06185-3}.

\bibitem[Bodnar et~al.(2024)Bodnar, Bruinsma, Lucic, Stanley, Brandstetter, Garvan, Riechert, Weyn, Dong, Vaughan, Gupta, Tambiratnam, Archibald, Heider, Welling, Turner, and Perdikaris]{bodnar2024aurora}
Cristian Bodnar, Wessel~P. Bruinsma, Ana Lucic, Megan Stanley, Johannes Brandstetter, Patrick Garvan, Maik Riechert, Jonathan Weyn, Haiyu Dong, Anna Vaughan, Jayesh~K. Gupta, Kit Tambiratnam, Alex Archibald, Elizabeth Heider, Max Welling, Richard~E. Turner, and Paris Perdikaris.
\newblock Aurora: A foundation model of the atmosphere, 2024.
\newblock URL \url{https://arxiv.org/abs/2405.13063}.

\bibitem[Lang et~al.(2024{\natexlab{a}})Lang, Alexe, Chantry, Dramsch, Pinault, Raoult, Clare, Lessig, Maier-Gerber, Magnusson, Bouallègue, Nemesio, Dueben, Brown, Pappenberger, and Rabier]{lang2024aifs}
Simon Lang, Mihai Alexe, Matthew Chantry, Jesper Dramsch, Florian Pinault, Baudouin Raoult, Mariana C.~A. Clare, Christian Lessig, Michael Maier-Gerber, Linus Magnusson, Zied~Ben Bouallègue, Ana~Prieto Nemesio, Peter~D. Dueben, Andrew Brown, Florian Pappenberger, and Florence Rabier.
\newblock {AIFS} -- {ECMWF}'s data-driven forecasting system.
\newblock \emph{arXiv preprint arXiv:2406.01465}, 2024{\natexlab{a}}.
\newblock URL \url{https://arxiv.org/abs/2406.01465}.

\bibitem[Hersbach et~al.(2020)Hersbach, Bell, Berrisford, et~al.]{hersbach2020era5}
H.~Hersbach, B.~Bell, P.~Berrisford, et~al.
\newblock The {ERA5} global reanalysis.
\newblock \emph{QJ R Meteorol Soc}, 146:\penalty0 1999--2049, 2020.
\newblock \doi{10.1002/qj.3803}.

\bibitem[Rabier et~al.(2000)Rabier, Järvinen, Klinker, Mahfouf, and Simmons]{rabier20004dvar}
F.~Rabier, H.~Järvinen, E.~Klinker, J.-F. Mahfouf, and A.~Simmons.
\newblock The {ECMWF} operational implementation of four-dimensional variational assimilation. {I}: Experimental results with simplified physics.
\newblock \emph{Quarterly Journal of the Royal Meteorological Society}, 126\penalty0 (564):\penalty0 1143--1170, 2000.
\newblock \doi{10.1002/qj.49712656415}.

\bibitem[ECMWF(2023)]{ecmwf4dvar48r1doc}
ECMWF.
\newblock \emph{{IFS} Documentation {CY48R1} - {P}art {II}: Data Assimilation}.
\newblock Number~2. ECMWF, 06/2023 2023.
\newblock \doi{10.21957/a744f32e74}.

\bibitem[{ECMWF}(2017)]{4dvarecmwfhistory}
{ECMWF}.
\newblock 20 years of {4D-Var}: {B}etter forecasts through a better use of observations.
\newblock https://www.ecmwf.int/en/about/media-centre/news/2017/20-years-4d-var-better-forecasts-through--better-use-observations, 2017.
\newblock Accessed: Dec 9, 2024.

\bibitem[McNally et~al.(2024{\natexlab{a}})McNally, Lessig, Lean, Boucher, Alexe, Pinnington, Chantry, Lang, Burrows, Chrust, Pinault, Villeneuve, Bormann, and Healy]{mcnally2024dop}
Anthony McNally, Christian Lessig, Peter Lean, Eulalie Boucher, Mihai Alexe, Ewan Pinnington, Matthew Chantry, Simon Lang, Chris Burrows, Marcin Chrust, Florian Pinault, Ethel Villeneuve, Niels Bormann, and Sean Healy.
\newblock Data driven weather forecasts trained and initialised directly from observations, 2024{\natexlab{a}}.
\newblock URL \url{https://arxiv.org/abs/2407.15586}.

\bibitem[Vaughan et~al.(2024)Vaughan, Markou, Tebbutt, Requeima, Bruinsma, Andersson, Herzog, Lane, Chantry, Hosking, and Turner]{vaughan2024aardvark}
Anna Vaughan, Stratis Markou, Will Tebbutt, James Requeima, Wessel~P. Bruinsma, Tom~R. Andersson, Michael Herzog, Nicholas~D. Lane, Matthew Chantry, J.~Scott Hosking, and Richard~E. Turner.
\newblock Aardvark weather: end-to-end data-driven weather forecasting, 2024.
\newblock URL \url{https://arxiv.org/abs/2404.00411}.

\bibitem[Tian et~al.(2024)Tian, Holdaway, and Kleist]{tian2024exploringusemachinelearning}
Xiaoxu Tian, Daniel Holdaway, and Daryl Kleist.
\newblock Exploring the use of machine learning weather models in data assimilation, 2024.
\newblock URL \url{https://arxiv.org/abs/2411.14677}.

\bibitem[Xu et~al.(2024{\natexlab{a}})Xu, Duan, and Xu]{Xu2024GSIpangu}
Hongxiong Xu, Yihong Duan, and Xiangde Xu.
\newblock Exploring the integration of a global ai model with traditional data assimilation in weather forecasting.
\newblock \emph{Environmental Research Letters}, 19\penalty0 (12):\penalty0 124079, nov 2024{\natexlab{a}}.
\newblock \doi{10.1088/1748-9326/ad93e8}.

\bibitem[Rozet and Louppe(2023)]{rozet2023scorebaseddataassimilation}
François Rozet and Gilles Louppe.
\newblock Score-based data assimilation, 2023.
\newblock URL \url{https://arxiv.org/abs/2306.10574}.

\bibitem[Huang et~al.(2024)Huang, Gianinazzi, Yu, Dueben, and Hoefler]{huang2024diffdadiffusionmodelweatherscale}
Langwen Huang, Lukas Gianinazzi, Yuejiang Yu, Peter~D. Dueben, and Torsten Hoefler.
\newblock Diffda: a diffusion model for weather-scale data assimilation, 2024.
\newblock URL \url{https://arxiv.org/abs/2401.05932}.

\bibitem[Li et~al.(2024)Li, Han, Li, Duan, Chen, Zhong, Wang, Liu, and Sun]{li2024fuxien4dvar}
Yonghui Li, Wei Han, Hao Li, Wansuo Duan, Lei Chen, Xiaohui Zhong, Jincheng Wang, Yongzhu Liu, and Xiuyu Sun.
\newblock {FuXi-En4DVar}: An assimilation system based on machine learning weather forecasting model ensuring physical constraints.
\newblock \emph{{G}eophysical {R}esearch {L}etters}, 51\penalty0 (22):\penalty0 e2024GL111136, 2024.
\newblock \doi{https://doi.org/10.1029/2024GL111136}.
\newblock URL \url{https://agupubs.onlinelibrary.wiley.com/doi/abs/10.1029/2024GL111136}.
\newblock e2024GL111136 2024GL111136.

\bibitem[Xu et~al.(2024{\natexlab{b}})Xu, Sun, Han, Zhong, Chen, and Li]{xu2024fuxida}
Xiaoze Xu, Xiuyu Sun, Wei Han, Xiaohui Zhong, Lei Chen, and Hao Li.
\newblock Fuxi-{DA}: A generalized deep learning data assimilation framework for assimilating satellite observations, 2024{\natexlab{b}}.
\newblock URL \url{https://arxiv.org/abs/2404.08522}.

\bibitem[Xiao et~al.(2024)Xiao, Bai, Xue, Chen, Han, and Ouyang]{xiao2024fengwu4dvarcouplingdatadrivenweather}
Yi~Xiao, Lei Bai, Wei Xue, Kang Chen, Tao Han, and Wanli Ouyang.
\newblock {F}eng{W}u-{4DV}ar: Coupling the data-driven weather forecasting model with {4D} variational assimilation, 2024.
\newblock URL \url{https://arxiv.org/abs/2312.12455}.

\bibitem[Sun et~al.(2024)Sun, Zhong, Xu, Huang, Li, Neelin, Chen, Feng, Han, Wu, and Qi]{sun2024fuxiweatherdatatoforecastmachine}
Xiuyu Sun, Xiaohui Zhong, Xiaoze Xu, Yuanqing Huang, Hao Li, J.~David Neelin, Deliang Chen, Jie Feng, Wei Han, Libo Wu, and Yuan Qi.
\newblock {FuXi} {W}eather: A data-to-forecast machine learning system for global weather, 2024.
\newblock URL \url{https://arxiv.org/abs/2408.05472}.

\bibitem[Xiang et~al.(2024)Xiang, Jin, Dong, Bai, Fang, Zhao, Sun, Thambiratnam, Zhang, and Huang]{xiang2024adafartificialintelligencedata}
Yanfei Xiang, Weixin Jin, Haiyu Dong, Mingliang Bai, Zuliang Fang, Pengcheng Zhao, Hongyu Sun, Kit Thambiratnam, Qi~Zhang, and Xiaomeng Huang.
\newblock Adaf: An artificial intelligence data assimilation framework for weather forecasting, 2024.
\newblock URL \url{https://arxiv.org/abs/2411.16807}.

\bibitem[Agrawal et~al.(2019)Agrawal, Barrington, Bromberg, Burge, Gazen, and Hickey]{Agrawal2019}
Shreya Agrawal, Luke Barrington, Carla Bromberg, John Burge, Cenk Gazen, and Jason Hickey.
\newblock Machine learning for precipitation nowcasting from radar images.
\newblock \emph{CoRR}, abs/1912.12132, 2019.
\newblock URL \url{http://arxiv.org/abs/1912.12132}.

\bibitem[S{\o}nderby et~al.(2020)S{\o}nderby, Espeholt, Heek, Dehghani, Oliver, Salimans, Agrawal, Hickey, and Kalchbrenner]{sonderby2020metnet}
Casper~Kaae S{\o}nderby, Lasse Espeholt, Jonathan Heek, Mostafa Dehghani, Avital Oliver, Tim Salimans, Shreya Agrawal, Jason Hickey, and Nal Kalchbrenner.
\newblock Metnet: {A} neural weather model for precipitation forecasting.
\newblock \emph{CoRR}, abs/2003.12140, 2020.
\newblock URL \url{https://arxiv.org/abs/2003.12140}.

\bibitem[Ravuri et~al.(2021)Ravuri, Lenc, Willson, Kangin, Lam, Mirowski, Fitzsimons, Athanassiadou, Kashem, Madge, Prudden, Mandhane, Clark, Brock, Simonyan, Hadsell, Robinson, Clancy, Arribas, and Mohamed]{ravuri2021nature}
Suman Ravuri, Karel Lenc, Matthew Willson, Dmitry Kangin, Remi Lam, Piotr Mirowski, Megan Fitzsimons, Maria Athanassiadou, Sheleem Kashem, Sam Madge, Rachel Prudden, Amol Mandhane, Aidan Clark, Andrew Brock, Karen Simonyan, Raia Hadsell, Niall Robinson, Ellen Clancy, Alberto Arribas, and Shakir Mohamed.
\newblock Skilful precipitation nowcasting using deep generative models of radar.
\newblock \emph{Nature}, 597\penalty0 (7878):\penalty0 672--677, September 2021.
\newblock \doi{10.1038/s41586-021-03854-z}.

\bibitem[Andrychowicz et~al.(2023)Andrychowicz, Espeholt, Li, Merchant, Merose, Zyda, Agrawal, and Kalchbrenner]{andrychowicz2023}
Marcin Andrychowicz, Lasse Espeholt, Di~Li, Samier Merchant, Alexander Merose, Fred Zyda, Shreya Agrawal, and Nal Kalchbrenner.
\newblock Deep learning for day forecasts from sparse observations, 2023.
\newblock URL \url{https://arxiv.org/abs/2306.06079}.

\bibitem[Zhang et~al.(2023)Zhang, Long, Chen, Xing, Jin, Jordan, and Wang]{Zhang2023}
Yuchen Zhang, Mingsheng Long, Kaiyuan Chen, Lanxiang Xing, Ronghua Jin, Michael~I. Jordan, and Jianmin Wang.
\newblock Skilful nowcasting of extreme precipitation with nowcastnet.
\newblock \emph{Nature}, 619\penalty0 (7970):\penalty0 526--532, Jul 2023.
\newblock ISSN 1476-4687.
\newblock \doi{10.1038/s41586-023-06184-4}.

\bibitem[McNally et~al.(2024{\natexlab{b}})McNally, Lessig, Lean, Chantry, Alexe, and Lang]{mcnally2024}
Tony McNally, Christian Lessig, Peter Lean, Matthew Chantry, Mihai Alexe, and Simon Lang.
\newblock Red sky at night... {P}roducing weather forecasts directly from observations.
\newblock \emph{{ECMWF} Newsletter No. 178}, 2024{\natexlab{b}}.
\newblock \doi{10.21957/1a8466ec2f}.

\bibitem[Lessig(2025)]{lessig2024dop}
Christian Lessig.
\newblock Manuscript in preparation, 2025.

\bibitem[Dee(2004)]{dee2004VarBC}
Dick~P. Dee.
\newblock Variational bias correction of radiance data in the {ECMWF} system.
\newblock \emph{ECMWF Workshop on Assimilation of high spectral resolution sounders in NWP, 28 June - 1 July 2004}, pages 97--112, 2004 2004.

\bibitem[Forsythe(2007)]{forsythe2007amv}
Mary Forsythe.
\newblock {A}tmospheric {M}otion {V}ectors: Past, present and future.
\newblock \emph{{ECMWF} Seminar on Recent development in the use of satellite observations in {NWP}}, 2007.

\bibitem[Fey and Lenssen(2019)]{fey2019pyg}
Matthias Fey and Jan~Eric Lenssen.
\newblock Fast graph representation learning with pytorch geometric.
\newblock \emph{CoRR}, abs/1903.02428, 2019.
\newblock URL \url{http://arxiv.org/abs/1903.02428}.

\bibitem[Keisler(2022)]{keisler2022forecasting}
Ryan Keisler.
\newblock Forecasting global weather with graph neural networks.
\newblock \emph{arXiv preprint arXiv:2202.07575}, Feb 2022.

\bibitem[Micikevicius et~al.(2018)Micikevicius, Narang, Alben, Diamos, Elsen, Garcia, Ginsburg, Houston, Kuchaiev, Venkatesh, and Wu]{micikevicius2018mixed}
Paulius Micikevicius, Sharan Narang, Jonah Alben, Gregory Diamos, Erich Elsen, David Garcia, Boris Ginsburg, Michael Houston, Oleksii Kuchaiev, Ganesh Venkatesh, and Hao Wu.
\newblock Mixed precision training.
\newblock \emph{arXiv preprint arXiv:1710.03740}, 2018.

\bibitem[Wedi(2014)]{Wedi2014}
N.~P. Wedi.
\newblock Increasing the horizontal resolution in numerical weather prediction and climate simulations: illusion or panacea?
\newblock \emph{Philosophical Transactions of the Royal Society A}, 372, 2014.
\newblock \doi{10.1098/rsta.2013.0289}.

\bibitem[Bauer et~al.(2015)Bauer, Thorpe, and Brunet]{Bauer2015}
Peter Bauer, Alan Thorpe, and Gilbert Brunet.
\newblock The quiet revolution of numerical weather prediction.
\newblock \emph{Nature}, 525\penalty0 (7567):\penalty0 47–55, September 2015.
\newblock ISSN 1476-4687.
\newblock \doi{10.1038/nature14956}.
\newblock URL \url{http://dx.doi.org/10.1038/nature14956}.

\bibitem[McNally et~al.(2014)McNally, Bonavita, and Thépaut]{McNally2014}
Tony McNally, Massimo Bonavita, and Jean-Noël Thépaut.
\newblock The role of satellite data in the forecasting of hurricane {S}andy.
\newblock \emph{Monthly Weather Review}, 142\penalty0 (2):\penalty0 634–646, January 2014.
\newblock ISSN 1520-0493.
\newblock \doi{10.1175/mwr-d-13-00170.1}.

\bibitem[Ben~Bouall\`egue et~al.(2024)Ben~Bouall\`egue, Adewoyin, Alexe, Chantry, Clare, Dramsch, Hahner, Lang, Lessig, Magnusson, Maier-Gerber, Mertes, Moldovan, Nemesio, O’Brien, Pinault, Raoult, Cruz, Theissen, and Tietsche]{blogaifs2}
Z.~Ben~Bouall\`egue, Rilwan Adewoyin, Mihai Alexe, Matthew Chantry, Mariana Clare, Jesper Dramsch, Sara Hahner, Simon Lang, Christian Lessig, Linus Magnusson, Michael Maier-Gerber, Gert Mertes, Gabriel Moldovan, Ana~Prieto Nemesio, Cathal O’Brien, Florian Pinault, Baudouin Raoult, Mario~Santa Cruz, Helen Theissen, and Steffen Tietsche.
\newblock A new {ML} model in the {ECMWF} web charts, 2024.

\bibitem[Hoffman et~al.(1995)Hoffman, Liu, Louis, and Grassoti]{Hoffman1995}
Ross Hoffman, Zheng Liu, Jean-Francois Louis, and Christopher Grassoti.
\newblock Distortion representation of forecast errors.
\newblock \emph{Monthly Weather Review}, 123\penalty0 (9):\penalty0 2758–2770, 1995.
\newblock ISSN 1520-0493.
\newblock \doi{10.1175/1520-0493(1995)123<2758:drofe>2.0.co;2}.

\bibitem[Ebert et~al.(2013)Ebert, Wilson, Weigel, Mittermaier, Nurmi, Gill, Göber, Joslyn, Brown, Fowler, and Watkins]{ebertdouble}
E.~Ebert, L.~Wilson, A.~Weigel, M.~Mittermaier, P.~Nurmi, P.~Gill, M.~Göber, S.~Joslyn, B.~Brown, T.~Fowler, and A.~Watkins.
\newblock Progress and challenges in forecast verification.
\newblock \emph{Meteorological Applications}, 20\penalty0 (2):\penalty0 130--139, 2013.
\newblock \doi{https://doi.org/10.1002/met.1392}.

\bibitem[Karras et~al.(2022)Karras, Aittala, Aila, and Laine]{karras2022elucidating}
Tero Karras, Miika Aittala, Timo Aila, and Samuli Laine.
\newblock Elucidating the design space of diffusion-based generative models.
\newblock \emph{arXiv preprint arXiv:2206.00364}, 2022.

\bibitem[Alexe et~al.(2024)Alexe, Lang, Clare, Leutbecher, Roberts, Magnusson, Matthew~Chantry, Prieto-Nemesio, Dramsch, Pinault, and Raoult]{alexe2024ecmwfnl}
Mihai Alexe, Simon Lang, Mariana Clare, Martin Leutbecher, Christopher Roberts, Linus Magnusson, Rilwan~Adewoyin Matthew~Chantry, Ana Prieto-Nemesio, Jesper Dramsch, Florian Pinault, and Baudouin Raoult.
\newblock Data-driven ensemble forecasting with the {AIFS}.
\newblock \emph{{ECMWF} Newsletter No. 181}, 2024.
\newblock \doi{10.21957/ma3p95hxe2}.

\bibitem[Lang et~al.(2024{\natexlab{b}})Lang, Alexe, Clare, Roberts, Adewoyin, Ben-Bouall\`egue, Chantry, Dramsch, D\"ueben, Hahner, Maciel, Prieto-Nemesio, O'Brien, Pinault, Polster, Raoult, Tietsche, and Leutbecher]{lang2024ensscore}
Simon Lang, Mihai Alexe, Mariana Clare, Christopher Roberts, Rilwan Adewoyin, Zied Ben-Bouall\`egue, Matthew Chantry, Jesper Dramsch, Peter D\"ueben, Sara Hahner, Pedro Maciel, Ana Prieto-Nemesio, Cathal O'Brien, Florian Pinault, Jan Polster, Baudouin Raoult, Steffen Tietsche, and Martin Leutbecher.
\newblock Ensemble forecasting with the {A}rtificial {I}ntelligence {F}orecasting {S}ystem trained with a loss function based on the continuous ranked probability score.
\newblock \emph{In preparation}, 2024{\natexlab{b}}.

\bibitem[Matricardi and McNally(2014)]{Matricardi2014}
Marco Matricardi and Anthony~P. McNally.
\newblock The direct assimilation of principal components of {IASI} spectra in the {ECMWF} {4D-Var}.
\newblock \emph{Quarterly Journal of the Royal Meteorological Society}, 140\penalty0 (679):\penalty0 573--582, 2014.
\newblock \doi{https://doi.org/10.1002/qj.2156}.

\bibitem[Owens and Hewson(2018)]{Owens18}
Robert Owens and Tim Hewson.
\newblock {ECMWF} forecast user guide.
\newblock Technical report, ECMWF, Reading, 05/2018 2018.

\bibitem[Baordo and Geer(2015)]{Baordo15}
Fabrizio Baordo and Alan Geer.
\newblock All-sky assimilation of {SSMIS} humidity sounding channels over land within the {ECMWF} system.
\newblock Technical Report~38, 07/2015 2015.

\bibitem[Duncan et~al.(2021)Duncan, Bormann, Geer, and Weston]{Duncan21}
David Duncan, Niels Bormann, Alan Geer, and Peter Weston.
\newblock Assimilation of {AMSU-A} in all-sky conditions.
\newblock Technical report, ECMWF, 10/2021 2021.

\bibitem[Haiden et~al.(2024)Haiden, Janousek, Vitart, Tanguy, Prates, and Chevalier]{Haiden24}
Thomas Haiden, Martin Janousek, Frédéric Vitart, Maliko Tanguy, Fernando Prates, and Matthieu Chevalier.
\newblock Evaluation of {ECMWF} forecasts.
\newblock Technical Report 918, ECMWF, 09/2024 2024.

\bibitem[Dahoui et~al.(2016)Dahoui, Radnoti, Healy, Isaksen, and Haiden]{Dahoui16}
Mohamed Dahoui, Gabor Radnoti, Sean Healy, Lars Isaksen, and Thomas Haiden.
\newblock Use of forecast departures in verification against observations.
\newblock ECMWF Newsletter 149, ECMWF, 2016.

\bibitem[Ingleby et~al.(2024)Ingleby, Arduini, Balsamo, Boussetta, Ochi, Pinnington, and de~Rosnay]{Ingleby24}
Bruce Ingleby, Gabriele Arduini, Gianpaolo Balsamo, Souhail Boussetta, Kenta Ochi, Ewan Pinnington, and Patricia de~Rosnay.
\newblock Improved two-metre temperature forecasts in the 2024 upgrade.
\newblock ECMWF Newsletter 178, 2024.

\bibitem[Lean et~al.(2021)Lean, Hólm, Bonavita, Bormann, McNally, and Järvinen]{Lean21}
Peter Lean, Elias Hólm, Massimo Bonavita, Niels Bormann, Anthony McNally, and Heikki Järvinen.
\newblock Continuous data assimilation for global numerical weather prediction.
\newblock \emph{Quarterly Journal of the Royal Meteorological Society}, 147\penalty0 (734):\penalty0 273--288, 2021.
\newblock \doi{10.1002/qj.3917}.

\bibitem[Bonavita and Laloyaux(2020)]{Bonavita20}
Massimo Bonavita and Patrick Laloyaux.
\newblock Machine learning for model error inference and correction.
\newblock \emph{Journal of Advances in Modeling Earth Systems}, 12\penalty0 (12), 2020.
\newblock \doi{https://doi.org/10.1029/2020MS002232}.

\bibitem[Farchi et~al.(2024)Farchi, Chrust, Bocquet, and Bonavita]{farchi24}
Alban Farchi, Marcin Chrust, Marc Bocquet, and Massimo Bonavita.
\newblock Online model error correction with neural networks: application to the integrated forecasting system, 2024.
\newblock URL \url{https://arxiv.org/abs/2403.03702}.

\bibitem[Bouallègue et~al.(2023)Bouallègue, Cooper, Chantry, Düben, Bechtold, and Sandu]{ziedT2Mpproc2023}
Zied~Ben Bouallègue, Fenwick Cooper, Matthew Chantry, Peter Düben, Peter Bechtold, and Irina Sandu.
\newblock Statistical modeling of 2-m temperature and 10-m wind speed forecast errors.
\newblock \emph{Monthly Weather Review}, 151\penalty0 (4):\penalty0 897 -- 911, 2023.
\newblock \doi{10.1175/MWR-D-22-0107.1}.

\bibitem[Geer et~al.(2022)Geer, Lonitz, Duncan, and Bormann]{Geer22}
Alan Geer, Katrin Lonitz, David Duncan, and Niels Bormann.
\newblock Developing an all-surface capability for all-sky microwave radiances.
\newblock ECMWF Newsletter 171, 2022.

\bibitem[Shepherd et~al.(2018)Shepherd, Polichtchouk, Hogan, and Simmons]{Sheperd18}
T.G. Shepherd, I.~Polichtchouk, Robin Hogan, and A.J. Simmons.
\newblock Report on stratosphere task force, 06/2018 2018.

\bibitem[Laloyaux et~al.(2020)Laloyaux, Bonavita, Dahoui, Farnan, Healy, Hólm, and Lang]{Laloyaux20}
P.~Laloyaux, M.~Bonavita, M.~Dahoui, J.~Farnan, S.~Healy, E.~Hólm, and S.~T.~K. Lang.
\newblock Towards an unbiased stratospheric analysis.
\newblock \emph{Quarterly Journal of the Royal Meteorological Society}, 146\penalty0 (730):\penalty0 2392--2409, 2020.
\newblock \doi{https://doi.org/10.1002/qj.3798}.

\bibitem[Scanlon et~al.(2024)Scanlon, Geer, Bormann, and Browne]{Scanlon24}
Tracy Scanlon, Alan Geer, Niels Bormann, and Philip Browne.
\newblock \emph{Improving Ocean Surface Temperature for {NWP} using All-Sky Microwave Imager Observations}, 08/2024 2024.

\bibitem[Rasp et~al.(2024)Rasp, Hoyer, Merose, Langmore, Battaglia, Russel, Sanchez-Gonzalez, Yang, Carver, Agrawal, Chantry, Bouallegue, Dueben, Bromberg, Sisk, Barrington, Bell, and Sha]{rasp2024weatherbench2benchmarkgeneration}
Stephan Rasp, Stephan Hoyer, Alexander Merose, Ian Langmore, Peter Battaglia, Tyler Russel, Alvaro Sanchez-Gonzalez, Vivian Yang, Rob Carver, Shreya Agrawal, Matthew Chantry, Zied~Ben Bouallegue, Peter Dueben, Carla Bromberg, Jared Sisk, Luke Barrington, Aaron Bell, and Fei Sha.
\newblock Weatherbench 2: A benchmark for the next generation of data-driven global weather models, 2024.
\newblock URL \url{https://arxiv.org/abs/2308.15560}.

\bibitem[Jung and Leutbecher(2008)]{Jung2008}
Thomas Jung and Martin Leutbecher.
\newblock Scale‐dependent verification of ensemble forecasts.
\newblock \emph{Quarterly Journal of the Royal Meteorological Society}, 134\penalty0 (633):\penalty0 973–984, 2008.
\newblock ISSN 1477-870X.
\newblock \doi{10.1002/qj.255}.

\bibitem[Mogensen et~al.(2018)Mogensen, Hewson, Keeley, and Magnusson]{Mogensen18}
Kristian~S. Mogensen, Tim Hewson, Sarah Keeley, and Linus Magnusson.
\newblock Effects of ocean coupling on weather forecasts.
\newblock ECMWF Newsletter 156, ECMWF, 2018.

\bibitem[de~Rosnay et~al.(2022)de~Rosnay, Browne, de~Boisséson, Fairbairn, Hirahara, Ochi, Schepers, Weston, Zuo, Alonso-Balmaseda, Balsamo, Bonavita, Borman, Brown, Chrust, Dahoui, Chiara, English, Geer, Healy, Hersbach, Laloyaux, Magnusson, Massart, McNally, Pappenberger, and Rabier]{Rosnay22}
Patricia de~Rosnay, Philip Browne, Eric de~Boisséson, David Fairbairn, Yoichi Hirahara, Kenta Ochi, Dinand Schepers, Peter Weston, Hao Zuo, Magdalena Alonso-Balmaseda, Gianpaolo Balsamo, Massimo Bonavita, Niels Borman, Andy Brown, Marcin Chrust, Mohamed Dahoui, Giovanna Chiara, Stephen English, Alan Geer, Sean Healy, Hans Hersbach, Patrick Laloyaux, Linus Magnusson, Sébastien Massart, Anthony McNally, Florian Pappenberger, and Florence Rabier.
\newblock Coupled data assimilation at {ECMWF}: current status, challenges and future developments.
\newblock \emph{Quarterly Journal of the Royal Meteorological Society}, 148\penalty0 (747):\penalty0 2672--2702, 2022.
\newblock \doi{https://doi.org/10.1002/qj.4330}.

\bibitem[Browne et~al.(2023)Browne, de~Rosnay, McNally, Massart, Semane, Healy, Anesiadou, Geer, and Scanlon]{browne2024interface}
Phil Browne, Patricia de~Rosnay, Tony McNally, Sebastien Massart, Noureddine Semane, Sean Healy, Katerina Anesiadou, Alan Geer, and Tracy Scanlon.
\newblock Exploiting interface observations in a coupled reanalysis system.
\newblock In \emph{ECMWF {A}nnual {S}eminar}, 2023.
\newblock URL \url{https://ecmwfevents.com/assets/presentations/as2023-browne1694175680.pdf}.

\bibitem[Diamond et~al.(2023)Diamond, Schreck, Allgood, Becker, Blake, Bringas, Camargo, Chen, Coelho, Fauchereau, Fogarty, Goldenberg, Goni, Harnos, He, Hu, Klotzbach, Knaff, Kumar, L’Heureux, Landsea, Lin, Lorrey, Luo, Magee, Pasch, Pezza, Rosencrans, Rozkošný, Trewin, Truchelut, Wang, Wang, and Wood]{osti_10492534}
Howard~J. Diamond, Carl~J. Schreck, Adam Allgood, Emily~J. Becker, Eric~S. Blake, Francis~G. Bringas, Suzana~J. Camargo, Lin Chen, Caio~A.S. Coelho, Nicolas Fauchereau, Chris Fogarty, Stanley~B. Goldenberg, Gustavo Goni, Daniel~S. Harnos, Qiong He, Zeng-Zhen Hu, Philip~J. Klotzbach, John~A. Knaff, Arun Kumar, Michelle L’Heureux, Chris~W. Landsea, I-I. Lin, Andrew~M. Lorrey, Jing-Jia Luo, Andrew~D. Magee, Richard~J. Pasch, Alexandre~B. Pezza, Matthew Rosencrans, Jozef Rozkošný, Blair~C. Trewin, Ryan~E. Truchelut, Bin Wang, Hui Wang, and Kimberly~M. Wood.
\newblock State of the climate in 2022. the {T}ropics.
\newblock \emph{Bulletin of the {A}merican {M}eteorological {S}ociety}, 104\penalty0 (9), 2023.
\newblock \doi{10.1175/BAMS-D-23-0078.1}.

\bibitem[Weston et~al.(2014)Weston, Bell, and Eyre]{weston2014obserrors}
P.~P. Weston, W.~Bell, and J.~R. Eyre.
\newblock Accounting for correlated error in the assimilation of high-resolution sounder data.
\newblock \emph{Quarterly Journal of the Royal Meteorological Society}, 140\penalty0 (685):\penalty0 2420--2429, 2014.
\newblock \doi{https://doi.org/10.1002/qj.2306}.

\end{thebibliography}
 
\end{document}